\documentclass[12pt]{article}
\usepackage{mymacros}

\title{{{{\rm Building the Holographic Dictionary of the DSSYK from Chords, Complexity \& Wormholes with Matter}}}}

\author[a]{Sergio E. Aguilar-Gutierrez}
\affiliation[a]{Qubits and Spacetime Unit, Okinawa Institute of Science and Technology Graduate University (\begin{CJK}{UTF8}{min}沖縄科学技術大学院大学\end{CJK}), 1919-1 Tancha, Onna, Okinawa 904 0495, Japan}
\emailAdd{sergio.ernesto.aguilar@gmail.com} 
\abstract{{{In this work, we formulate the holographic dictionary for the double-scaled SYK (DSSYK) model with matter operators.}} Based on the {{two-sided}} Hartle-Hawking (HH) state, we derive several properties of the DSSYK model, without making assumptions about the specific dual theory, including its semiclassical thermodynamics, correlation functions, and Krylov complexity. {{We derive}} these quantities from the saddle points of the DSSYK path integral preparing the HH state. We also construct a Lanczos algorithm that simultaneously evaluates Krylov state and operator complexity in the two-sided Hamiltonian system including finite temperature effects. In the semiclassical limit, both measures are encoded in the saddle points of the path integral. They have a bulk interpretation in terms of minimal geodesic lengths in an effective AdS$_2$ space with matter backreaction. Different saddle points correspond to geodesic distances with different evolution, and they display different scrambling properties. {{We also discuss about the quantization of}} the bulk theory dual to the DSSYK model. {{At last, we formulate the double-scaled algebra \cite{Lin:2022rbf} in bulk terms,}} and the dual entry of the proper radial momentum of a bulk probe.}
\begin{document}

\maketitle

\section{Introduction}\label{sec:intro}
\paragraph{Motivation}
Bulk matter fields/ dual boundary operators are responsible for rich dynamics in holographic dictionary. These are indispensable components to build interesting models that resemble some aspects of our universe. It is thus important to thoroughly study dynamical observables that encode how bulk matter fields backreact in the geometry. A promising testing ground to carry out a comprehensive study involving bulk matter from a boundary perspective is the double-scaled SYK (DSSYK) model \cite{Berkooz:2018qkz,Berkooz:2018jqr}. It has a remarkable balance between interesting features for quantum gravity, and some degree of simpleness of working in lower dimensions.\footnote{A review of the DSSYK model can be found in App. \ref{sapp:DSSYK review}, and we also refer the reader to \cite{Berkooz:2024lgq} for a modern introduction to this topic.} The model is UV finite and the rules for general correlation functions within the model have been derived by \cite{Berkooz:2018jqr}. Meanwhile, at very low energies, the DSSYK model can describe the Schwarzian mode (as well as certain deformations \cite{Berkooz:2024ifu}) of Jackiw-Teitelboim (JT) \cite{JACKIW1985343,TEITELBOIM198341} gravity; and de-Sitter JT gravity in a similar but opposite regime \cite{Okuyama:2025hsd}. {{This connects to the}} several approaches to holography {{within}} this model, including dS$_3$ space \cite{Susskind:2021esx,Susskind:2022bia,Lin:2022nss,Rahman:2022jsf,Rahman:2023pgt,Rahman:2024iiu,Rahman:2024vyg,Narovlansky:2023lfz,Verlinde:2024znh,HVtalks,Verlinde:2024zrh,Gaiotto:2024kze,Tietto:2025oxn,Milekhin:2023bjv,Yuan:2024utc,Aguilar-Gutierrez:2024nau} (see also \cite{Sekino:2025bsc,Miyashita:2025rpt}) which depends on gauge conditions, the physical operators and energy sectors of the DSSYK spectrum; and sine dilaton gravity \cite{Blommaert:2023opb,Blommaert:2024whf,Blommaert:2024ymv,Blommaert:2025avl}.\footnote{In \cite{Aguilar-Gutierrez:2025hty} we explain how these approaches are connected to each other building up on this work.} It is expected in all of them that the so-called matter chords in this model play the role of matter fields in the bulk dual. However, while there has been great progress in deriving the holographic dictionary of the DSSYK model without matter in these mentioned approaches, there is currently no study on how the double-scaled algebra \cite{Lin:2022rbf} enters in their proposals.

\paragraph{The double-scaled algebra} As originally proposed in \cite{Lin:2022rbf}, and later developed in \cite{Xu:2024hoc}, the double-scaled algebra is a type II$_1$ Von Neumann algebra of the DSSYK model defined by strings of operators (i.e. a finite linear span involving a polynomial of the generators \cite{Xu:2024hoc}) of the form
\begin{equation}\label{eq:operators DS algebra}
    \mathcal{A}_{\rm DS}=\expval{\hH_{L/R},~\hat{\mathcal{O}}^{L/R}_{\Delta}}''~,
\end{equation}
where $\expval{\cdot,\cdot}$ indicates an algebra of bounded operators, which in this case includes (i) the two-sided DSSYK Hamiltonians $\hH_{L/R}$ (\ref{eq:Hamiltonians LR}), and (ii) matter chord operators $\hat{\mathcal{O}}^{L/R}_{\Delta}$ (\ref{eq:matter ops DSSYK}) with a conformal weight $\Delta$, and $''$ indicates the {{double commutant}} of the algebra.

In order to identify the dual bulk theory of the DSSYK model, it is crucial to formulate the holographic dictionary for the boundary operators (\ref{eq:operators DS algebra}) in terms of the bulk-to-boundary map {{in}} \cite{Lin:2022rbf}. For this reason, we need to deduce what quantum observables encode the dynamics of the dual bulk geometry with matter fields. For instance, as noticed in \cite{Berkooz:2022fso}, two-sided two-point functions are useful dynamical probes from the boundary side that encode length in the bulk with matter fields. However, we want to emphasize on its connection with holographic complexity (treated from a boundary perspective) to derive the holographic dictionary. 

\paragraph{Krylov complexity}We will work with a boundary measure of bulk geodesics lengths, Krylov complexity \cite{Balasubramanian:2022tpr,Parker:2018yvk} (which we review in App. \ref{sapp:Krylov review}). It has been shown that a wormhole (i.e. an Einstein-Rosen bridge) geodesic distance in an AdS$_2$ black hole background matches with Krylov complexity for the thermofield double (TFD) state at infinite temperature of the DSSYK model \cite{Rabinovici:2023yex}. The authors make a connection of their results with JT gravity, which is recovered in the very low temperature regime  {{in the canonical ensemble (corresponding to very low-energies)}} instead. This is an interesting observation. 
\textit{How can the infinite temperature TFD of the DSSYK model describe a geodesic length in JT gravity?}\footnote{We thank Adrián Sánchez-Garrido for comments on this.} Given that the triple-scaling limit involves a shift $\ell\rightarrow\ell+2\log\lambda$, where $\ell\sim\mathcal{O}(1)$ while $\lambda\rightarrow0$, the renormalized bulk length in the JT gravity (i.e. subtracting $2\log\lambda$) should also be reproduced by the bulk theory dual to the DSSYK (i.e. without a triple-scaling limit). In the sine dilaton gravity, \cite{Blommaert:2023opb,Blommaert:2024whf,Blommaert:2024ymv,Blommaert:2025avl,Bossi:2024ffa}, there is an effective AdS$_2$ geometry, valid even at infinite temperatures. Krylov complexity is a natural measure of distance in the dual effective geometry, as observed in \cite{Heller:2024ldz}.

There have been several other relevant developments. The properties of the DSSYK model with particle insertions have been better understood \cite{Lin:2022rbf,Lin:2023trc,Xu:2024hoc,Xu:2024gfm}. Krylov complexity for states and operators in the DSSYK model with and without one-particle insertions has been investigated in \cite{Xu:2024gfm,Ambrosini:2024sre}. \cite{Xu:2024gfm} derived several dynamical observables relating the zero and one-particle chord space of the DSSYK model. Shortly after, an explicit Krylov basis and the Lanczos coefficients for one-particle chord states were derived by \cite{Ambrosini:2024sre}. The authors showed that the total chord number with a one chord matter operator insertion {{determines the}} Krylov complexity in the semiclassical limit. {{This limit is defined as \cite{Lin:2022rbf}
\begin{equation}\label{eq:semiclassical}
    \lambda\rightarrow0^+~,\quad \text{and}\quad\lambda ~n~:~~\text{fixed}~, 
\end{equation}
where $\lambda>0$ is a parameter of the theory which measures quantum effects (as the would-be Planck's constant of this model or its bulk dual \cite{Blommaert:2024whf}), $0^+$ indicates that $\lambda\rightarrow0$ from above; while $n\in\mathbb{Z}_{\geq0}$ is a quantum number, called the chord number, labeling states in the Hilbert space (reviewed in App. \ref{sapp:DSSYK review}). Moreover, the combination $\lambda n$ is interpreted as a wormhole length in the bulk dual theory \cite{Lin:2022rbf}. This means that the semiclassical limit can be interpreted as a regime where quantum corrections are suppressed while bulk wormhole lengths take finite values.} Furthermore, if there are matter operators, with conformal dimension $\Delta$. We say that the corresponding operator is \emph{heavy} if $\Delta\lambda$ is a fixed number when \eqref{eq:semiclassical}, or it is a \emph{light} operator when $\Delta_{\rm S}\sim\mathcal{O}(1)$.} In this limit {{the distribution of}} chords gets very dense, providing an emergent bulk geometry \cite{Berkooz:2022fso}. This fact allowed \cite{Ambrosini:2024sre} to evaluate the Krylov complexity for the matter chord operator, and for a TFD state with a matter chord insertion. Soon afterwards, it was reported in \cite{Heller:2024ldz} that, in the case without matter, one could interpret spread complexity as a wormhole distance in an (effective) AdS$_2$ spacetime prepared in the Hartle-Hawking (HH) state at finite temperature. It has also been realized and stressed by \cite{Heller:2024ldz} that spread complexity is a natural candidate dual to the complexity equals volume (CV) proposal \cite{Susskind:2014rva,Stanford:2014jda} (at least in this model) if the switchback effect is also satisfied. While it is known that Krylov complexity for states and operators for particular states of the DSSYK model experience late-time linear growth \cite{Rabinovici:2023yex,Ambrosini:2024sre,Xu:2024gfm}, it remains to be seen if the switchback effect can also be realized by Krylov complexity. Therefore, Krylov complexity might be promoted to a full-fledged entry in the holographic dictionary of the DSSYK model. We address this in \cite{Aguilar-Gutierrez:2025mxf}.

Other relevant developments in Krylov space methods for the DSSYK model can be found in \cite{Bhattacharjee:2022ave,Aguilar-Gutierrez:2024nau,Anegawa:2024yia,Nandy:2024zcd,Balasubramanian:2024lqk,Miyaji:2025ucp}.

To summarize, we are motivated to compare the following observables to identify the holographic dual of the DSSYK and its holographic dictionary:
\begin{itemize}
    \item \emph{On the boundary side}: There is no unique definition of quantum complexity that should be used. Krylov complexity is a valid candidate. Since it is defined in a gauge invariant matter, it should have a geometric description in the holographic dictionary.
    \item \emph{On the bulk side}: Previous works indicate that geodesic lengths connecting the asymptotic boundaries in an AdS black hole geometry can be matched to Krylov complexity \cite{Rabinovici:2023yex,Xu:2024gfm,Heller:2024ldz}. We will show that the boundary-to-bulk map of the DSSYK \cite{Lin:2022rbf} is manifested in terms of Krylov complexity.
\end{itemize}
\paragraph{Double-scaled PETS} To investigate the holographic dictionary, we need to prepare states with operator insertions of the double-scaled algebra \cite{Lin:2022rbf} on the unique tracial state \cite{Xu:2024hoc}.\footnote{In Fig. \ref{fig:composite_op} (a) we illustrate the chord diagram when all the particles are inserted at the same location, which we refer to a composite operator, and also an example of a non-composite matter chord diagram with many particles is shown Fig. \ref{fig:composite_op} (b).} This results in a two-sided Hamiltonian system, with a temperature associated to each side due to modification of the would-be TFD state. In other words, we construct a double-scaled version of the \emph{partially entangled thermal states} (PETS), originally defined for the finite N SYK model in \cite{Goel:2018ubv}.\footnote{Our notion of double-scaled SYK PETS is different {{but related to}} \cite{Goel:2023svz,Berkooz:2022fso}. We define the PETS on the chord Hilbert space, instead of constructing the PETS from the energy basis of a finite $N$ SYK and taking its double-scaled limit. The two-sided correlation functions in these procedures are equivalent (see Sec. \ref{ssec:twopoints}).} The relative temperature difference corresponds to the angle of insertion of the operator along the thermal circle in the chord diagram. 
These types of states in the one-particle space have appeared in a recent study \cite{Xu:2024gfm}. Our work instead explores how the light or heavy chord operators can be used to derive explicit observables that encode the dual minimal length geodesics.

Our analysis of the properties of the two-sided HH PETS (shown explicitly in (\ref{eq:HH state tL tR})) is also motivated by {{the following}} factors.
\begin{itemize}
    \item The two-sided HH state contains information about the conformal dimensions and the evolution of matter chord operators acting on the unique tracial state of the DSSYK model. This makes it a natural reference state to study dynamical observables at finite temperature. 
    \item The classical phase space variables in the DSSYK model in the HH state is closely related with geodesic lengths in the semiclassical limit. The phase space analysis of the DSSYK model has allowed much progress in the bulk dual proposals \cite{Narovlansky:2023lfz,Blommaert:2024ymv,Blommaert:2024whf}. 
    \end{itemize}

\subsection{Questions and findings}
{{Our main question is
\begin{quote}
\emph{How do we incorporate matter operators in the holographic dictionary of the DSSYK model, and in the bulk dual of the double-scaled algebra?}
\end{quote}
We deduced the holographic dictionary with matter (see Tab. \ref{tab:holographic_dictionary}) by incorporating bulk free scalar fields backreacting in an effective AdS$_2$ black hole geometry. Based on the holographic dictionary, we proposed how to formulate (\ref{eq:operators DS algebra}) in the terms of bulk operators with a one-particle insertion}}
\begin{equation}\label{eq:bulk algebra}
    \expval{\hH^{(\rm ADM)}_{L/R},~\hat{\xi}^{L/R}_{\mathfrak{R}}}''~,
\end{equation}
where $\hH^{(\rm ADM)}_{L/R}$ are ADM Hamiltonians, $\hat{\xi}^{L/R}_{\mathfrak{R}}(x)$ are quantized minimally coupled scalar fields. {{Here the subindex $(\mathfrak{R})$ is meant to indicate that the fields are dressed with respect to a reference frame $\mathfrak{R}$ (so that $\hat{\xi}^{L/R}_{\mathfrak{R}}$ are diffeomorphism invariant operators), corresponding to either of the asymptotic boundaries}} in a two-sided AdS$_2$ effective background (\ref{eq:effective geometry}){{. We elaborate more about this in Sec. \ref{ssec:canonical quantization}}}. The bulk theory that reproduces these results obeys certain constraints {{(as in more general settings, e.g. \cite{Brown:1994py,Giesel:2007wi,Kuchar:1993ne,Lapchinsky:1979fd})}}. Namely, as observed in the original works \cite{Blommaert:2024whf,Blommaert:2024ymv} (without matter), the bulk states obey a Schrödinger equation, whose energy spectrum coincides with those of the boundary theory, and {{a momentum shift symmetry. We explain how to incorporate the constraints in the canonical quantization of the bulk theory to}} reproduce a two-sided Hamiltonian with matter {{dual to \eqref{eq:pair DSSYK Hamiltonians 1 particle}. Other constraints in the chord Hilbert space results in other bulk duals \cite{Aguilar-Gutierrez:2025hty}. However, our work in this manuscript focuses mostly on the boundary observables, there are technical steps in the bulk evaluation that need further elaboration, as we describe in Sec. \ref{ssec:canonical quantization} to describe a precise correspondence between the bulk and the DSSYK model with matter chords beyond the semiclassical limit, including their algebras of observables}}.

\paragraph{Krylov complexity as a probe of bulk/boundary dynamics}
As highlighted above, there is not a sufficient understanding about the Krylov complexity with matter chords in the literature, especially when considering more than one-particle chords, in a finite temperature ensemble, and beyond the triple-scaling limit. For this reason, we will deduce a Lanczos algorithm for two-sided Hamiltonians that improves on these aspects.\footnote{This algorithm evaluates Krylov complexity for one-particle irreducible representations (irreps.), which means that while it works for arbitrary many matter operators, they need to inserted in the same location in the thermal circle, as illustrated in Fig. \ref{fig:composite_op}.} The reference state. It then natural to ask:
\begin{quote}
    \emph{What properties of double-scaled matter operators are encoded in the evolution of Krylov complexity for a the two-sided HH state? What is its bulk interpretation? How does it compare with holographic complexity conjectures in other settings?}
\end{quote}
The conformal dimensions and the operator insertion times are encoded in the Krylov operator or spread complexity of the two-sided HH state. By {{deriving}} the saddle point solutions {{of the path integral preparing the two-sided HH state, we recover the semiclassical}} Krylov complexity {{as the}} minimal length geodesic in an AdS$_2$ black hole with massive particles or shockwave insertions, {{as well as}} the holographic dictionary {{shown in Tab. \ref{tab:holographic_dictionary}}}. The Krylov operator and spread complexity correspond to geodesic lengths with different boundary time evolution. Krylov operator complexity displays several features (such as scrambling of information, hyperfast growth, and it obeys certain bounds) that have appeared in other contexts in the literature, e.g. \cite{Susskind:2021esx,Milekhin:2024vbb,Susskind:2014rva}.

Lastly, we also connect our results to recent a proposal relating the proper momentum of a probe particle in asymptotically AdS geometries with the time derivative of Krylov complexity \cite{Caputa:2024sux,Caputa:2025dep}.\footnote{See other works relating proper radial momentum with Krylov complexity \cite{Fan:2024iop,He:2024pox} and holographic complexity in, e.g. \cite{Susskind:2018tei,Susskind:2019ddc,Susskind:2020gnl,Jian:2020qpp,Barbon:2019tuq,Barbon:2020olv,Barbon:2020uux}.} We show that the radial proper momentum {{(defined in Sec. \ref{ssec:proper momentum})}} corresponds to the growth of Krylov operator complexity. The holographic dictionary is consistent with our findings in Tab. \ref{tab:holographic_dictionary}. The answer then indicates a {{direct connection}} between the wormhole distance (encoded in Krylov operator complexity) and the radial proper momentum of a probe particle.

\paragraph{Path integral, Krylov space and chord diagram approaches}
The path integral approach is useful to prepare the two-sided HH state. However, the exact way to make the path integral preparation of state is not clearly addressed for two-sided Hamiltonian systems, where one has more freedom in how to generate the Hamiltonian evolution. For this reason,
\begin{quote}
    \emph{How should one systematize the procedure to uniquely determine the path integral that prepares a given state in the two-sided Hamiltonian system with matter?}
\end{quote}
We will use the Heisenberg picture to formalize this point above and derive what conditions one needs to impose in the saddle point solution preparing the HH state. The saddle point solution for the total chord number is equivalent to the semiclassical Krylov complexity of the HH state. As for the chord diagram perspective, it has been previously realized that two-sided two-point functions in the DSSYK model (corresponding to crossed four-point functions) also reproduce bulk geodesics in shockwave AdS$_2$ black hole geometries \cite{Berkooz:2022fso}. However, these developments are also limited to the triple-scaling limit and have a different formulation where there is no clear connection with the path integral nor Krylov complexity results in the more recent literature.
\begin{quote}
    \emph{Do these different approaches, namely Krylov complexity, path integral and chord diagrams, encode equivalent information about the bulk dual geometry once we incorporate matter?}
\end{quote}
We find that all these methods encode a minimal length geodesic in the bulk. This is not unexpected since the path integral and Krylov complexity approaches relay on chord diagram Hamiltonian \cite{Lin:2022rbf}. However, the way that this information is encoded in each measure is in principle different. While all these approaches reproduce the same minimal geodesic length in semiclassical limit; the relationship between these methods is more involved away from this regime. For instance, as previously reported in \cite{Ambrosini:2024sre}, and as we also show under more general considerations in Sec. \ref{sec:deriving Krylov} and App. \ref{app:details Lanczos Krylov}, the total chord number is equivalent to Krylov complexity operator only in the $\lambda\rightarrow0$ limit. Meanwhile, as reported in \cite{Xu:2024gfm}, in general the Krylov basis is a function of the total chord number, even away from the semiclassical limit. Nevertheless, each method offers different technical and conceptual advantages.

\subsection{Summary}\label{ssec:summary}
A table summarizing the holographic dictionary of the DSSYK with matter chords and a compatible bulk dual theory (sine dilaton gravity \cite{Blommaert:2023opb,Blommaert:2024whf,Blommaert:2024ymv,Blommaert:2025avl} with matter) is displayed in Table \ref{tab:holographic_dictionary}.

\begin{table}[t!]
    \centering
    \begin{tabular}{|C{7.5cm}|C{7.5cm}|}
    \hline\textbf{Boundary (DSSYK)}&\textbf{Bulk (AdS$_2$ black hole)}\vspace{0.2cm}\\\hline
    \textbf{\footnotesize Two-sided HH state with an operator insertion,} $\ket{\Psi_\Delta}$ (\ref{eq:HH state tL tR})&\textbf{\footnotesize Black hole in the HH state with a particle insertion} (\ref{eq:effective geometry})\vspace{0.2cm}\\
    \footnotesize{Hamiltonian time} $t_{L/R}$ & \footnotesize{Asymptotic boundary time} $u_{L/R}$\\
      {\footnotesize Rescaled total chord number} &{\footnotesize Geodesic distance between the asymptotic}\\
     $\lambda~\expval{\hat{N}(t_L,t_R)}$ \eqref{eq:womrhole semiclassical answer many} &{\footnotesize boundaries, $L_{\rm AdS}(u_L,-u_R)$ (\ref{eq:equality propagators})}\vspace{0.2cm}\\
     \footnotesize{Inverse fake temperature} $\beta_{\rm fake}^{L/R}$ (\ref{eq:fake temperature result})&\footnotesize{Inverse bulk temperature} $\beta_{\rm AdS}$\vspace{0.2cm}\\\hline
     \multicolumn{2}{|c|}{Case 1: massive scalar (\ref{eq:holographic dictionary massive particle})}\\\hline $4\pi\frac{\sin^2\theta+\eta q^{\Delta}\rme^{-\ell^{(-)}_*(\theta)}}{\sin^2\theta-\eta q^{\Delta}\rme^{-\ell^{(-)}_*(\theta)}}$&\footnotesize{Backreaction parameter} $\frac{8\pi G_N L_b}{\Phi_b}m_\xi$\vspace{0.2cm}\\
    $\ell^{(-)}_*(\theta)+2\log\frac{\sin^2\theta-\eta q^{\Delta}\rme^{-\ell^{(-)}_*(\theta)}}{2\sin^2\theta}$ &\footnotesize{Regulator length} $L_{\rm reg}-\frac{8 G_Nm_\xi L_b}{\Phi_b}$\vspace{0.2cm}\\\hline
    
     \multicolumn{2}{|c|}{Case 2: single shockwave \eqref{eq:holographic dictionary shockwave}}\\\hline 
     \footnotesize{Shift parameter} $\alpha_{\rm sw}$ (\ref{eq:shockwave no approx})&$\frac{1}{2}\qty(1-\frac{q^{\Delta_{\rm S}}\rme^{-\ell_*^{(+)}}}{\sin^2\theta})\rme^{2J\sin\theta~t_w}$\\
     \footnotesize{Regularization length} $L_{\rm reg}$&$\ell^{(+)}_*(\theta)$\\\hline
    \end{tabular}
    \caption{Holographic dictionary obtained from the match of Krylov complexity in the DSSYK with matter chords and wormhole lengths in an AdS$_2$ black hole geometry. {{$\ell^{(+)}_*(\theta)$ and $\ell^{(-)}_*(\theta)$ are}} the initial total chord number {{in (\ref{eq:ell * state}) and (\ref{eq:ell * operator}) respectively}}; $L_b/\Phi_b$ a regularized ratio between bulk parameters; $t_w$ the insertion time of the precursor operator $\hat{\mathcal{O}}_{\Delta} (t_w)$. App \ref{app:notation} provides a detailed overview of the notation implemented in the table.}
    \label{tab:holographic_dictionary}
\end{table}
In this work, we first introduce the concept of a double-scaled PETS as a natural generalization of the HH state in the chord Hilbert space DSSYK model with arbitrarily many particles. This provides valuable information for defining dynamical observables in finite temperature ensembles.\footnote{The saddle point solutions in our work are evaluated in a microcanonical ensemble. However, the temperature of the saddle point solutions is completely determined by the energy spectrum in this model, which allows us to work with both the canonical and microcanonical ensemble at once.} 
We construct the partition function, and thermal two-sided two-point correlation functions by taking expectation values with respect to the two-sided HH state. Both are evaluated using chord diagrams in the App. \ref{app:partition1p}, \ref{app:twopoint oneparticle}. We will then derive the path integral that prepares the two-sided HH state (deriving the equations of motion (EOM) and the initial conditions for the canonical variables from the Heisenberg picture). We then solve for the saddle points of the path integral, and the evolution of the total chord number in the HH state. From this calculation, one can easily deduce thermal two-point correlation functions for heavy or light operators in the semiclassical limit. 

We will then define a double-scaled version of precursor operators (which originate from circuit complexity and the CV conjecture \cite{Susskind:2014rva}). This will allow us to study the saddle point solutions in a particular limit where the expectation value of the total chord number evolves in the same way as minimal length shockwave geodesics \cite{Shenker:2013pqa}.

Later, we construct the Krylov basis of the DSSYK model with the two-sided HH state as reference in the Lanczos algorithm for two-sided Hamiltonians. The key advantage of this formulation is that we can evaluate both Krylov operator or spread complexity at once and including finite temperatures. We show that the total chord number is the Krylov complexity operator in the semiclassical limit (using similar arguments as \cite{Ambrosini:2024sre}, while allowing for finite temperature effects). This also implies that the thermal two-point function (\ref{eq:generating function 1 particle}) takes the role of the generating function of Krylov complexity (and higher moments) in the semiclassical limit. Furthermore, both types of Krylov complexity correspond to geodesic lengths in the bulk with different relative {{boundary}} times due to the two-sided Hamiltonian evolution. Our results reproduce those in the literature as special cases. (i) Taking an infinite temperature limit \cite{Ambrosini:2024sre}, and (ii) considering a vanishing conformal weight of the matter operator \cite{Heller:2024ldz}. 
 
Importantly, the Krylov operator complexity displays a transition from exponential to linear growth. The results show that the scrambling time in Krylov operator complexity corresponds to the time scale of decay in the corresponding out-of-time-ordered correlator (OTOC). 
The latter is a different and more commonly used definition of scrambling time \cite{Xu:2022vko}. We find that it does not depend on the system size, which is a property we will refer to as hyperfast growth \cite{Anegawa:2024yia}, in contrast with the bound on the fastest information scrambling in \cite{Sekino:2008he} (which is proportional to an exponential of the system's entropy). The OTOC obeys general bounds on fermionic systems \cite{Milekhin:2024vbb}, and its exponent satisfies the chaos bound \cite{Maldacena:2015waa}.

Our results on thermal two-sided two-point functions and Krylov complexity, and more details on the holographic dictionary are summarized in App. \ref{app:results}.

Later on, we elaborate on the bulk description of our results. We first study JT and sine dilaton gravity with a minimally coupled scalar field in an (effective) AdS$_2$ black hole background. We show that a wormhole geodesic distance matches with Krylov operator complexity of the two-sided HH state when the backreaction in the bulk is perturbatively small. Then, we take a {{different approach using}} shockwaves {{(see e.g. \cite{Shenker:2013pqa,Shenker:2013yza})}}. We match our results on classical phase space from the path integral with precursor operator insertions to geodesic lengths in an AdS$_2$ shockwave geometry \cite{Shenker:2013pqa}. From our results, {{we present implications for the}} quantization of the bulk theory (sine dilaton gravity) and {{the}} holographic description of the double-scaled algebra.

Thus, since Krylov complexity measures a wormhole in the bulk (i.e. one of the classical phase space variables in JT and sine dilaton gravity), this is a crucial element in the holographic dictionary of the DSSYK model. We hope this reveals useful lessons in more generic systems, even if Krylov is not always an appropriate measure of quantum chaos. We also stress that \emph{we do not assume a holographic correspondence between a specific bulk theory with the DSSYK model}, we simply show that our boundary results are compatible with sine dilaton gravity with matter. We later canonically quantize the bulk theory to show that it is indeed dual to the DSSYK model with one-particle, and to {{identify a}} dual double-scaled algebra.

Other technical results are included in the appendices. In particular, we derive the semiclassical thermodynamics of the system from the partition function, to determine the thermal stability of the system. We also evaluate two-sided two-point (i.e. crossed four) correlation functions from chord diagrams. This serves as a consistency check of our results from classical phase space methods. The evaluation involves a new R-matrix calculation in the DSSYK model (also seen as a $6j$-symbol of $\mathcal{U}_{q}(\text{su}(1,1))$ \cite{Berkooz:2018jqr,Blommaert:2023opb,Belaey:2025ijg}) in the semiclassical limit. Some of the technical steps in the calculation follow similarly from evaluating OTOCs in JT gravity \cite{Mertens:2017mtv,Lam:2018pvp,Mertens:2022irh}. However, while the phase space answer is valid for heavy and light particles, the chord diagram calculation {{in App. \ref{app:twopoint oneparticle}}} uses an ansatz valid only for light operators. Due to this technical limitation we have not crossed verified the calculation for the heavy operator case from the chord diagram methods. Nevertheless, we find agreement in the expected regime of validity with our path integral results.

The reader interested in a \emph{quick overview} of the \emph{main results} may check App. \ref{app:results} (complemented by a summary of the notation in App. \ref{app:notation}).

\paragraph{Plan of the manuscript}In \textbf{Sec}. \ref{sec:saddle points}, we define the two-sided HH state (\ref{eq:HH state tL tR}) in the one-particle chord space, we study the path integral preparing this state, and its saddle points. In \textbf{Sec}. \ref{sec:deriving Krylov} we derive Krylov operator and spread complexity in the irrep. one-particle chord space from a Lanczos algorithm for two-sided Hamiltonians with the two-sided, finite temperature HH state as reference state. We also discuss specifically about Krylov operator and spread complexity. In \textbf{Sec}. \ref{sec:bulk}, we study bulk geodesics with backreaction, which we match to the Krylov operator and spread complexity to derive the holographic dictionary, {{and its implications for}} the canonical quantization of the bulk theory, {{as well as to express the}} double-scaled algebra in bulk terms. We also comment on the proper momentum of a point particle in an AdS$_2$ black hole. \textbf{Sec}. \ref{sec:dis} concludes with a discussion and an outlook.

For the ease of reading, \textbf{App}. \ref{app:notation} includes a list of the notation (including acronyms) used in the manuscript. \textbf{App}. \ref{app:results} is a list of the main outcomes from our study. To keep the presentation self-contained, \textbf{App}. \ref{sec:background material} reviews useful results on (i) the DSSYK model (\textbf{App}. \ref{sapp:DSSYK review}), and Krylov complexity for states and operators (\textbf{App}. \ref{sapp:Krylov review}). Then, \textbf{App}. \ref{app:isometric map} contains details about the isometric linear map between one- and zero-particle states, which we use in order to carry out evaluations in later sections, and to prove (\ref{eq:initial velocity}) in the main text. Meanwhile, \textbf{App}. \ref{app:partition1p} contains new results regarding the semiclassical thermodynamics in the two-sided HH state, including derived quantities in the (micro)canonical ensemble (i.e. temperature, entropy, and heat capacity). In \textbf{App}. \ref{app:twopoint oneparticle} we calculate the thermal two-point correlation function in the one-particle chord space (a crossed four-point function in $\mathcal{H}^{L}_0\otimes\mathcal{H}_0^{R}$, which we illustrate in Fig. \ref{fig:OTOC_chord_diagram}).
Meanwhile, in \textbf{App}. \ref{app:properties OTOC} we analyze the properties of the semiclassical OTOC in (\ref{eq:twosided twopoint OTOC}). In \textbf{App}. \ref{app:conversion}, we discuss a criterion to determine if states in $\mathcal{H}_m$ have a smooth geometry in the bulk dual theory \cite{Berkooz:2022fso}. \textbf{App.} \ref{app:details Lanczos Krylov} contains technical details about the Lanczos coefficients and Krylov basis. At last, \textbf{App.} \ref{app:more general states} shows how to decompose more general states in $\mathcal{H}_m$ into $\mathcal{H}_0$ and $\mathcal{H}_1$ irreps.

\section{The path integral of the DSSYK with matter chords}\label{sec:saddle points}
In this section, we explain how to prepare the two-sided HH state from the path integral of the DSSYK, and we study its saddle point solutions. 

\paragraph{Outline}In Sec. \ref{ssec:PETS from DS} we explain basic properties about the two-sided HH state, its partition function (discussed in detail in App. \ref{app:partition1p}), and how to simplify expressions with composite chord matter operators. In Sec. \ref{ssec:many chords} we discuss the path integral that prepares the two-sided HH state. We first derive the Heisenberg equations for canonical operators in the two-sided theory, and the initial conditions for their expectation values. This allows us to determine the correct path integral for the two-sided HH state. The real part of the complexified time is related to the temperature of the system, and the imaginary part with real-time evolution.
In Sec. \ref{ssec:twopoints} we study two-sided two-point correlation functions using the classical phase space results in the path integral. In Sec. \ref{ssec:shockwaves} we point out significant modifications in the state preparation by allowing for time dependence in the Schrödinger picture operators (known as precursors \cite{Susskind:2014rva}) which are inserted on the two-sided HH state. 

\subsection{Partially entangled thermal states from double-scaled operators}\label{ssec:PETS from DS}
To set the stage, we first explain how to build states on the chord Hilbert space, $\hat{\mathcal{H}}_m$ ($m$ labels the total number of particle chords) where the algebra (\ref{eq:operators DS algebra}) closes. The Hilbert space $\mathcal{H}_m$ takes the form \cite{Lin:2022rbf}
\begin{equation}\label{eq:states notation matter}
\mathcal{H}_m=\bigoplus_{n_0,n_1,\dots,n_m=0}^\infty\mathbb{C}\ket{\tilde{\Delta};n_0,n_1,\dots,n_m}~,
\end{equation}
where $\tilde{\Delta}=\qty{\Delta_1,\cdots,\Delta_m}$, represents a string of matter operator insertions
\begin{equation}\label{eq:reference pets}
\hat{\mathcal{O}}_{\tilde{\Delta}}=\qty{\hat{\mathcal{O}}_{\Delta_{1}},\dots,~\hat{\mathcal{O}}_{\Delta_{m}}}~,  \quad m\in\mathbb{Z}_{\geq0}~;
\end{equation}
while $n_0$ is the number of DSSYK chords (called H-chords) to the left of all matter chords, $n_1$ the number between the first two particles; all the way up to the number of chords between all the $m$ particles. We review this in more detail in App. \ref{sapp:DSSYK review}. As shown in \cite{Xu:2024hoc}, one can generate any element in the DSSYK Hilbert space $\mathcal{H}_m$ by acting with a string of bounded double-scaled operators in (\ref{eq:operators DS algebra}) with appropriate operator ordering on the unique \emph{tracial state} $\ket{0}$.

The evolution of states in $\mathcal{H}_m$ is generated by the DSSYK two-sided Hamiltonian, which was originally constructed in \cite{Lin:2022rbf} in terms of creation ($\hat{a}_i^\dagger$) and annihilation ($\hat{\alpha}_i$) operators as
\begin{subequations}\label{eq:Hmultiple}
    \begin{align}
\hH_L&=\frac{J}{\sqrt{\lambda}}\qty(\hat{a}^\dagger_0+\sum_{i=0}^m\hat{\alpha}_i\qty(\frac{1-q^{\hat{n}_i}}{1-q})q^{\hat{n}_i^<})\quad\text{where}\quad\hat{n}_i^<=\sum_{j=0}^{i-1}\qty(\hat{n}_j+\Delta_{j+1})~,\label{eq:HLmultiple}\\
\hH_R&=\frac{J}{\sqrt{\lambda}}\qty(\hat{a}^\dagger_m+\sum_{i=0}^m\hat{\alpha}_i\qty(\frac{1-q^{\hat{n}_i}}{1-q})q^{\hat{n}_i^>})\quad\text{where}\quad\hat{n}_i^>=\sum_{j={i+1}}^{m}\qty(\hat{n}_j+\Delta_{j})~,\label{eq:HRmultiple}
\end{align}
\end{subequations}
where $J$ is a coupling constant, and $q=\rme^{-\lambda}\in[0,1)$ is a parameter of the model (reviewed in App. \ref{sapp:DSSYK review}). Note that the triple-scaling limit of (\ref{eq:Hmultiple}) has been previously studied in \cite{Lin:2022rbf,Xu:2024hoc}. The action of the operators in $\mathcal{H}_m$ is displayed in (\ref{eq:Fock Hm 1}, \ref{eq:Fock Hm 2}). Note that $\hat{\alpha}_{0/m}$ and $\hat{a}^\dagger_{0/m}$ are not Hermitian conjugate operators of each other with respect to the inner product for in the $\ket{\Delta;n_0,\dots,n_m}$ basis \cite{Lin:2022rbf,Lin:2023trc,Ambrosini:2024sre}. However, the Hamiltonians (\ref{eq:Hamiltonians LR}) can be brought to the same form as the auxiliary Hamiltonian of the DSSYK model with matter chords, see (\ref{eq:two-sided new}).

\paragraph{Double scaled PETS}We will now introduce a double-scaled version of the PETS, which was originally defined for the SYK model in \cite{Goel:2018ubv} (and further studied in \cite{Goel:2023svz,Jian:2020qpp}). In our construction, we insert two or more matter chord operators $\hat{\mathcal{O}}_{\Delta_i}^{L/R}$ of the double-scaled algebra (\ref{eq:operators DS algebra}) (which we review inn App \ref{sapp:DSSYK review}) in the thermal circle of the DSSYK model. We allow for unequal temperatures between the different subregions of the circle, and a two-sided Hamiltonian ($\hH_{L/R}$) evolution in complex time, $\tau\in\mathbb{C}$ (where the imaginary part corresponds to the real time and the real one to the temperature dependence). This takes the form:\footnote{Note that we can displace the last term $\rme^{-\hH_L\tau_m}$ to the front using $(\hH_L-\hH_R)\ket{0}=0$.}
\begin{equation}\label{eq:PETS general precursor}
\begin{aligned}
    \ket{\Psi^{L/R}_{\tilde{\Delta}}(\tau_0,\dots,\tau_m)}=&\rme^{-\hH_{L/R}\tau_0}\hat{\mathcal{O}}^{L/R}_{\Delta_1}\rme^{-\hH_{L/R}\tau_1}\dots\rme^{-\hH_{L/R}\tau_{m-1}}\hat{\mathcal{O}}^{L/R}_{\Delta_m}\rme^{-\hH_{L/R}\tau_{m}}\ket{0}~.
\end{aligned}
\end{equation}
For explicit computations, we will mostly be interested in the special case below.

\paragraph{The two-sided Hartle-Hawking (HH) state}
We define a two-sided HH state in $\mathcal{H}_m$ as a PETS (\ref{eq:PETS general precursor}) where the operators are arranged in form
\begin{equation}\label{eq:HH state tL tR}
\begin{aligned}
    \ket{\Psi_{\tilde{\Delta}}(\tau_L,\tau_R)}&=\rme^{-\tau_L\hH_L-\tau_R\hH_R}\ket{\tilde{\Delta};0,\dots,0}\\
    &=\rme^{-\tau_L\hH_L}\hat{\mathcal{O}}_{\Delta_1}\hat{\mathcal{O}}_{\Delta_2}\cdots \hat{\mathcal{O}}_{\Delta_{m-1}}\hat{\mathcal{O}}_{\Delta_m}\rme^{-\tau_R \hH_R}\ket{0}~,
    \end{aligned}
\end{equation}
where $\tau_{L/R}=\rmi t_{L/R}+\beta_{L/R}/2$. Here, the left and right-sided real time evolution is parametrized by $t_{L/R}$; while $\beta_{L/R}$ represents the left and right inverse temperatures in the thermal circle (see Fig. \ref{fig:one_particle_Euclidean}). 

\begin{figure}
\centering
\subfloat[]{\includegraphics[height= 0.42\textwidth]{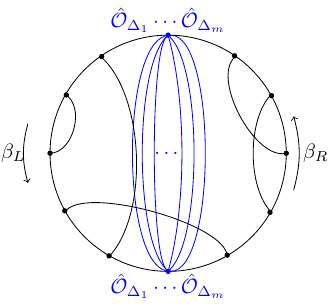}}\hfill\subfloat[]{\includegraphics[height=0.42\textwidth]{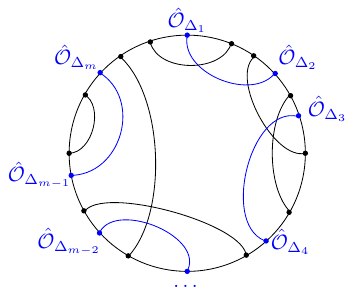}}
\caption{Chord diagram of (\ref{eq:PETS general precursor}) with (a) one composite operator insertion $\prod_{i=1}^m\hat{\mathcal{O}}_{\Delta_i}$ (blue); (b) Several non-composite operators $\hat{\mathcal{O}}_{\Delta_i}$. The thermal circle is divided into two portions with a periodicity associated to the left and right sectors $\beta_L,~\beta_R$ with respect to the operator insertion. The black curves present Hamiltonian (H) chords.}\label{fig:composite_op}
\end{figure}

We stress that the two-sided HH state in (\ref{eq:HH state tL tR}) is a composite operator propagating between two points in the thermal circle of chord diagrams, as displayed in Fig. \ref{fig:composite_op} (a).
 This can be seen from the partition function:
\begin{equation}\label{eq:overall partition function}
\begin{aligned}
Z_{\tilde{\Delta}}(\beta_L,\beta_R)&=    \eval{\bra{\Psi_{\tilde{\Delta}}(\tau_L,\tau_R)}\ket{\Psi_{\tilde{\Delta}}(\tau_L,\tau_R)}}_{\tau_{L/R}=\frac{\beta_{L/R}}{2}}\\
    &=\bra{0}\hat{\mathcal{O}}^{L}_{\Delta_1}\cdots\hat{\mathcal{O}}^{L}_{\Delta_m}\rme^{-\beta_L\hH_L}\hat{\mathcal{O}}^{L}_{\Delta_1}\cdots\hat{\mathcal{O}}^{L}_{\Delta_m}\rme^{-\beta_R\hH_L}\ket{0}~,
\end{aligned}
\end{equation}
where we used the boost invariance: $(\hH_R-\hH_L)\ket{0}=0$; and also $\hat{\mathcal{O}}_\Delta^\dagger=\hat{\mathcal{O}}_\Delta$.

We see from (\ref{eq:overall partition function}) that the total periodicity of the thermal circle must be $\beta_L+\beta_R$. Technical details on the evaluation are presented in App. \ref{app:partition1p}.

One may also study more observables whose expectation value is evaluated in the more general state (\ref{eq:PETS general precursor}) that has a relative phase (which is important for the switchback effect \cite{Aguilar-Gutierrez:2025mxf}). However, this introduces technical difficulties in our analysis, as we explain below. We provide more details on how to treat this case in App. \ref{app:more general states} (see also \cite{Aguilar-Gutierrez:2025mxf}).

To make simplifications in the (\ref{eq:HH state tL tR}) we begin showing that
\begin{equation}\label{eq:important product to sum weights}
    \prod_i\hat{\mathcal{O}}_{\Delta_i}\ket{0}=\hat{\mathcal{O}}_{\Delta_{\rm S}}\ket{0}~,\quad\text{where}\quad \Delta_{\rm S}=\sum^m_{i=1}{\Delta_i}~.
\end{equation}
This can be proved recursively using (82) in \cite{Lin:2023trc}, which states that states in $\mathcal{H}_2$ can be decomposed into $\mathcal{H}_1$ irreps.:
\begin{equation}\label{eq:LinStanfordIdentity}
    \left|\Delta_1,\Delta_2;n_L, n_1, n_R\right\rangle=\sum_{m_L+m_R+l=n_1} \chi_{l, m_L, m_R}\left|\Delta_1+\Delta_2+l ; n_L+m_L, n_R+m_R\right\rangle~,
\end{equation}
where $\chi_{k, m_L, m_R}$ are Clebsch-Gordan coefficients. In particular, this implies
\begin{equation}\label{eq:Additive property}
    \ket{\Delta_1,\Delta_2;0,0,0}=\hat{\mathcal{O}}_{\Delta_1}\hat{\mathcal{O}}_{\Delta_2}\ket{0}=\hat{\mathcal{O}}_{\Delta_1+\Delta_2}\ket{0}=\ket{\Delta_1+\Delta_2;0,0}~.
\end{equation}
(\ref{eq:important product to sum weights}) follows straightforwardly. Then, $\prod_i\hat{\mathcal{O}}_{\Delta_i}$ acts as a composite operator in a generalized free field (GFF) theories in the large $N$ limit, where (\ref{eq:important product to sum weights}) also holds \cite{Duetsch:2002hc}. 
This means that while (\ref{eq:HH state tL tR}) lives in $\mathcal{H}_m$, it can be reduced to a state in $\mathcal{H}_1$, which is then an irrep.

We also remark that
\begin{equation}\label{eq:conversion LR}
\rme^{-\frac{\beta_R}{2}\hH_R}\hat{\mathcal{O}}^{R}_{\Delta_{\rm S}}(t_R)\rme^{-\frac{\beta_R}{2}\hH_R}\ket{0}=\rme^{-\frac{\beta_R}{2}\hH_L}\hat{\mathcal{O}}^{L}_{\Delta_{\rm S}}(t_R)\rme^{-\frac{\beta_R}{2}\hH_L}\ket{0}=\ket{\Psi_{\tilde{\Delta}}(\tau_R^*,\tau_R)}~,
\end{equation}
so that we can choose to express everything purely in terms of $\hat{\mathcal{O}}^{L}_{\Delta_{\rm S}}$ and the associated Hamiltonians, which is similar to the ``\textit{matching}'' property in \cite{Berkooz:2022fso}. We will denote $\hat{\mathcal{O}}_\Delta^{L}$ as $\hat{\mathcal{O}}_\Delta$ from now on, unless otherwise specified, to simplify the notation. Thus, the two-sided correlator can be re-expressed as
\begin{equation}\label{eq:def two-sided LR density matrices}
\begin{aligned}
    &\expval{\rme^{-\frac{\beta_L}{2}\hH_L}\hat{\mathcal{O}}_{\Delta_{\rm S}}(t_L)\rme^{-\frac{\beta_L}{2}\hH_L}\hat{\mathcal{O}}_{\Delta_w}\rme^{-\frac{\beta_R}{2}\hH_R}\hat{\mathcal{O}}_{\Delta_{\rm S}}\rme^{-\frac{\beta_R}{2}\hH_R}\hat{\mathcal{O}}_{\Delta_w}(t_R)}_{J,M}\\
    &=\bra{\Psi_{\tilde{\Delta}}(\tau_L^*,\tau_L)}q^{\Delta_w\hat{N}}\ket{\Psi_{\tilde{\Delta}}(\tau_R^*,\tau_R)}~.
\end{aligned}
\end{equation}
From the bulk side, there is an analogous conformal factor problem in the Euclidean path integral, similar to what has been pointed out in higher dimensional settings (see e.g.\cite{Gibbons:1978ac,Monteiro:2009tc,Monteiro:2008wr,Prestidge:1999uq,Marolf:2021kjc,Marolf:2022ybi}), which is dealt with by requiring finiteness of the dilaton \cite{Blommaert:2024whf}. {From the boundary side, this is equivalent to allowing the range of the energy parametrization, $\theta_{L/R}$, to extend beyond $[0,\pi]$, which leads to a thermodynamical instability in the saddle point solutions (see App. \ref{app:partition1p}). Well-definiteness of the Euclidean path integral and thermal stability are also linked together in higher dimensional examples (e.g. \cite{Monteiro:2008wr,Prestidge:1999uq})}.\footnote{It would be useful to develop a Lorentzian path integral formulation of our analysis where the {{analogous}} conformal mode problem from the bulk side does not arise.}

\subsection{Phase space from the path integral}\label{ssec:many chords}
In this subsection, we solve for the classical phase space of the DSSYK in the two-sided HH state from path integral methods. First, we will derive the canonical variables in the quantum Hamiltonian of the theory, and the Heisenberg equations that they must satisfy.

We now redefine the creation and annihilation operators in (\ref{eq:Hmultiple}) to its canonical form by identifying the distance $\hat{\ell}_i$, and momenta $\hat{P}_i$ operators, in analogy with the $\mathcal{H}_{0}$ case \cite{Lin:2022rbf}\footnote{Note that $\hat{P}$ is not a Hermitian operator since $\hat{\alpha}_{i}$ and $\hat{a}_{i}^\dagger$ are not conjugate operators with respect to the inner product introduced by \cite{Lin:2023trc}.}
\begin{equation}\label{eq:non-conjugate ops many particles}
\hat{a}_{i}^\dagger=\frac{\rme^{-\rmi \hat{P}_{i}}}{\sqrt{1-q}}~,\quad \hat{\alpha}_{i}=\sqrt{1-q}~\rme^{\rmi \hat{P}_{i}}~,\quad q^{\hat{n}_{i}}=\rme^{-\hat{\ell}_{i}}~,
\end{equation}
and we will denote 
\begin{equation}
\hat{\ell}\equiv\sum_i\hat{\ell}_i=\lambda\hat{N}~,\quad \hat{N}\equiv\sum_i\hat{n}_i    ~,
\end{equation}
which we denote as the total chord number (\ref{eq:total chord number basis}).

We can now express (\ref{eq:Hmultiple}) as
\begin{subequations}\label{eq:Hamiltonians LR}
\begin{align}\label{eq:H LR many particles1}
&\hH_{L}=\frac{J}{\sqrt{\lambda(1-q)}}\qty(\rme^{-\rmi \hat{P}_0}+\sum_{i=0}^m\rme^{\rmi \hat{P}_i}(1-\rme^{-\hat{\ell}_i})\rme^{-\hat{\ell}_i^<})~,\\
&\hH_{R}=\frac{J}{\sqrt{\lambda(1-q)}}\qty(\rme^{-\rmi \hat{P}_m}+\sum_{i=0}^m\rme^{\rmi \hat{P}_i}(1-\rme^{-\hat{\ell}_i})\rme^{-\hat{\ell}_i^>})~,\label{eq:H LR many particles2}
\end{align}    
\end{subequations}
where we define
\begin{equation}
    \hat{\ell}_i^<=\sum_{j=0}^{i-1}\qty(\hat{\ell}_j+\lambda\Delta_{j+1})~,\quad \hat{\ell}_i^>=\sum_{j={i+1}}^{m}\qty(\hat{\ell}_j+\lambda\Delta_{j})~.
\end{equation}
{{For instance, in the case $m=1$ and relabeling $0\rightarrow L$ and $m\rightarrow R$
\begin{equation}\label{eq:pair DSSYK Hamiltonians 1 particle}
\begin{aligned}
    \hH_{L/R}=\frac{J}{\sqrt{\lambda(1-q)}}\qty(\rme^{-\rmi \hat{P}_{L/R}}+\rme^{\rmi \hat{P}_{L/R}}\qty(1-\rme^{-\hat{\ell}_{L/R}})+q^{\Delta}\rme^{\rmi \hat{P}_{R/L}}\rme^{-\hat{\ell}_{L/R}}\qty(1-\rme^{-\hat{\ell}_{R/L}}))~.
    \end{aligned}
\end{equation}
}}In the following, we first solve for the dynamics of the model in the Heisenberg picture, which will allow us to derive the path integral that describes the two-sided system (\ref{eq:Hamiltonians LR}).

\paragraph{Quantum dynamics}We want to specify the quantum dynamics of the canonical operators in this theory. It is convenient to change from the Schrödinger to Heisenberg picture, by defining the operators
\begin{equation}\label{eq:Heisenberg operators}
\begin{aligned}
    \hat{\ell}(\tau_L,\tau_R)&\equiv\rme^{-\tau_L\hH_L-\tau_R^*\hH_R}~\hat{\ell}~\rme^{-\tau_L\hH_L-\tau_R\hH_R}~,\\
   \rme^{-\rmi\hat{P}_{0/m}(\tau_L,\tau_R)}&\equiv\rme^{-\tau_L\hH_L-\tau_R^*\hH_R}\rme^{-\rmi\hat{P}_{0/m}}\rme^{-\tau_L\hH_L-\tau_R\hH_R}~.
\end{aligned}
\end{equation}
Note that expectation values in both pictures agree with each other, i.e.
\begin{equation}\label{eq:Equivalence of pictures}
\begin{aligned}
       &\bra{\tilde{\Delta};0,\dots,0}\hat{\ell}(\tau_L,\tau_R)\ket{\tilde{\Delta};0,\dots,0}= \bra{\Psi_{\tilde{\Delta}}(\tau_L,\tau_R)}\hat{\ell}\ket{\Psi_{\tilde{\Delta}}(\tau_L,\tau_R)}~,\\
   &\bra{\tilde{\Delta};0,\dots,0}\rme^{-\rmi\hat{P}_{0/m}(\tau_L,\tau_R)}\ket{\tilde{\Delta};0,\dots,0}= \bra{\Psi_{\tilde{\Delta}}(\tau_L,\tau_R)}\rme^{-\rmi\hat{P}_{0/m}}\ket{\Psi_{\tilde{\Delta}}(\tau_L,\tau_R)}~,
\end{aligned}
\end{equation}
where the two-sided HH state is defined (\ref{eq:HH state tL tR}).

The Heisenberg EOM become
\begin{subequations}\label{eq:Heisenberg}
    \begin{align}\label{eq:Heisenberg1}
        -\frac{\rmi}{\lambda}\partial_{t_{L/R}}\hat{\ell}(\tau_L,\tau_R)&=[\hH_{L/R},~\hat{\ell}(\tau_L,\tau_R)]=\hH_{L/R}-\frac{2J}{\sqrt{\lambda(1-q)}}\rme^{-\rmi\hat{P}_{0/m}(\tau_L,\tau_R)}~,\\
    -\rmi\partial_{t_{R/L}}\rme^{-\rmi\hat{P}_{m/0}(\tau_L,\tau_R)}&=[\hH_{L/R},~\rme^{-\rmi\hat{P}_{m/0}(\tau_L,\tau_R)}]=J\sqrt{\frac{1-q}{\lambda}}q^{\Delta_{\rm S}}\rme^{-\hat{\ell}(\tau_L,\tau_R)}~,\label{eq:Heisenberg2}\\
    -\rmi\partial_{t_{R/L}}\rme^{-\rmi\hat{P}_{m/0}(\tau_L,\tau_R)}&=[\hH_{L/R},~\rme^{-\rmi \hat{P}_{0/m}}]_q-(1-q)\rme^{-\rmi \hat{P}_{0/m}}\hH_{L/R}\\
    &=J\sqrt{\frac{1-q}{\lambda}}\qty(1+\rme^{-2\rmi P_{0/m}(\tau_L,\tau_R)})-(1-q)\rme^{-\rmi \hat{P}_{0/m}}\hH_{L/R}~,\nonumber
    \end{align}
\end{subequations}
where we used the commutation relations for $\hat{a}_{L/R}$ and $\hH_{L/R}$ in \cite{Lin:2023trc} (28) together with (\ref{eq:non-conjugate ops many particles}).

Combining (\ref{eq:Heisenberg1}) and (\ref{eq:Heisenberg2}) we get a Liouville-like equation of motion:
\begin{equation}\label{eq:quantum Liouville like}
    -\partial_{t_L}\partial_{t_R}\hat{\ell}(\tau_L,\tau_R)=[\hH_L,~[\hH_R,~\hat{\ell}]]=-2J^2q^{\Delta_{\rm S}}\rme^{-\ell}~,
\end{equation}
which is a dynamical version of (30) in \cite{Lin:2023trc} (theirs use Euclidean time $\partial_{L/R}\equiv\partial_{\tau_{L/R}}$ instead). Note (\ref{eq:quantum Liouville like}) is consistent with the \emph{Ehrenfest theorem} for Krylov complexity in (2.23) of \cite{Erdmenger:2023wjg} when $\Delta=0$ so that $\hH_L=\hH_R$ (and $t_L=t_R$) where the Krylov complexity operator is quantum mechanically equal to the chord number operator (with the HH state as reference state).\footnote{\label{fnt:generalized Ehrenfest}(\ref{eq:quantum Liouville like}) suggests that there might be an extension of the Ehrenfest theorem in Krylov complexity \cite{Erdmenger:2023wjg} for two sided Hamiltonians, which would be noteworthy to study in the future.}

Next, we can also specify the initial conditions for the operator valued expectation values in the two-sided HH state (\ref{eq:Equivalence of pictures}),
\begin{subequations}\label{eq:initial quantum picture}
    \begin{align}\label{eq:initial length exp val}
    &\frac{1}{Z_{\Delta_{\rm S}}(\beta_L,\beta_R)}\bra{\tilde{\Delta};0,\dots,0}\hat{\ell}\qty(\tfrac{\beta_L}{2},\tfrac{\beta_R}{2})\ket{\tilde{\Delta};0,\dots,0}= \ell_*~,\\
    &\begin{aligned}\label{eq:initial velocity}
    &\eval{\partial_{t_{L/R}}\bra{\tilde{\Delta};0,\dots,0}\hat{\ell}(\tau_L,\tau_R)\ket{\tilde{\Delta};0,\dots,0}}_{t_{L/R}=0}\\
    &=\bra{\tilde{\Delta};0,\dots,0}\rme^{-\frac{\beta_L}{2}\hH_L-\frac{\beta_R}{2}\hH_R}[\hH_{L/R},~\hat{\ell}]\rme^{-\frac{\beta_L}{2}\hH_L-\frac{\beta_R}{2}\hH_R}\ket{\tilde{\Delta};0,\dots,0}=0~.
    \end{aligned}
\end{align}
\end{subequations}
First, $\ell_*$ in (\ref{eq:initial length exp val}) is just a constant determined by the temperatures. We will show that the specific value (\ref{eq:ell cond thetaLR}) can be determined in terms of the conserved energy spectrum of the system (\ref{eq:conserved energies}). For instance, one can see that in the infinite temperature limit $\beta_{L/R}\rightarrow0$, then $\ell_*=0$ (because $\hat{N}\ket{\tilde{\Delta};0,\dots,0}=0$). However, we will work with arbitrary values of the two-sided temperatures. Meanwhile, we \emph{prove} (\ref{eq:initial velocity}) in App. \ref{app:isometric map}. Thus, the evolving PETS HH state (\ref{eq:HH state tL tR}) generates expectation values satisfying the EOM (\ref{eq:Heisenberg}) and the initial conditions (\ref{eq:initial quantum picture}).

\paragraph{Hartle-Hawking saddle point}
We will now study the path integral of (\ref{eq:Hamiltonians LR}) that prepares the two-sided HH state (\ref{eq:HH state tL tR}). While there are other possible approaches (one can evaluate the expectation values directly in the $\lambda\rightarrow0$ limit), the path integral is convenient as we can think about the solutions as saddle points.

\begin{figure}
    \centering
    \includegraphics[width=0.5\linewidth]{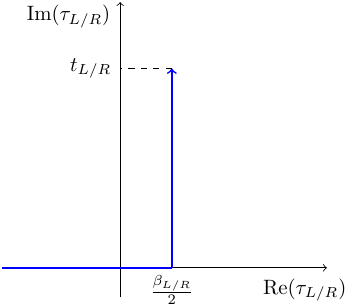}
    \caption{{{HH contour (in blue) in the complex plane for $\tau_{L/R}$, which we apply to solve for the saddle points of \eqref{eq:Hamilton PI} evaluated at $\tau_{L/R}=\beta_{L/R}/2+\rmi t_{L/R}$.}}}
    \label{fig:HH_contour}
\end{figure}

Based on our analysis of the Heisenberg picture operators, we should use
\begin{equation}\label{eq:Hamilton PI}
    \int\prod_{i=0}^m[\rmd \ell_i][\rmd P_i]\exp\qty[\int\rmd\tau_L\rmd\tau_R\qty(\frac{\rmi}{\lambda}\sum_{i=0}^m\qty(P_i(\partial_{\tau_L}+\partial_{\tau_R})\ell_i)+H_L+H_R)]~,
\end{equation}
where {{the expectation values}} operators in the two-sided Hamiltonian (\ref{eq:Hamiltonians LR}) have been exchanged for field variables, {{i.e.}}
{{\begin{equation}\label{eq:key eqs}
\begin{aligned}
    \ell:=\frac{\bra{\Psi_{\Delta_S}(\tau_L,\tau_R)}\hat{\ell}\ket{\Psi_{\Delta_S}(\tau_L,\tau_R)}}{\bra{\Psi_{\Delta_S}(\tau_L,\tau_R)}\ket{\Psi_{\Delta_S}(\tau_L,\tau_R)}}~,\\
    P_{i}:=\frac{\bra{\Psi_{\Delta_S}(\tau_L,\tau_R)}\hat{P}_i\ket{\Psi_{\Delta_S}(\tau_L,\tau_R)}}{\bra{\Psi_{\Delta_S}(\tau_L,\tau_R)}\ket{\Psi_{\Delta_S}(\tau_L,\tau_R)}}~.
\end{aligned}
\end{equation}}}
{{Furthermore, we solve the saddle point of the path integral using the HH preparation of state \cite{Hartle:1983ai}} with the contour in the complex plane for $\tau_{L/R}$ shown in Fig. \ref{fig:HH_contour}.}
The saddle point solution {{of}} the path integral (i.e. $\lambda\rightarrow0$) satisfy 
\begin{equation}\label{eq:more general EOM}
\begin{aligned}
        (\partial_{t_{L}}+\partial_{t_{R}})\ell_i=\dv{(H_L+H_R)}{P_i}~,\qquad (\partial_{t_{L}}+\partial_{t_{R}})P_i=-\dv{(H_L+H_R)}{\ell_i}~,
\end{aligned}
\end{equation}
where, {{based on Fig. \ref{fig:HH_contour}, we have used the analytic continuation}} $\tau_{L/R}=\frac{\beta_{L/R}}{2}+\rmi t_{L/R}$, and $\beta_{L/R}$ are constant{{s}}. We want to study the saddle points describing the HH state, which obey the Heisenberg equations (\ref{eq:Heisenberg}). This means that we study a family of solutions of (\ref{eq:more general EOM}) that split into the left/right chord sectors
\begin{align}\label{eq:evol many}
\frac{1}{\lambda}\pdv{\ell}{t_{L/R}}=\sum_{i=0}^m\pdv{H_{L/R}}{P_{i}}~,\qquad
\frac{1}{\lambda}\pdv{P_{i}}{t_{L/R}}=-\pdv{H_{L/R}}{\ell_{i}}~,
\end{align}
where $\ell\equiv\sum_i\ell_i$. This indeed reproduces (\ref{eq:Heisenberg1}, \ref{eq:quantum Liouville like}) in the $\lambda\rightarrow0$ limit.
Following from our derivation in (\ref{eq:initial quantum picture}), we impose as initial condition the classical version of (\ref{eq:initial quantum picture}) for one-particle irreps
\begin{equation}\label{eq:in cond many}
\ell{(t_L=t_R=0)}=\ell_*(\theta_L,\theta_R)\in\mathbb{R}~,\quad \eval{\pdv{\ell}{t_{L/R}}}_{t_L=t_R=0}=0~,
\end{equation}
where we have parametrized the conserved energy spectrum with an angle $\theta_{L/R}$,\footnote{There are additional saddle points when we extend the range of $\theta_{L/R}$, but they are thermodynamically unstable, as we show in App. \ref{app:partition1p}.}
\begin{equation}\label{eq:conserved energies}
   E_{L/R}=\frac{2J}{\sqrt{\lambda(1-q)}}\cos\theta_{L/R}~.
\end{equation}
The range $\theta_{L/R}\in[0,\pi]$ is unchanged by the particle insertion \cite{Xu:2024hoc}.

These initial conditions correspond to the two-sided HH state preparation (\ref{eq:HH state tL tR}) which also allows for the total length-type of solution to be a real-valued distance in the bulk theory. Once we modify the state, the initial value of the classical phase space solutions will change accordingly.
The solutions (\ref{eq:key eqs}) become\footnote{We present a comprehensive derivation of these equations in \cite{Aguilar-Gutierrez:2025mxf}. We also discuss the bulk interpretation of these solutions in Sec. \ref{ssec:canonical quantization}.}
\begin{subequations}\label{eq:solutions HH}
    \begin{align}\label{eq:womrhole semiclassical answer many}
&\ell=2\log \qty(\cosh(\gamma_L t_L)\cosh(\gamma_R t_R)+\frac{J^2q^{\Delta_{\rm S}}\rme^{-\ell_*}}{\gamma_L\gamma_R}\sinh(\gamma_L t_L)\sinh(\gamma_R t_R))+\ell_*~,\\
&P_{0}=\rmi\log(\cos\theta_{L}+\rmi\sin\theta_{L}\frac{\rme^{\ell_*}\sin\theta_L\sin\theta_R\tanh(\gamma_{L}t_{L})+q^{\Delta_{\rm S}}\tanh(\gamma_{R}t_{R})}{\rme^{\ell_*}\sin\theta_L\sin\theta_R+q^{{\Delta_{\rm S}}}\tanh(\gamma_Rt_R)\tanh(\gamma_Lt_L)})~,\label{eq:canonical momenta0 many}\\
&P_{m}=\rmi\log(\cos\theta_{R}+\rmi\sin\theta_{R}\frac{\rme^{\ell_*}\sin\theta_L\sin\theta_R\tanh(\gamma_{R}t_{R})+q^{\Delta_{\rm S}}\tanh(\gamma_{L}t_{L})}{\rme^{\ell_*}\sin\theta_L\sin\theta_R+q^{{\Delta_{\rm S}}}\tanh(\gamma_Rt_R)\tanh(\gamma_Lt_L)})~,\label{eq:canonical momentam many}
\end{align} 
\end{subequations}
{{Note that in this derivation, we are keeping $q^{\Delta_{\rm S}}$ fixed as $\lambda\rightarrow0$; and}} we defined 
\begin{equation}\label{eq:gamma parameter}
\gamma_{L/R}=\sqrt{J^2-\qty(\frac{E_{L/R}}{2})^2}=J\sin\theta_{L/R}~,\quad \theta_{L/R}\in[0,\pi]~.
\end{equation}
The classical fields {{in \eqref{eq:womrhole semiclassical answer many}}} can be then promoted to quantum expectation values by canonical quantization, resulting in our starting quantum theory (\ref{eq:Hamiltonians LR}).

Next, we would like to determine the initial total distance $\ell_*$ from the energy conservation in the left and right Hamiltonians (\ref{eq:Hamiltonians LR}). Since we are working with one-particle irreps due to (\ref{eq:Additive property}) (see \cite{Aguilar-Gutierrez:2025mxf} for the more general case), we impose as initial conditions $P_{0< i< m}(t_L=t_R=0)=0$ and $\ell_{0< i< m}(t_L=t_R=0)=0$. One could impose other initial conditions; where the additional values of the other length variables and their momenta would appear in {{the explicit solution $\ell_*(\theta_L,\theta_R)$}}. However, these would not correspond to one-particle irreps. In other words, the dynamical information of the system is contained in the parameters $\ell$, $P_m$, $P_0$ instead of all the canonical coordinates when we restrict the analysis to composite operators to construct the state  (\ref{eq:HH state tL tR}). This remains true even when the Schrödinger picture operators are time dependent, as we discuss later, see below (\ref{eq:double scaled precursor shockwave}).
{{This reduces the problem to the two-sided Hamiltonians (\ref{eq:Hamiltonians LR}) with a one-particle state and a conserved energy spectrum $E(\theta_{L/R})$ in (\ref{eq:conserved energies}) that we can express as
\begin{equation}\label{eq:conserved 1 particle Hamiltonians}
    2\cos\theta_{L/R}=\rme^{-\rmi P_{0/m}}+\rme^{\rmi P_{0/m}}\qty(1-\rme^{-\ell_{0/m}})+q^{\Delta}\rme^{\rmi P_{m/0}}\qty(\rme^{-\ell_{0/m}}-\rme^{-\ell})~,
\end{equation}
where $\ell=\ell_0+\ell_m$. \eqref{eq:conserved 1 particle Hamiltonians} can also be written as
\begin{equation}\label{eq:ell0m}
\begin{aligned}
\rme^{-\ell_{0}}&=\frac{-\rme^{-\ell+\rmi P_m} q^{\Delta _S}+\rme^{-\rmi P_0}+\rme^{i
   P_0}-2 \cos \left(\theta _L\right)}{\rme^{\rmi P_0}-\rme^{\rmi P_m} q^{\Delta _S}}~,\\
    \rme^{-\ell_{m}}&=\frac{-\rme^{-\ell+\rmi P_0} q^{\Delta _S}+\rme^{-\rmi P_m}+\rme^{\rmi P_m}-2 \cos \left(\theta
   _R\right)}{e^{\rmi P_m}-\rme^{\rmi P_0} q^{\Delta _S}}~.
\end{aligned}
\end{equation}
Plugging in the explicit solutions for $\ell$, $P_{0/m}$ in \eqref{eq:solutions HH} and using that $\ell=\ell_0+\ell_m$}} leads to $\ell_*(\theta_L,\theta_R)$ {{(i.e. the initial condition (\ref{eq:in cond many}))}}\footnote{See additional details in \cite{Aguilar-Gutierrez:2025mxf}.}
\begin{align}\label{eq:ell cond thetaLR}
    &\rme^{-\ell_*(\theta_L,\theta_R)}=\frac{q^{-2\Delta_{\rm S}}}{2}\biggl(1+q^{2\Delta_{\rm S}}-2q^{\Delta_{\rm S}}\cos\theta_L\cos\theta_R\\
    &-\sqrt{1+q^{4\Delta_{\rm S}}+2 q^{2\Delta_{\rm S}}(1+\cos(2\theta_L)+\cos(2\theta_R))-4q^{\Delta_{\rm S}}(1+q^{2\Delta_{\rm S}})\cos\theta_L\cos\theta_R}\biggr)~.\nonumber
\end{align}
{{Here, \eqref{eq:ell cond thetaLR} represents the initial length of the dual wormhole an operator insertion with fixed two-sided energies $E_{L/R}$ \eqref{eq:conserved energies} as boundary condition.}} We study its bulk interpretation in more detail in Sec. \ref{ssec:towardsquantization}. Note that the $\Delta_{\rm S}$ factor applies to both heavy and light DSSYK operators. This is reminiscent of higher dimensional CFTs where both heavy, and smeared light operator insertion in the boundary can generate shockwave backreaction in the bulk dual \cite{Afkhami-Jeddi:2017rmx}.

\subsection{Two-sided two-point (crossed four-point) correlation functions}\label{ssec:twopoints}
\paragraph{On the most general semiclassical answer}
In the classical regime we used to derive (\ref{eq:womrhole semiclassical answer many}), $\expval{\hat{N}^k}=\expval{\hat{N}}^k$,\footnote{{{To verify that $\expval{\hat{N}^k}=\expval{\hat{N}}^k$ at leading order in the $\lambda\rightarrow0$ limit, as it should, we evaluate $\expval{q^{\Delta N}}$ in App. \ref{app:twopoint oneparticle} and compare it to $q^{\Delta \expval{N}}$ that we obtain through (\ref{eq:womrhole semiclassical answer many}); as we explain in the following paragraphs.}}} where $\expval{\hat{N}}$ represents the average position of a classical particle in an ordered lattice given by the Krylov basis in Sec. \ref{ssec:total chord as Krylov}. This allows us to deduce the corresponding generating function. In the specific case of (\ref{eq:womrhole semiclassical answer many}), the thermal two-point function for arbitrary $\tau_L$ and $\tau_R$ can be constructed from the moments as
{{\begin{equation}\label{eq:generating function explicit DSSYK 1 particle}
\begin{aligned}
        G^{(\Delta_w)}_{\tilde{\Delta}}&(\tau_L,\tau_R):=\frac{\bra{\Psi_{\Delta_{\rm S}}(\tau_L,\tau_R)}q^{\Delta_w\hat{N}}\ket{\Psi_{\Delta_{\rm S}}(\tau_L,\tau_R)}}{\bra{\Psi_{\Delta_{\rm S}}(\tau_L,\tau_R)}\ket{\Psi_{\Delta_{\rm S}}(\tau_L,\tau_R)}}\\
        &\eqlambda \exp(-\Delta_{w}\frac{\bra{\Psi_{\Delta_{\rm S}}(\tau_L,\tau_R)}\lambda\hat{N}\ket{\Psi_{\Delta_{\rm S}}(\tau_L,\tau_R)}}{\bra{\Psi_{\Delta_{\rm S}}(\tau_L,\tau_R)}\ket{\Psi_{\Delta_{\rm S}}(\tau_L,\tau_R)}})\\
        &\eqlambda
    \qty(\frac{\rme^{-\ell_*/2}}{\cosh(\gamma_Lt_L)\cosh(\gamma_Rt_R)+\frac{J^2\rme^{-\ell_*}q^{{\Delta_{\rm S}}}}{\gamma_L\gamma_R}\sinh(\gamma_Lt_L)\sinh(\gamma_Rt_R)})^{2\Delta_w}~,
\end{aligned}
\end{equation}}}
where {{we used $q=\rme^{-\lambda}$};} $\tau_{L/R}=\rmi t_{L/R}+\frac{\beta_{L/R}}{2}$; {{$\hat{\ell}=\lambda\hat{N}$,}} and $\gamma_{L/R}=J\sqrt{\frac{\lambda}{1-q}}\sin\theta_{L/R}$ {{in (\ref{eq:womrhole semiclassical answer many})}}. It is, arguably, not surprising that the thermal two-point function follows from solving EOM of the corresponding saddle points in the DSSYK path integral (as it often happens in the $G\Sigma$ formalism \cite{Jia:2025tvn,Berkooz:2024evs,Berkooz:2024ifu,Berkooz:2024lgq,Berkooz:2024ofm}).\footnote{It would be interesting to show if the path integral in terms of the canonical variables (\ref{eq:Hamilton PI}) can be recovered from the $G\Sigma$ formalism. While in the $G\Sigma$ formalism the fields are bilocal, these two approaches reproduce the same classical expressions for the two-point correlation function. A difficulty to make this connection in our study is that the Hamiltonians (\ref{eq:Hamiltonians LR}) is derived from the auxiliary chord Hilbert space \cite{Lin:2022rbf}. However, this is usually derived before taking ensemble averaging \cite{Cotler:2016fpe,Goel:2023svz}.}

Since we have an explicit result for the two-sided two-point function from the classical phase space solutions, (\ref{eq:generating function explicit DSSYK 1 particle}), we also performed a consistency check in App. \ref{app:twopoint oneparticle} to confirm that (\ref{eq:generating function explicit DSSYK 1 particle}) follows from a chord diagram evaluation in the two-sided HH state (\ref{eq:HH state tL tR}). This calculation corresponds to a crossed four-point function in $\mathcal{H}^{L}_0\otimes\mathcal{H}_0^{R}$, as we detail in App. \ref{app:twopoint oneparticle}. The resulting correlation function (\ref{eq:result chord amplitude}) is in agreement with (\ref{eq:generating function explicit DSSYK 1 particle}) at leading order in the $\lambda\rightarrow 0$ regime, and assuming that the operators are light (i.e. the conformal dimension $\Delta$ is fixed as $\lambda\rightarrow0$). In this regime, we know what ansatz to use for the chord diagram calculation. {In the semiclassical limit with light operators, $q^{\Delta_{\rm S}}\rightarrow1$ at leading order, and thus, the derivation from the chord diagram method with the light operators is not sensitive to the specific functional dependence on $\Delta_{\rm S}$ in (\ref{eq:womrhole semiclassical answer many}). However, we expect to reproduce (\ref{eq:generating function explicit DSSYK 1 particle}) exactly by refining the technical steps in the computation. We hope to develop technical tools to confirm the full semiclassical answer (\ref{eq:womrhole semiclassical answer many}) in the future. Nevertheless, the match from these two evaluations in the appropriate regime 
represents non-trivial evidence for self-consistency in the derivation. Thus, (\ref{eq:generating function explicit DSSYK 1 particle}) should also be seen as a prediction for the crossed four-point function of the DSSYK model in the semiclassical regime (which we partially verify with our calculation in \ref{app:twopoint oneparticle}). The corresponding chord diagram is displayed in Fig. \ref{fig:OTOC_chord_diagram}.
\begin{figure}
    \centering
    \includegraphics[width=0.5\linewidth]{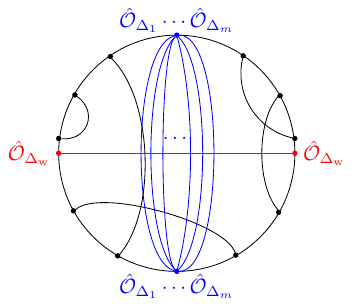}
    \caption{Example of the crossed chord diagram we evaluate in (\ref{eq:generating function explicit DSSYK 1 particle}). The blue chords correspond to the composite operator $\prod_{i=1}^m\hat{\mathcal{O}}_{\Delta_i}$, and the red chord to a pair of probe operators $\hat{\mathcal{O}}_{\Delta_w}$.}
    \label{fig:OTOC_chord_diagram}
\end{figure}

Moreover, we can verify that the two-point correlation function corresponding to the two-sided HH state (\ref{eq:HH state tL tR}) in the one-particle irrep. is equivalent to a crossed four-point function with respect to $\mathcal{H}^{L}_0\otimes\mathcal{H}_0^{R}$. Consider $\tau_L=-\tau_R=\rmi t$,
\begin{equation}\label{eq:twosided twopoint OTOC}
\begin{aligned}
    \bra{\Psi_{\tilde{\Delta}}(-\rmi t,\rmi t)}q^{\Delta_w \hat{N}}\ket{\Psi_{\tilde{\Delta}}(-\rmi t,\rmi t)}&=\bra{0}\hat{\mathcal{O}}_{\Delta_{\rm S}}(t)q^{\Delta_w \hat{N}}\hat{\mathcal{O}}_{\Delta_{\rm S}}(t)\ket{0}\\
    &=\wick[offset=1.2em,sep=0.4em]{\langle0\vert\hat{\mathcal{O}}_{\Delta_{\rm S}}(t)\c{\hat{\mathcal{O}}}_{\Delta_w}(0)\hat{\mathcal{O}}_{\Delta_{\rm S}}(t)\c{\hat{\mathcal{O}}}_{\Delta_w}(0)\vert{0}\rangle}~,
\end{aligned}
\end{equation}
where the bar denotes Wick contraction. Thus, we see that the two-sided two-point function is indeed an OTOC. A plot of the OTOC for all times is displayed in Fig. \ref{fig:OTOC evol}.
\begin{figure}
    \centering
    \includegraphics[width=0.6\linewidth]{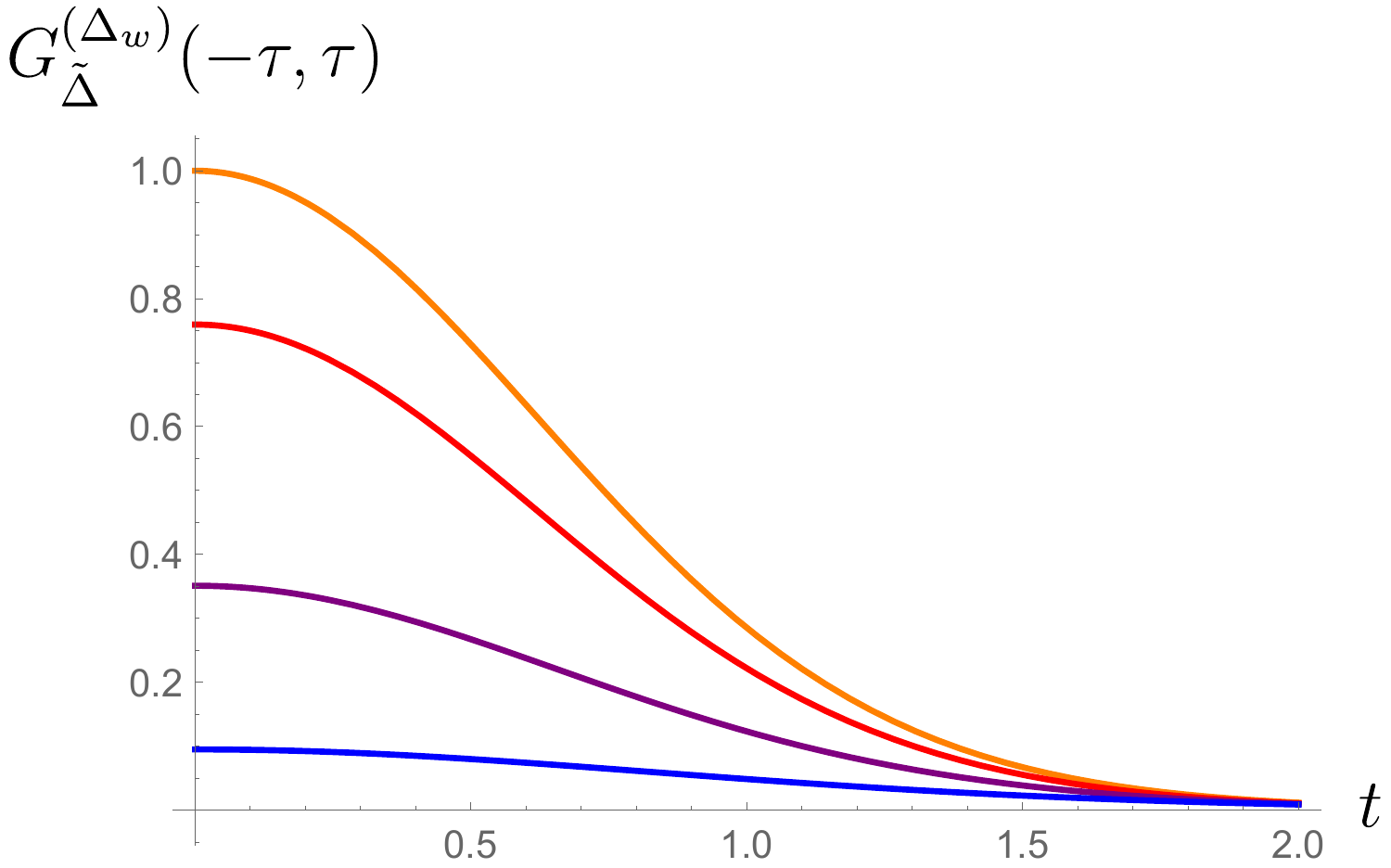}
    \caption{Evolution of the OTOC (\ref{eq:generating function explicit DSSYK 1 particle}) for $\tau_R=-\tau_L=\frac{\beta(\theta)}{2}+\rmi t$, {{where $\beta(\theta)$ (the semiclassical microcanonical inverse temperature) appears in \eqref{eq:entropy temp}}; see also App. \ref{app:partition1p}}. We are taking $\lambda\Delta_{\rm S}=1$, $J=1$, $\lambda=10^{-6}$, $\Delta_w=1$, and different values of $\theta$ starting from $\frac{\pi}{2}$ (orange) and decreasing in steps of $0.3$ (the corresponding curves are displayed in descending order).}
    \label{fig:OTOC evol}
\end{figure}

We study different properties of the OTOC in App. \ref{app:properties OTOC}.

\subsection{Double-scaled precursor operators}\label{ssec:shockwaves}
The {{wormhole length in a shockwave geometry}} have {{been previously matched to chord diagram computations}} in \cite{Berkooz:2022fso} (albeit in the triple-scaling limit). In this section, we explain how to derive the same type of distances from the path integrals without the triple-scaling limit. In the following, we consider how to recover single shockwave geometries from the one-particle HH state. We have a comprehensive study with multiple shockwaves in \cite{Aguilar-Gutierrez:2025mxf}.

\paragraph{Double-scaled precursors}Motivated by the original definition in \cite{Stanford:2014jda} (see also \cite{Berkooz:2022fso}), we introduce double-scaled precursor operators in the Schrödinger picture, as
\begin{equation}\label{eq:double scaled precursor shockwave}
    \begin{aligned}
        &\hat{\mathcal{O}}^{L}_\Delta(t_w)\equiv\rme^{\rmi t_w \hH_{L}}\hat{\mathcal{O}}^{L}_\Delta\rme^{-\rmi t_w\hH_{L}}~,\\
        &\hat{\mathcal{O}}^{R}_\Delta(t_w)\equiv\rme^{-\rmi t_w \hH_{R}}\hat{\mathcal{O}}^{R}_\Delta\rme^{\rmi t_w\hH_{R}}~,
    \end{aligned}
\end{equation}
where $t_{w}\in\mathbb{R}$ is the insertion time of the operator. The definition (\ref{eq:double scaled precursor shockwave}) from the right chord sector corresponds to evolving an state in the past, inserting an operator and evolving by the same time scale. Note that transforming the matter chord operators for the precursor operators amounts to an overall time shift in factors of the form
\begin{equation}
\begin{aligned}
\rme^{-\tau_{L}\hH_{L}-\tau_{R}\hH_{R}}\hat{\mathcal{O}}_{\Delta}(t_w)\ket{0}&=\rme^{-\qty(\frac{\beta_{L}}{2}-\rmi (t_{L}-t_w))\hH_{L}}\hat{\mathcal{O}}^{L}_{\Delta}\rme^{-\qty(\frac{\beta_{R}}{2}+\rmi (t_{R}+t_w))\hH_{L}}\ket{0}\\
&=\rme^{-\qty(\frac{\beta_{R}}{2}-\rmi (t_{R}+t_w))\hH_{R}}\hat{\mathcal{O}}_{\Delta}^{R}\rme^{-\qty(\frac{\beta_{L}}{2}+\rmi (t_{L}-t_w))\hH_{R}}\ket{0}~.
\end{aligned}
\end{equation}
This means that the double-scaled precursor operators (\ref{eq:double scaled precursor shockwave}) generate an overall time shift $t_L\rightarrow t_L-t_w$ and $t_R\rightarrow t_R+t_w$. In terms of the classical phase space solutions (\ref{eq:womrhole semiclassical answer many}), the replacement of $\hat{\mathcal{O}}_{\Delta}$ for the precursor (\ref{eq:double scaled precursor shockwave}) shifts the initial conditions in (\ref{eq:in cond many}) as
\begin{equation}
\eval{\ell}_{t_L=-t_R=t_w}=\ell_*(\theta_L,\theta_R)~,\quad \eval{\dv{\ell}{t_{L/R}}}_{t_L=-t_R=t_w}=0~,
\end{equation}
where $\ell_*(\theta_L,\theta_R)$ was determined in (\ref{eq:ell cond thetaLR}) from the EOM and energy conservation. We can then express (\ref{eq:womrhole semiclassical answer many}) as
\begin{equation}\label{eq:almost shockwave}
\begin{aligned}
\ell=2\log \biggl(&\cosh(\gamma_L(t_L-t_w)+\gamma_R (t_R+t_w))\\
&+\qty(\frac{ J^2q^{\Delta_{\rm S}}\rme^{-\ell_*}}{\gamma_L\gamma_R}-1)\sinh(\gamma_L (t_L-t_w))\sinh(\gamma_R (t_R+t_w))\biggr)+\ell_*~.
\end{aligned}
\end{equation}
\paragraph{Shockwave limit}We would like to study the shockwave limit, i.e. the regime where (\ref{eq:almost shockwave}) has the same evolution as the wormhole geodesic distances as \cite{Shenker:2013pqa} (we discuss the bulk interpretation in Sec. \ref{ssec:towardsquantization}). In this picture, one has to consider a very early (meaning $t_w\rightarrow\infty$) perturbation of the TFD state. This implies that the system is always in an equal temperature ensemble, $\beta_L=\beta_R$, so that $\gamma_L=\gamma_R\equiv J\sin\theta$. {{In this case, we denote $\ell_*^{(+)}(\theta):=\ell_*(\theta_L=\theta,\theta_R=\theta)$, which we identify from \eqref{eq:ell cond thetaLR} as
\begin{equation}\label{eq:ell * state}
\rme^{-\frac{1}{2}\ell^{(+)}_*(\theta)}=\frac{q^{-\Delta_{\rm S}}}{2}\qty(-1+q^{\Delta_{\rm S}}+\sqrt{1+q^{2{\Delta_{\rm S}}}-2q^{ \Delta_{\rm S}} \cos2\theta})~.
\end{equation}
}}
Thus, the evolution of (\ref{eq:almost shockwave}) for $t_w\gg t_{L/R}$ becomes
\begin{equation}\label{eq:shockwave approx}
\begin{aligned}
\ell(t_L,t_R)\simeq&2\log \biggl(\cosh(J\sin\theta(t_L+t_R))+\qty(\frac{1}{4}-\frac{q^{\Delta_{\rm S}}\rme^{-\ell^{(+)}_*(\theta)}}{4\sin^2\theta})\rme^{-J\sin\theta (t_L-t_R-2t_w)}\biggr)\\
&+\ell^{(+)}_*(\theta)~.
\end{aligned}
\end{equation}
and the corresponding momenta (\ref{eq:womrhole semiclassical answer many}):
\begin{equation}\label{eq:Pom shockwave}
    {P_{L/R}=\rmi\log(\cos\theta+\rmi\sin\theta\frac{{\sin^2\theta}\tanh(\delta t_{L/R})+{q^{\Delta}\rme^{-\ell_*^{(+)}}}\tanh(\delta t_{R/L})}{{\sin^2\theta}+{q^{\Delta}\rme^{-\ell_*^{(+)}}}\tanh(\delta t_{L})\tanh(\delta t_{R})})~,}
\end{equation}
where
\begin{equation}
    \delta t_{L}=J\sin\theta(t_L-t_w)~,\quad \delta t_{R}=-J\sin\theta(t_R+t_w)~.
\end{equation}
Note that the limiting behavior in (\ref{eq:shockwave approx}) is only valid for $t_w\gg \abs{t_{L/R}}$. This is what allows us to consistently discard exponentially suppressed terms to arrive at (\ref{eq:shockwave approx}).

Moreover, \eqref{eq:shockwave approx} displays a scrambling behavior (i.e. the transition from exponential to linear growth) when we consider the time regimes
\begin{equation}\label{eq:scrambling shockwave}
\ell(t_L,t_R)\simeq\begin{cases}
    \qty(\frac{1}{2}-\frac{q^{\Delta_{\rm S}}\rme^{-\ell_*^{(+)}(\theta)}}{2\sin^2\theta})\rme^{-J\sin\theta (t_L-t_R-2t_w)}+\ell^{(+)}_*(\theta)~,& t_L\gg \abs{t_R}\\
    2\log\qty(\frac{1}{4}-\frac{q^{\Delta_{\rm S}}\rme^{-\ell_*^{(+)}(\theta)}}{4\sin^2\theta})+J\sin\theta (t_R+2t_{w}-t_L)+\ell^{(+)}_*(\theta)~,& t_R\gg \abs{t_L}~.
\end{cases}
\end{equation}
We will discuss more about this limit and its place in the holographic dictionary in Sec. \ref{ssec:towardsquantization}. For now, note that $t_L=t_R$ (associated with spread complexity) does not display scrambling (as observed in our discussion in Sec. \ref{ssec:special}). Additionally, the last entry in (\ref{eq:scrambling shockwave}) obeys (4.7) in \cite{Stanford:2014jda} for $n=1$.

\section{The Krylov space of the two-sided Hartle-Hawking state}\label{sec:deriving Krylov}
In this section, we connect our previous results on saddle points of the path integral with the Krylov complexity of the two-sided HH state. 

\paragraph{Outline}In Sec. \ref{ssec:total chord as Krylov} we formulate the Lanczos algorithm for two-sided Hamiltonians that evaluates the Krylov operator complexity of the operator $\hat{\mathcal{O}}_\Delta$, or the spread complexity of the two-sided HH state (\ref{eq:HH state tL tR}), depending on the relative {{two-sided}} times in the Hamiltonian evolution, and that incorporates finite temperature effects. We derive the Lanczos coefficients and Krylov basis from the chord Hamiltonians. 
In Sec. \ref{ssec:special} we apply the classical phase space results to deduce the semiclassical Krylov complexity. Then, we {{study}} the properties of spread and Krylov operator complexity. Special cases in our results match with existing literature \cite{Ambrosini:2024sre,Heller:2024ldz}. We find that an OTOC is the generator of Krylov operator complexity in the semiclassical limit. It displays a scrambling behavior, i.e. a transition between exponential to late-time linear growth (in a similar way as in circuit complexity \cite{Chapman:2021jbh,Susskind:2014rva}), inherited from scrambling in the OTOC. It is hyperfast, in the sense of \cite{Susskind:2021esx}, and it is consistent with several bounds in the literature \cite{Maldacena:2015waa,Anegawa:2024yia,Milekhin:2024vbb}. 
Lastly, in Sec. \ref{ssec:interpretation shockwav krylov} we discuss how the Lanczos algorithm changes when we insert double-scaled precursor operators (\ref{eq:double scaled precursor shockwave}) to build the two-sided HH state, and its shockwave limit, introduced in Sec. \ref{ssec:shockwaves}.

\subsection{Lanczos algorithm for two-sided Hamiltonians}\label{ssec:total chord as Krylov}
In this subsection, we will show that the Krylov operator or spread complexity for matter chord operators or the two-sided HH state (\ref{eq:HH state tL tR}) respectively equals the expectation value of the total chord number with different {{two-sided time relations}} in the semiclassical limit. We implement the Lanczos algorithm with two-sided Hamiltonians and finite temperatures based on the Lanczos algorithm for one-sided Hamiltonian with complex time evolution in \cite{Erdmenger:2023wjg}, which we summarize in App. \ref{sapp:Krylov review}. We first express the two-sided HH state (\ref{eq:HH state tL tR}) as
\begin{equation}
    \ket{\Psi_{\tilde{\Delta}}(\tau_L,\tau_R)}=\rme^{-\tau_L\hH_L-\tau_R\hH_R}\ket{\tilde{\Delta};0\dots,0}=\rme^{-\tau_R \mathcal{L}_\eta}\ket{\Delta_{\rm S};0,0}~,
\end{equation}
where $\Delta_{\rm S}\equiv\sum_{i=1}^m\Delta_i$, and 
\begin{equation}
\hat{\mathcal{L}}_\eta\equiv \hH_R+\eta\hH_L~,\quad (\eta\equiv\tau_L/\tau_R)    
\end{equation}
is a generalized Liouvillian operator for two-sided Hamiltonians. In principle, $\eta$ does not need to be a constant, and it could be complex. However, in order for the complex time generator $\mathcal{L}_\eta$ to be independent of $\tau_{L/R}$, \emph{we set $\eta$ to be constant}; i.e. when
\begin{equation}\label{eq:eta restriction}
t_R=\eta t_L~,\quad\beta_R=\eta \beta_L~.
\end{equation}
The above definitions reproduce the usual Schrödinger and Heisenberg picture evolution for states and operators used in the Lanczos algorithm when we restrict the problem to $\eta=+1$ and $-1$ respectively {{for the {{relations}} in (\ref{eq:eta restriction})}}.

\textbf{State case}: Given that we work with the two-sided HH state at an initial time
\begin{equation}
    \ket{\Psi_\Delta\qty(\tfrac{\beta_L}{2},\tfrac{\beta_R}{2})}=\rme^{-\frac{\beta_L}{2}\hH_L-\frac{\beta_R}{2}\hH_R}\hat{\mathcal{O}}_{\tilde{\Delta}}\ket{0}~,
\end{equation}
then, this state is in the Schrödinger picture with a total Hamiltonian $\hH_L+\hH_R$, meaning
\begin{equation}
    \rmi\partial_t\ket{\Psi_\Delta}=(\hH_L+\hH_R)\ket{\Psi_\Delta}~,
\end{equation}
which is solved by (\ref{eq:HH state tL tR}) with $t_L=t_R=t$. For this reason, the Lanczos algorithm describes spread complexity (corresponding to the Schrödinger picture) when $t_L=t_R=t$. In principle, we do not require $\beta_L=\beta_R$; this is simply a simplification to threat the problem with a single complex time parameter $\tau$.

\textbf{Operator case}:
As reviewed in App. \ref{sapp:Krylov review}, the Liovillian operator, generating the evolution of an operator $\hat{\mathcal{O}}_\Delta^{L}\equiv \hat{\mathcal{O}}_\Delta\otimes \mathbb{1}$ in a two-sided Hamiltonian system can be expressed as $\hat{\mathcal{L}}=\hH_{R}-\hH_{L}$ (and $\hat{\mathcal{L}}=\hH_{L}-\hH_{R}$ for $\hat{\mathcal{O}}_\Delta^{R}\equiv \mathbb{1}\otimes\hat{\mathcal{O}}_\Delta$). We will use the Choi–Jamiołkowski isomorphism to represent the Heisenberg picture operator {{with respect to one-sided Hamiltonians $\hH_{L/R}$:}}
\begin{equation}
    \hat{\mathcal{O}}^{L/R}_\Delta(t)\equiv\rme^{-\rmi\mathcal{L}t}\hat{\mathcal{O}}^{L/R}_\Delta=\rme^{\rmi \hH_{L/R}t}\hat{\mathcal{O}}^{L/R}_\Delta\rme^{-\rmi \hH_{L/R}t}~,
\end{equation}
into a state $\hat{\mathcal{O}}_\Delta(t)\ket{0}$. In order to define the Hilbert space of the states resulting from the above operator-state map, we need to specify their inner product, which in this case involves a thermal ensemble with arbitrary inverse temperatures $\beta_L$, $\beta_R$. There are different choices satisfying specific axioms, as reviewed in \cite{Nandy:2024htc,Sanchez-Garrido:2024pcy} in the one-sided Hamiltonian case. Given that there is no unique way to generalize the different choices of thermal ensemble inner product in two-sided Hamiltonian systems with different temperatures, we are free to choose any that satisfies the inner product space axioms \cite{Nandy:2024htc,Sanchez-Garrido:2024pcy} at this point, so for technical convenience we adopt $\rme^{-\frac{\beta_L}{2}\hH_L-\frac{\beta_R}{2}\hH_R}\hat{\mathcal{O}}_{\tilde{\Delta}}$ as a single or composite operator that is mapped through the Choi–Jamiołkowski isomorphism to a state:\footnote{{{This is compatible with (45) in \cite{Lin:2022rbf}, a natural choice of inner product in the double scaled algebra.}}}
\begin{equation}\label{eq:thermal ensemble inner product two sided}
\begin{aligned}
\rme^{-\frac{\beta_L}{2}\hH_L-\frac{\beta_R}{2}\hH_R}\hat{\mathcal{O}}^{L/R}_{\Delta}(t)\rightarrow \rme^{-\frac{\beta_L}{2}\hH_L-\frac{\beta_R}{2}\hH_R}\hat{\mathcal{O}}^{L/R}_{\Delta}(t)\ket{0}~,
\end{aligned}
\end{equation}
which comes equipped with the inner product in $\mathcal{H}_m$ (which in this case a $\mathcal{H}_1$ irrep).

Thus, we can treat the Krylov space problem for operators (\ref{eq:Lanczos alg Operator comp}), (\ref{eq:Krylov complexity}) in the same way as for states in (\ref{eq:Schro}, \ref{eq:spread Complexity}) {{with}} the replacement $\hH\rightarrow\hat{\mathcal{L}}$.

\paragraph{Both cases}
We now work with a Krylov complexity measure labeled by $\eta$ for $t_R=\eta t_L$ (corresponding to spread or Krylov operator complexity when $\eta=+1$ or $-1$ respectively). Note that the requirement $\beta_R=\eta \beta_L$ is just a matter of technical convenience that allows us to find closed form expressions for the Lanczos coefficients and Krylov basis. One could use arbitrary values of $\beta_R$ and $\beta_L$ that are not necessarily related to each other, compared to (\ref{eq:eta restriction}). This essentially corresponds to a different choice of thermal ensemble in the evaluation of the inner product, as seen in the previous cases. On the other hand, one is tempted to formulate new measures of Krylov complexity that interpolate between state and operator complexity by allowing $\eta\in\mathbb{R}$; however, we have not found a Krylov basis for $\eta\neq\pm 1$ that solves the Lanczos algorithm below. For this reason, we restrict the analysis to $\eta=\pm 1$.

We perform the decomposition of the two-sided HH state into its Krylov basis
\begin{equation}\label{eq:decomposition Krylov}
    \ket{\Psi_{\tilde{\Delta}}(\tau_L=\eta\tau,\tau_R=\tau)}=\sum_{n=0}^\infty\Psi^{(\eta)}_n(\tau)\ket{K^{(\eta)}_n}~,
\end{equation}
where $\Psi^{(\eta)}_n(\tau)\equiv\bra{K_n^{(\eta)}}\ket{\Psi_{\tilde{\Delta}}(\tau_L=\eta\tau,\tau_R=\tau)}$, and the orthonormal basis $\qty{\ket{K_n^{(\eta)}}}$ is generated from the Lanczos algorithm
\begin{equation}\label{eq:Lanczos eta}
    \begin{aligned}
        \hat{\mathcal{L}}_\eta\ket{K^{(\eta)}_n}&\equiv(\hH_R+\eta\hH_L)\ket{K^{(\eta)}_n}\\
        &=b^{(\eta)}_{n+1}\ket{K_{n+1}^{(\eta)}}+b^{(\eta)}_{n}\ket{K_{n-1}^{(\eta)}}+a_n^{(\eta)}\ket{K_{n}^{(\eta)}}~,
    \end{aligned}
\end{equation}
where $\ket{K_0^{(\eta)}}=\ket{\tilde{\Delta};0,\dots,0}=\ket{{\Delta_{\rm S}};0,0}$ (by (\ref{eq:important product to sum weights})). By demanding that $\eta$ in (\ref{eq:eta restriction}) is a fixed number, $\hat{\mathcal{L}}_\eta$ is $\tau$-independent. The amplitudes $\Psi^{(\eta)}_n(\tau)$ solve a Lanczos algorithm of the form:
\begin{equation}\label{eq:Lanczos algorithm eta}
    -\partial_\tau\Psi^{(\eta)}_n(\tau)=b^{(\eta)}_{n+1}\Psi^{(\eta)}_{n+1}(\tau)+b^{(\eta)}_n\Psi^{(\eta)}_{n-1}(\tau)+a^{(\eta)}_n\Psi^{(\eta)}_{n}(\tau)~,
\end{equation}
where $\Psi_n^{(\eta)}\qty(\tau=\frac{\beta_R}{2})=Z_{\tilde{\Delta}}(\eta\beta,\beta)$.

The {{(generalized)\footnote{{{\eqref{eq:Krylov complexity eta} is a notion of generalized Krylov complexity with two-sided Hamiltonian. It is related to recent work in \cite{FarajiAstaneh:2025thi}. However, we are most interested in the case $\eta=\pm1$ since our derivation of the explicit Krylov basis in the semiclassical limit is only valid for $\eta=\pm1$, as we elaborate in App. \ref{app:details Lanczos Krylov}. Nevertheless we expect that our result for $\ell(t_L,t_R)$ $t_L\neq\pm t_R$ \eqref{eq:womrhole semiclassical answer many} has a Krylov interpretation based on the generalized notion of Krylov complexity.}}}}} Krylov complexity operator, and its expectation value are just
\begin{equation}\label{eq:Krylov complexity eta}
\begin{aligned}
        \hat{\mathcal{C}}&=\sum_n n\ket{K^{(\eta)}_n}\bra{K^{(\eta)}_n}~,\\
    \mathcal{C}^{(\eta)}(t)&=\eval{\frac{\bra{\Psi_{\tilde{\Delta}}(\eta\tau,\tau)}\hat{\mathcal{C}}\ket{\Psi_{\tilde{\Delta}}(\eta\tau,\tau)}}{\bra{\Psi_{\tilde{\Delta}}(\eta\frac{\beta}{2},\frac{\beta}{2})}\ket{\Psi_{\tilde{\Delta}}\qty(\eta\frac{\beta}{2},\frac{\beta}{2})}}}_{\tau=\frac{\beta}{2}+\rmi t}~.
\end{aligned}
\end{equation}
To identify the specific Krylov basis corresponding to the two-sided HH state (\ref{eq:HH state tL tR}) with $\tau_L=\eta\tau_R$, we can make an ansatz where the total chord number is fixed by expanding the term $\rme^{-\tau\hat{\mathcal{L}}_\eta}$ with the binomial theorem:
\begin{equation}\label{eq:Psi tauLtauR}
    \ket{\Psi_{\tilde{\Delta}}(\tau_L=\eta\tau,\tau_R=\tau)}=\sum_{n=0}^\infty \frac{(-\tau)^n}{n!}\sum_{l=0}^n\begin{pmatrix}
        n\\l
    \end{pmatrix}\eta^l\hH_L^l\hH_R^{n-l}\ket{\Delta_{\rm S};0,0}~,
\end{equation}
where, again, we denote $\Delta_{\rm S}=\sum_{i=1}^m\Delta_i$. Using the normal ordering prescription of \cite{Xu:2024hoc},
\begin{equation}\label{eq:normal order basis}
    \ket{\Delta;n_1,n_2}=:\hH_L^{n_1}\hat{\mathcal{O}}_\Delta\hH_R^{n_2}:\ket{0}~,
\end{equation}
one can then build a fixed total chord number basis corresponding to (\ref{eq:Psi tauLtauR}) (which in does not necessarily need to be a Krylov basis):
\begin{equation}\label{eq:normal ordered total chord number}
    \begin{aligned}
        \ket{k^{(\eta)}_n}&=c^{(\eta)}_n\sum_{l=0}^n\eta^k\begin{pmatrix}
        n\\
        l
    \end{pmatrix}\ket{\Delta_{\rm S};l,n-l}~,
    \end{aligned}
\end{equation}
where $c_n^{(\eta)}$ is a normalization constant, shown explicitly in (\ref{eq:normalization Krylov}). The basis becomes orthonormal in the semiclassical limit, in the sense of $\bra{k^{(\eta)}_n}\ket{k^{(\eta)}_m}\eqlambda\delta_{nm}$ (as shown in App. \ref{app:details Lanczos Krylov}).

However, it is not guaranteed that $\qty{\ket{k^{(\eta)}_n}}$ solves the Lanczos algorithm (\ref{eq:Lanczos eta}) which defines the Krylov basis $\qty{\ket{K^{(\eta)}_n}}$ (hence the different notation). Thus, \begin{equation}
\begin{aligned}
    \label{eq:chord number as Krylov complexity op}
    \frac{\bra{\Psi_{\tilde{\Delta}}(\eta\tau,\tau)}\hat{N}\ket{\Psi_{\tilde{\Delta}}(\eta\tau,\tau)}}{\braket{\Psi_{\tilde{\Delta}}(\eta\frac{\beta}{2},\frac{\beta}{2})}{\Psi_{\tilde{\Delta}}(\eta\frac{\beta}{2},\frac{\beta}{2})}}&=(Z_{\tilde{\Delta}}(\eta\beta,\beta))^{-1}\sum_{n=0}^\infty n\abs{\braket{k^{(\eta)}_n}{\Psi_{\tilde{\Delta}}(\eta\tau,\tau)}}^2~,
\end{aligned}
\end{equation} 
does not need to represent Krylov complexity. It was shown in \cite{Ambrosini:2024sre} that only in the exact semiclassical regime the total chord number basis corresponds to the Krylov basis for $\ket{\Delta;0,0}$ as the initial state in the Lanczos algorithm with $t_L=\pm t_R$ and at infinite temperature. In App. \ref{app:details Lanczos Krylov}, we show that the Krylov basis for the two-sided HH state in (\ref{eq:HH state tL tR}) is given by the normal ordering prescription basis $\qty{\ket{k^{(\eta)}_n}}$ in (\ref{eq:normal ordered total chord number}) in the semiclassical limit $\lambda\rightarrow0^+$ (while $n\lambda$ is held fixed,{{ see \eqref{eq:semiclassical}}}). This means that $\ket{K_n^{(\eta)}}\eqlambda\ket{k_n^{(\eta)}}$ up to the addition of a state whose norm is negligible in the semiclassical limit.

The Lanczos coefficients are found in (\ref{eq:Lanczos many}). Thus, the Krylov complexity for $t_L=\pm t_R$ and temperature $\beta_L=\pm\beta_R$ in the $\mathcal{H}_m$ double-scaled PETS (\ref{eq:HH state tL tR}) indeed is given by the expectation value in (\ref{eq:chord number as Krylov complexity op}) in the $\lambda\rightarrow0$ regime.\footnote{{{It would be interesting to derive the first non-trivial correction in $\lambda$ for (\ref{eq:chord number as Krylov complexity op}), as we elaborate in Sec. \ref{ssec:outlook}.}}} {{Note that $\eta\neq\pm1$ might require a different ansatz, as we explain below \eqref{eq:results Lanczos eta}.}} We also deduced simplified the amplitudes entering in the Lanczos algorithm (\ref{eq:Lanczos eta}); see (\ref{eq:Explicit amplitude}) for the details.

\subsection{Krylov complexity from the two-sided Hartle-Hawking state}\label{ssec:special}

Following the discussion in Sec. \ref{ssec:total chord as Krylov}, (\ref{eq:womrhole semiclassical answer many}) represents spread complexity (i.e. $\tau_L=\tau_R$) and Krylov operator complexity (i.e. $\tau_L=-\tau_R$). This means:
\begin{equation}\label{eq:relation length and Krylov}
    \lambda~\mathcal{C}^{(\eta)}(t)\equiv\ell(t_L=\eta t,t_R=t)~.
\end{equation}
We notice that the non-zero initial value for Krylov complexity $\mathcal{C}^{(\eta)}(t=0)=\ell_*/\lambda$ in a HH preparation of state can be interpreted as the survival amplitude for $\ket{0}$ to remain unchanged during its evolution. This has been discussed in \cite{Erdmenger:2023wjg,Balasubramanian:2022tpr} for spread complexity in finite dimensional quantum systems.

Furthermore, since the chord number operator corresponds to Krylov complexity operator in the semiclassical regime ($\lambda\rightarrow0$), it follows that the Krylov complexity generating function is exactly given by the thermal two-point correlation function in (\ref{eq:2point correlation function m particles}) when $t_L=\eta t_R$, $\beta_L=\eta \beta_R$, 
\begin{align}\label{eq:generating function 1 particle}
&G^{(\Delta_w)}_{\tilde{\Delta}}(\tau_L=\eta\tau,\tau_R=\tau)=\frac{\bra{\Psi_{\tilde{\Delta}}(\eta\tau,\tau)}q^{\Delta_w \hat{N}}\ket{\Psi_{\tilde{\Delta}}(\eta\tau,\tau)}}{\bra{\Psi_{\tilde{\Delta}}(\eta\frac{\beta}{2},\frac{\beta}{2})}\ket{\Psi_{\tilde{\Delta}}(\eta\frac{\beta}{2},\frac{\beta}{2})}}~.
\end{align}
Here $\hat{N}=\hat{n}_L+\hat{n}_1+\dots+\hat{n}_{m-1}+\hat{n}_R$. We can look for the Krylov complexity moments encoded in (\ref{eq:generating function 1 particle}) (where Krylov complexity corresponds to the first moment) using
\begin{equation}\label{eq:rule 1particle}
    \expval{\hat{\mathcal{C}}^k}\eqlambda\eval{\qty(\frac{-1}{\lambda})^{k}\frac{\partial^k}{\partial\Delta_w^k}G^{(\Delta_w)}_{\tilde{\Delta}}(\eta\tau,\tau)}_{\Delta_w=0}~.
\end{equation}
Now, we would like to determine Krylov complexity for states and operators with temperature dependence from the two-sided HH state (\ref{eq:HH state tL tR}). Following our discussion in Sec. \ref{ssec:total chord as Krylov}, we stress that the dynamical distance variable (\ref{eq:womrhole semiclassical answer many}) that we have derived from classical phase space \emph{only} has a clear interpretation as semiclassical Krylov complexity (i.e. it has a Krylov basis obeying the Lanczos algorithm (\ref{eq:Lanczos eta})) when we demand that both $t_L=\eta t_R\equiv\eta t$ and $\beta(\theta_L)=\eta\beta(\theta_R)\equiv\eta \beta(\theta)$ {{(\ref{eq:entropy temp})}} \emph{for $\eta=\pm1$}. This places restrictions on the relationship between the angular parameterization of the energy spectrum of the left and right chord sectors (i.e. in $E_{L/R}=E(\theta_{L/R})$ (\ref{eq:energy spectrum})).

\paragraph{Spread complexity}
In the particular case where $t_L=t_R$, we require $\beta(\theta_L)=\beta(\theta_R)$, where the inverse temperature ((\ref{eq:entropy temp}))  with $n=0$)
\begin{equation}\label{eq:beta stable saddle point}
    \beta(\theta)=2\frac{\pi-2\theta}{J\sin\theta}~,
\end{equation}
which for $\theta_{L/R}\in[0,\pi]$ (\ref{eq:beta stable saddle point}) implies $\theta_L=\theta_R\equiv\theta$. This condition in (\ref{eq:ell cond thetaLR}) {{says}} that $\ell_{t_L=t_R=0}(0)=\ell_{m}(t_L=t_R=0)\equiv\ell^{(+)}_*(\theta)/2$ (\ref{eq:ell * state}) since they satisfy the same equation. 

Thus, the corresponding spread complexity for states of the form
\begin{equation}
    \rme^{-\qty(\frac{\beta(\theta)}{2}+\rmi t)(\hH_L+\hH_R)}\hat{\mathcal{O}}_{\Delta_1}\dots\hat{\mathcal{O}}_{\Delta_m}\ket{0}~,
\end{equation}
at \emph{leading order} as $\lambda\rightarrow0$ ($\frac{\lambda}{1-q}\rightarrow1$) is then given by (\ref{eq:womrhole semiclassical answer many}) with $t_L=t_R$ and $\beta_L=\beta_R$, i.e.:
\begin{equation}\label{eq:Krylov complexity State many}
\begin{aligned}
    &\mathcal{C}^{(\eta=+1)}(t)=\frac{2}{\lambda}\log\qty(1+\qty(1+\frac{\rme^{-\ell^{(+)}_*(\theta)}q^{\Delta_{\rm S}}}{\sin^2\theta})\sinh^2\qty(J\sin\theta ~t))+\frac{\ell^{(+)}_*(\theta)}{\lambda}\\
    &=\frac{2}{\lambda}\log(A(\theta,q^{\Delta_{\rm S}})+B(\theta,q^{\Delta_{\rm S}})\cosh(2J\sin\theta t))~,
\end{aligned}
\end{equation}
where $\ell^{(+)}_*(\theta)$ was defined in (\ref{eq:ell * state}), and
\begin{align}
    &A(\theta,q^\Delta)=\frac{(\cos \theta
   -\cos 2 \theta ) q^{\Delta
   }+\left(1-q^{\Delta }\right) \left(q^{\Delta
   }+\sqrt{q^{2 \Delta }-2 \cos \theta 
   q^{\Delta }+1}-1\right)}{2
   \sin ^2\theta \left(q^{\Delta }+\sqrt{q^{2 \Delta }-2 \cos
   \theta  q^{\Delta }+1}-1\right)}~,\label{eq:A theta}\\
   &B(\theta,q^\Delta)=\frac{q^{2 \Delta
   }-\left((\cos \theta +\cos 2 \theta )
   q^{\Delta }\right)-\left(1-q^{\Delta
   }\right) \sqrt{q^{2 \Delta }-2 \cos \theta
    q^{\Delta }+1}+1}{2
   \sin ^2\theta \left(q^{\Delta }+\sqrt{q^{2 \Delta }-2 \cos
   \theta  q^{\Delta }+1}-1\right)}~.\label{eq:B theta}
\end{align}
We can notice that (\ref{eq:Krylov complexity State many}) displays a transition between parabolic to late-time linear growth.

\paragraph{Comparison with the literature}We can now compare the result with the existing literature. (\ref{eq:Krylov complexity State many}) agrees with (4.16) in \cite{Ambrosini:2024sre} when we consider a one-particle chord and $\theta=\pi/2$ (corresponding to an infinite temperature limit, where $\ell^{(+)}_*=0$ from (\ref{eq:ell * state})). This means that (\ref{eq:womrhole semiclassical answer many}) is also supported by the numerical evaluation of Krylov complexity for perturbations of the TFD state at infinite temperature in \cite{Ambrosini:2024sre}. Furthermore, the expression (\ref{eq:Krylov complexity State many}) with ${\Delta}_1=\cdots=\Delta_m=0$ and $J=1$ also matches \cite{Heller:2024ldz}.\footnote{Note that \cite{Heller:2024ldz} adopts $\lq=\lambda=p^2/N$, instead of $\lambda=2p^2/N$ in our case. This produces the overall coefficient $\frac{1}{2}\sin\theta$ accompanying $t$.} where $\rme^{-\ell^{(+)}_*}\rightarrow\sin^2\theta$. {{Furthermore, we can compare wormhole with the same initial length with and without matter insertions, from (\ref{eq:ell * state}) and $\rme^{-\ell^{(+)}_*}\rightarrow\sin^2\theta$, which have different boundary energies, $\theta_{\rm matter}$ and $\theta_{\rm empty}$ respectively. We find the relationship, valid at the semiclassical level of the derivation:
\begin{equation}\label{eq:new eqq}
    \frac{2J}{\lambda}\cos\theta_{\rm matter}=\frac{2J}{\lambda}\sqrt{1+q^{-2\Delta_{\rm S}}(1+q^{\Delta_{\rm S}})\sin(\theta_{\rm empty})-q^{-3\Delta_{\rm S}}\sin^2\theta_{\rm empty}}~.
\end{equation}
This indicates a modification in the boundary energy spectrum (for fixed initial wormhole length). On the other hand, the authors in \cite{Ambrosini:2024sre} noticed in the evaluation of Krylov complexity that one can recast the problem in terms of an effective potential energy term encoding information about the conformal dimension of the chord matter operator. The authors associated the difference between the potential for $\Delta\neq0$ and $\Delta=0$ (Fig. 9 \cite{Ambrosini:2024sre}) to the insertion of shockwaves in the bulk. \eqref{eq:new eqq} is reminiscent of their observation, although we compare different quantities. Nevertheless, their conclusions are compatible with the bulk interpretation that we carry on in Sec. \ref{ssec:towardsquantization}.}}

\paragraph{Krylov operator complexity}\label{ssec:Krylov operator}
For Krylov operator complexity, we recover a Lanczos algorithm of the form (\ref{eq:Lanczos algorithm eta}) when $t_L=-t_R\equiv -t$ and\footnote{The reader is referred to App. \ref{app:partition1p} for a proof that this relation holds regardless whether heavy or light operators enter in the two-sided HH state \ref{eq:HH state tL tR}.}
\begin{equation}
    \frac{2(\pi-2\theta_L)}{\sin\theta_L}=-\frac{2(\pi-2\theta_R)}{\sin\theta_R}~,
\end{equation}
i.e. $\beta(\theta_L)=-\beta(\theta_R)$. This condition is satisfied by $\theta_R=\pi-\theta_L$; which implies that the parameter $\sin\theta_L=\sin\theta_R\equiv\sin\theta$ (i.e. $\gamma_L=\gamma_R$ (\ref{eq:gamma parameter})). This means that 
\begin{equation}
\ell_{0}(t_L=t_R=0)=\ell_{m}(t_L=t_R=0)=\ell^{(-)}_*/2    
\end{equation}
as the initial condition in (\ref{eq:ell cond thetaLR}) (which relates the initial total distance variable with the energy of the left/right chord sectors). Denoting $\theta_R=\pi-\theta_L\equiv\theta$, (\ref{eq:ell cond thetaLR}) becomes
\begin{equation}\label{eq:ell * operator}
\rme^{-\frac{1}{2}\ell^{(-)}_*(\theta)}=\frac{q^{-\Delta_{\rm S}}}{2}\qty(1+q^{\Delta_{\rm S}}-\sqrt{1+q^{2\Delta_{\rm S}}+2q^{\Delta_{\rm S}} \cos2\theta})~.
\end{equation}
Thus, using the Choi–Jamiołkowski isomorphism (App. \ref{sapp:Krylov review}), we can feed the Lanczos algorithm (\ref{eq:Lanczos eta}) with reference (Heisenberg picture) operators of the form
\begin{equation}
    \rme^{-\qty(\frac{\beta(\theta)}{2}+\rmi t)(\hH_R-\hH_L)}\hat{\mathcal{O}}_{\Delta_1}\dots\hat{\mathcal{O}}_{\Delta_m}~,
\end{equation}
to recover the Krylov operator complexity (\ref{eq:womrhole semiclassical answer many}) at \emph{leading order} in the $\lambda\rightarrow0$ limit
\begin{equation}
\begin{aligned}
    \mathcal{C}^{(\eta=-1)}(t)&=\frac{2}{\lambda}\log\qty(1+\qty(1-\frac{\rme^{-\ell^{(-)}_*(\theta)}q^{\Delta_{\rm S}}}{\sin^2\theta})\sinh^2\qty(J\sin\theta~ t))+\frac{\ell^{(-)}_*(\theta)}{\lambda}\\
    &=\frac{2}{\lambda}\log\qty(A(\theta,-q^{\Delta_{\rm S}})+B(\theta,-q^{\Delta_{\rm S}})\cosh(2J\sin\theta~t))~,
\end{aligned}\label{eq:Krylov complexity Operator many}
\end{equation}
where $A(\theta,q^\Delta)$, $B(\theta,q^{\Delta})$ are shown in (\ref{eq:A theta}), (\ref{eq:B theta}). {{As in \eqref{eq:solutions HH}, we emphasize that this relation assumes $q^{\Delta_{\rm S}}$ is a fixed parameter in the semiclassical regime (\ref{eq:semiclassical}). We note that (\ref{eq:Krylov complexity Operator many}) takes the same form as (\ref{eq:shockwave approx}) after replacing $\ell^{(-)}_*\rightarrow\ell^{(+)}_*$ in (\ref{eq:ell * operator}). Furthermore, }}one can see that (\ref{eq:Krylov complexity Operator many}) experiences parabolic to exponential growth as captured by
\begin{subequations}\label{eq:early KOC}
\begin{align}\label{eq:early K operator}
\mathcal{C}^{(\eta=-1)}(t)\simeq& \frac{1}{\lambda}\frac{\sin^2\theta-q^{\Delta_{\rm S}}\rme^{-\ell_*^{(-)}}}{2\sin^2\theta}(2J\sin\theta~t)^2+\frac{\ell^{(-)}_*(\theta)}{\lambda}\\
&\text{when}\quad t\ll(J\sin\theta)^{-1}~,\nonumber\\
\label{eq:intermediate time COp}
    \mathcal{C}^{(\eta=-1)}(t)\simeq&\frac{1}{\lambda}\frac{\sin^2\theta-\rme^{-\ell^{(-)}_*(\theta)}q^{\Delta_{\rm S}}}{\sin^2\theta+\rme^{-\ell^{(-)}_*(\theta)}q^{\Delta_{\rm S}}}\rme^{2J\sin\theta~t}+\frac{2}{\lambda}\log(\frac{\sin^2\theta+\rme^{-\ell^{(-)}_*(\theta)}q^{\Delta_{\rm S}}}{2\sin^2\theta})~\\
    &+\frac{\ell^{(-)}_*(\theta)}{\lambda}~\quad \text{when}\quad\frac{1}{J\sin\theta}\ll t\ll  t_{\rm sc}~,\nonumber
\end{align}
\end{subequations}
where $t_{\rm sc}$ is the transition time such that $\frac{\sin^2\theta-\rme^{-\ell^{(-)}_*}q^{\Delta_{\rm S}}}{\sin^2\theta+\rme^{-\ell^{(-)}_*}q^{\Delta_{\rm S}}}\rme^{2J\sin\theta~t}\approx1$, i.e.
\begin{equation}\label{eq:scrambling time}
 t_{\rm sc}=-\frac{1}{2J\sin\theta}\log\frac{{\sin^2\theta}-{\rme^{-\ell^{(-)}_*(\theta)}}q^{\Delta_{\rm S}}}{{\sin^2\theta}+{\rme^{-\ell^{(-)}_*(\theta)}}q^{\Delta_{\rm S}}}~,
\end{equation}
where $\ell_*^{(-)}(\theta)$ is in (\ref{eq:ell * operator}). At later times than \eqref{eq:scrambling time}, there is a transition to linear growth at
\begin{equation}\label{eq:late time COp}
\begin{aligned}
    \mathcal{C}^{(\eta=-1)}(t)\simeq&\frac{1}{\lambda}\qty(\ell^{(-)}_*(\theta)+{4}J\sin\theta~t+2\log(\frac{\sin^2\theta-\rme^{-\ell^{(-)}_*(\theta)}q^{\Delta_{\rm S}}}{4\sin^2\theta}))~,\quad t\gg  t_{\rm sc}~.
\end{aligned}
\end{equation}
We display a plot of Krylov operator complexity for heavy operators in Fig. \ref{fig:plot_Krylov_operator}.\footnote{Note that $\mathcal{C}^{(\eta=-1)}(t=0)\neq0$ in the examples in the figure, except for the $\theta=\pi/2$ case. The reason for this is that we study Krylov complexity in a finite temperature ensemble (see e.g. \cite{Jian:2020qpp,Heller:2024ldz,Erdmenger:2023wjg}), i.e.\begin{equation}
    \bra{\Psi_\Delta(\tfrac{\beta_L}{2},\tfrac{\beta_R}{2})}\hat{\mathcal{C}}\ket{\Psi_\Delta(\tfrac{\beta_L}{2},\tfrac{\beta_R}{2})}\neq 0~,
\end{equation}
($\hat{\mathcal{C}}$ defined in (\ref{eq:Krylov complexity eta})) which otherwise vanishes when $\beta_L=\beta_R=0$.}
\begin{figure}
    \centering
\includegraphics[width=0.6\linewidth]{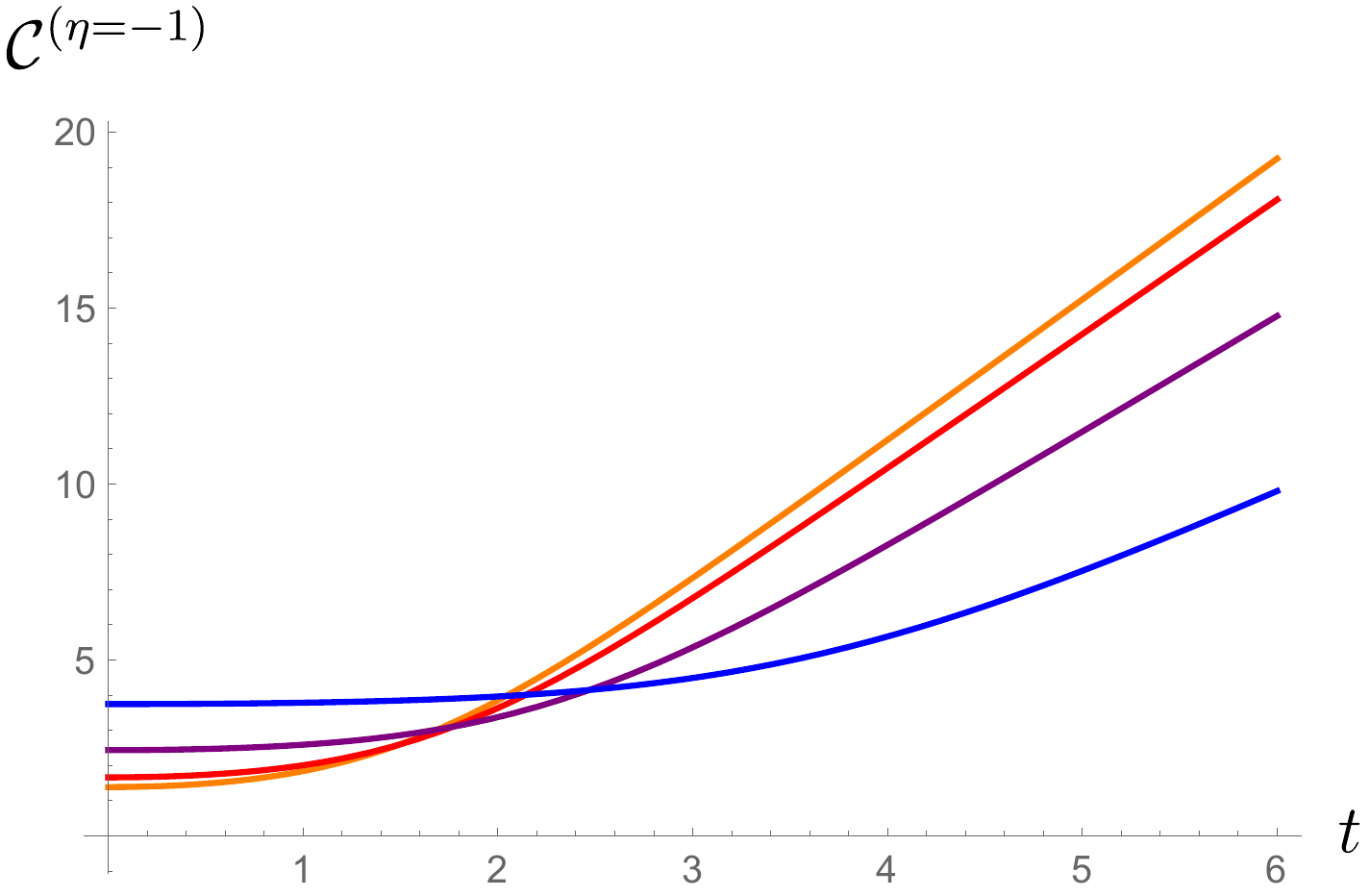}
    \caption{Krylov operator complexity (\ref{eq:Krylov complexity Operator many}). We are taking $\lambda=0.001$, $\lambda\Delta_{\rm S}=1$, $J=1$, and $\theta$ decreases from $\theta=\pi/2$ (orange) in multiples of $0.3$ from top to bottom at late times. Notice the exponential growth (\ref{eq:intermediate time COp}) at the time scale below the scrambling time (\ref{eq:scrambling time}), followed by late time linear growth.}
    \label{fig:plot_Krylov_operator}
\end{figure}

{{
 On the other hand, if one instead takes $\beta_L=+\beta_R$, then the scrambling time becomes (3.38) in \cite{Aguilar-Gutierrez:2025mxf},\footnote{{{\cite{Aguilar-Gutierrez:2025mxf} considers a wormhole density matrix instead of the two-sided HH \eqref{eq:HH state tL tR} to evaluate the expectation values of the total chord number; this amounts to changing $t_L^{(\rm here)}=t_L^{(\rm there)}$ and $t_R^{(\rm here)}=-t_R^{(\rm there)}$ in \ref{eq:womrhole semiclassical answer many} of this work in comparison with (3.25) in \cite{Aguilar-Gutierrez:2025mxf} where $\beta_L=\beta_R$ is assumed. Thus, the result for the scrambling time in \eqref{eq:plus case} applies, which is derived in \cite{Aguilar-Gutierrez:2025mxf} for $t^{(\rm there)}_L=t^{(\rm there)}_R$.}}} which we display here for further discussion below
\begin{equation}\label{eq:plus case}
t_{\rm sc}^{(+)}=\frac{1}{2J\sin\theta}\log\frac{{\sin^2\theta}+{\rme^{-\ell^{(+)}_*(\theta)}}q^{\Delta_{\rm S}}}{{\sin^2\theta}-{\rme^{-\ell^{(+)}_*(\theta)}}q^{\Delta_{\rm S}}}~, 
\end{equation}
where $\ell^{(+)}_*(\theta)$ appears in \eqref{eq:ell * state}.
\paragraph{Validity of the approximation}
Let us now discuss about $\lambda\rightarrow0$ in the scrambling time. One can wonder whether the approximation $\expval{\hN^k}\eqlambda\expval{\hN}^k$ is valid until the scrambling time in \eqref{eq:scrambling time}, or \eqref{eq:plus case}, since in principle the quantum corrections to answer \eqref{eq:womrhole semiclassical answer many} can also depend with time. For instance, if one were to find that the first quantum correction to \eqref{eq:womrhole semiclassical answer many} is of the type $\lambda\rme^t$, then one would need to consider this in the answer after a scrambling time proportional to $\log\lambda^{-1}$. Below we examine the scaling of the scrambling time \eqref{eq:scrambling time} and \eqref{eq:plus case} in the $\lambda\rightarrow0$ limit:
\begin{itemize}
    \item When $\hmO_{\tilde{\Delta}}$ is a (composite or single) \textbf{heavy operator} (i.e. $q^{\Delta_{\rm S}}$ is fixed as $\lambda\rightarrow0$ (\ref{eq:heavy and light})) the scrambling time in both \eqref{eq:scrambling time} and \eqref{eq:plus case} are fixed as $\lambda\rightarrow0$ since $q^{\Delta_{\rm S}}$ and therefore also $\rme^{-\ell_*^{(\pm)}}$ (above) are fixed, meaning that the scrambling time is $\mathcal{O}(1)$ as $\lambda\rightarrow0$. Indeed, the derivation of \eqref{eq:solutions HH} assumes that $q^{\Delta_{\rm S}}$ is fixed as $\lambda\rightarrow0$. So, while there could be quantum corrections for \eqref{eq:womrhole semiclassical answer many} growing with time (e.g. $\lambda~\rme^t$); as long as we work in the semiclassical regime $\lambda\rightarrow0$, then $\ell(t_L,t_R)$ is described at leading order by \eqref{eq:womrhole semiclassical answer many} in the limit $\lambda\rightarrow0$  past beyond the scrambling time \eqref{eq:scrambling time} till when the quantum corrections become $\mathcal{O}(1)$ (which is left for future directions, see Sec. \ref{ssec:outlook}).
    \item When $\hmO_{\tilde{\Delta_{\rm S}}}$ is (a composite or single) \textbf{light operator}  (where $\Delta_{\rm S}$ is fixed as $\lambda\rightarrow0$ \eqref{eq:heavy and light}) the analysis is more subtle. Let us analyze the cases \eqref{eq:scrambling time} ($\beta_L=-\beta_R$) and \eqref{eq:plus case} ($\beta_L=+\beta_R$) separately.
    \begin{enumerate}
        \item When $\beta_L=-\beta_R$, a Taylor expansion of \eqref{eq:scrambling time} shows that 
        \begin{equation}
        t_{\rm scr}\eqlambda-\frac{1}{2J\sin\theta}\log(\abs{\cos\theta})~.    
        \end{equation}
        Thus, the scrambling time that we consider in this work remains $\mathcal{O}(1)$ as $\lambda\rightarrow0$ for light operators; similar to the case with heavy operators before.
        \item In contrast when $\beta_L=+\beta_R$, we see again by doing a Taylor expansion of the scrambling time in \eqref{eq:plus case} that $t_{\rm scr}^{(+)}\eqlambda-\frac{1}{2J\sin\theta}\log\frac{\Delta_{\rm S}\lambda}{2\sin\theta}$, as also seen from \cite{Aguilar-Gutierrez:2025mxf} (3.38). In principle, one needs to show whether there are $\lambda~\rme^t$ corrections in (\ref{eq:womrhole semiclassical answer many}) to verify that it evolution is still approximated by the semiclassical answer. We only discuss about the scrambling time \eqref{eq:scrambling time} ($\beta_L=-\beta_R$) in this work.
        \item An interesting limiting case is $\beta_L=\beta_R=0$ where $t_{\rm scr}\eqlambda-\frac{1}{2J}\log(\frac{\Delta_{\rm S}\lambda}{2})$. One can find numerical results about this case including quantum corrections in \cite{Ambrosini:2024sre} (see their Fig 6).\footnote{{{Note however that \cite{Ambrosini:2024sre} uses a slightly different definition of scrambling time that differs from ours \eqref{eq:scrambling time} by an additive constant $(2J)^{-1}\ln2$, which is not relevant in the semiclassical limit of $t_{\rm scr}\eqlambda\frac{1}{2J}\log(\frac{\Delta_{\rm S}\lambda}{2})$.}}} Their evaluation shows that even when $q\rightarrow 1^-$ the analytic answer for Krylov operator complexity (with $\lambda\rightarrow0$), which agrees with our \eqref{eq:womrhole semiclassical answer many} when $\beta_L=\beta_R=0$ and $t_{L}=-t_R$, is very well approximated by the numerical answer. This indicates that there are no $\mathcal{O}(1)$ corrections to the leading order answer \eqref{eq:womrhole semiclassical answer many} in $\lambda\rightarrow0$, at least when $\beta_L=\beta_R=0$ (see Sec. \ref{ssec:outlook} for more comments).
    \end{enumerate}
\end{itemize}}}

\paragraph{Comparison with the literature}
It can be seen that (\ref{eq:Krylov complexity State many}) agrees with \cite{Ambrosini:2024sre} when $\theta=\pi/2$ (corresponding to an infinite temperature limit, where $\ell^{(-)}_*=0$ from (\ref{eq:ell * operator})). This means that (\ref{eq:Krylov complexity Operator many}) (and spread complexity in the previous section) at infinite temperature is also supported by their numerics. (\ref{eq:Krylov complexity Operator many}) exhibits an early time parabolic growth, followed by exponential growth and late-time linear growth. In fact, the results can be mapped to (3.85) in \cite{Ambrosini:2024sre} with replacements $J\rightarrow J\sin\theta$, $\tilde{q}\rightarrow\rme^{-\ell^{(-)}_*(\theta)}\frac{q^{\Delta_{\rm S}}}{\sin^2\theta}$.

\paragraph{Relation with quantum ergodic theory}We remark that the generator of Krylov operator complexity in (\ref{eq:rule 1particle}) is the OTOC we investigated in (\ref{eq:classical phase space OTOC}). As we saw, the OTOC experiences a scrambling transition, which is indeed the scrambling time in Krylov operator complexity. Both can be used as notions of quantum chaos in this system. Moreover, there is a related notion, mixing in quantum ergodic theory (see e.g. \cite{Camargo:2025zxr,Gesteau:2023rrx,Furuya:2023fei}), which quantifies the degree of independence between elements in the operator algebra of a quantum system. In the simplest case, given two operators in the Heisenberg picture $\hat{A}$ and $\hat{B}(t)$, a quantum system is said to be 2-mixing when the cumulant (in this case a connected two-point function)
\begin{equation}
    F_2(t)\equiv \expval{\hat{A}\hat{B}(t)}_\varphi-\expval{\hat{A}}_\varphi\expval{\hat{B}(t)}_\varphi~
\end{equation}
vanishes at late times, i.e. 
\begin{equation}\label{eq:F2 cumulant}
    \lim_{t\rightarrow\infty}F_2(t)=0~,
\end{equation}
and where $\ket{\varphi}$ is a normal state, which allows for continuous expectation values with respect to ultraweak topology \cite{Camargo:2025zxr}.

In the case of the DSSYK model, we note that if we select matter chord operators to generate the connected two-point functions, i.e. $\hat{A}=\hat{\mathcal{O}}_{\Delta_1}$, and $\hat{B}(t)=\hat{\mathcal{O}}_{\Delta_2}(t)$, then $\expval{\hat{A}}_\varphi=\expval{\hat{B}(t)}_\varphi=0~\forall\ket{\varphi}\in\mathcal{H}_m$, while
\begin{equation}\label{eq:F2 statement}
    \lim_{t\rightarrow\infty}F_2(t)=0~\forall~\varphi\in\mathcal{H}_{0}~\text{or}~\mathcal{H}_{1}~.
\end{equation}
The case $\mathcal{H}_{0}$ follows immediately from known results in the literature (see e.g. \eqref{eq:correlator DSSYK}). Meanwhile, the $\mathcal{H}_1$ case follows from the late time regime of (\ref{eq:classical phase space OTOC}). In fact the mixing of the system applies more generally to $\mathcal{H}_m$, as we will see in \cite{Aguilar-Gutierrez:2025mxf}. {{Again, we remark that $F_2$ when $\varphi\in\mathcal{H}_1$ is a four point function OTOC (\ref{eq:classical phase space OTOC}) when $\varphi\in\mathcal{H}_1$.}} Therefore, despite the DSSYK model having a continuous energy spectrum, it displays signatures of quantum chaos, in the sense of Krylov complexity, OTOCs, {{which are related to}} quantum mixing {{in this system}}. Therefore, the DSSYK provides an analytically solvable example where the interplay between these different notions of chaos for $N\rightarrow\infty$ systems at finite temperature are concretely and directly related one to the other.

\paragraph{Universal operator growth hypothesis}
We also emphasize that the Krylov operator complexity in (\ref{eq:Krylov complexity Operator many}) is in agreement with the \emph{universal operator growth hypothesis} (UOGH) \cite{Parker:2018yvk}. The UOGH states that the maximal growth of the Krylov operator complexity is of the type $\exp(\frac{4\pi}{\beta}t)$, with $\beta$ being the inverse temperature of the system, $t$ the time, similar to the bound on the Lyapunov exponent of OTOCs \cite{Maldacena:2015waa}. As we saw (\ref{eq:Krylov complexity Operator many}) can at most grow exponentially (before the scrambling time in  (\ref{eq:intermediate time COp})). In this regime, the UOGH is explicitly saturated when we consider the fake temperature of the DSSYK model (defined as the decay rate of two-point correlation functions \cite{Blommaert:2024ymv}), and undersaturated for the physical temperature (except at $\theta=0$),
\begin{equation}\label{eq:fake temperature result}
    \frac{2\pi}{\beta_{\rm fake}(\theta)}={2J}\sin\theta~,\quad\text{vs}\quad    \frac{2\pi}{\beta(\theta)}=\frac{2J}{2\theta/\pi-1}\sin\theta~.
\end{equation}
{Thus, the Lyapunov exponent for Krylov operator complexity, as well as its generating function (an OTOC with respect to $\mathcal{H}_0$) displays sub-maximal or maximal scrambling depending on whether we compare the Lyapunov exponent to the physical or fake temperature respectively.} In \cite{Lin:2023trc}, it was also found that the Lyapunov exponent was control by a parameter that may generate sub-maximal chaos. This seems to be captured by the factor $J\sin\theta$ in our work.

\paragraph{Later evolution}Lastly, while (\ref{eq:scrambling time}) displays eternal late-time linear growth, it is also expected that at very late times Krylov complexity and other measure of quantum complexity will saturate, when considering finite-dimensional quantum systems instead of the ensemble theory we considered.\footnote{\label{fnt:matrix models}For instance; it has been argued that sine dilaton gravity \cite{Blommaert:2024whf,Blommaert:2024ymv,Blommaert:2025avl} (related to the DSSYK model at the disk level) has a higher genus topology expansion and it can be described by a dual finite cut matrix integral \cite{Blommaert:2025avl}. {{The parameter $\lambda$ in the DSSYK model corresponds to Newton's constant in sine dilaton gravity, as we elaborate in Sec. \ref{sec:bulk}, see (\ref{eq:constants relation}).}} We expect spread complexity for states in that type of model to display a peak behavior  \cite{Balasubramanian:2022tpr,Miyaji:2025ucp}. It would be useful to study the behavior of the different Krylov complexity proposals (\ref{eq:Lanczos eta}) that evaluates either state and operator complexity, in finite matrix models (e.g. if a peak behavior can also be recovered). This might help us to connect with a recent proposal by \cite{Balasubramanian:2024lqk} for recovering the peak behavior in a modification of the DSSYK model, and related systems.}

\subsection{Krylov complexity for precursor operators}\label{ssec:interpretation shockwav krylov}
As we remarked in Sec. \ref{ssec:shockwaves} the addition of the double-scaled precursor (\ref{eq:double scaled precursor shockwave}) translates into a time shift $t_{L}\rightarrow t_{L}-t_w$ and $t_{R}\rightarrow t_{R}+t_w$; which means that all the results we have derived in Sec. \ref{ssec:total chord as Krylov} are unaffected under the replacement of a matter chord to precursor operator: $\hat{\mathcal{O}}^{L/R}_\Delta\rightarrow\hat{\mathcal{O}}^{L/R}_\Delta(t_w)$ in (\ref{eq:double scaled precursor shockwave}). This implies that the two-sided HH state (\ref{eq:HH state tL tR}) transforms as
\begin{equation}\label{eq:shifted HH state}
    \ket{\Psi_{\tilde{\Delta}}(\tau_L,\tau_R)}\rightarrow\ket{\Psi_{\tilde{\Delta}}(\tau_L-\rmi t_w,\tau_R+\rmi t_w)}~.
\end{equation}
Thus, to apply the Lanczos algorithm (\ref{eq:Lanczos algorithm eta}), we simply need to work with shifted variables, such as
\begin{equation}
 \tilde{\tau}_{L}=\frac{\beta_{L}}{2} +\rmi(t_L-t_w)~,\quad \tilde{\tau}_{R}=\frac{\beta_{R}}{2} +\rmi(t_R+t_w)~,
\end{equation}
and impose that $\tilde{\eta}\equiv\tilde{\tau}_L/\tilde{\tau}_R$ is fixed to recover the semiclassical Lanczos coefficients in (\ref{eq:Lanczos eta}), Krylov basis (\ref{eq:normal ordered total chord number}) , and Krylov complexity (\ref{eq:womrhole semiclassical answer many}) with $t_{L}\rightarrow t_{L}-t_w$ and $t_{R}\rightarrow t_{R}+t_w$.

Note however, that in the case of interest in \cite{Shenker:2013pqa}, seen as a early time perturbation of the TFD state, we need to set $\beta_L=\beta_R$ ($E(\theta_L)=E(\theta_R)$), and $t_w\rightarrow\infty$. So the results of the Lanczos algorithm (\ref{eq:Lanczos algorithm eta}) in (\ref{eq:Lanczos eta}, \ref{eq:normal ordered total chord number}, \ref{eq:womrhole semiclassical answer many}) 
apply immediately only when $\beta_L=\beta_R=0$ for $t_L\neq t_R$; or when $\eta=+1$ (the spread complexity case). However, we stress that is is not necessary to fix $\beta_L=\eta\beta_R=\beta(\theta)$. The Lanczos coefficients (\ref{eq:Lanczos eta}), Krylov basis (\ref{eq:normal ordered total chord number}), and Krylov complexity will depend on the parameter $\beta(\theta)$, as we find in (\ref{ssec:interpretation shockwav krylov}).

\section{The holographic dictionary and the dual double-scaled algebra}\label{sec:bulk}
In this section, we derive the holographic dictionary for the DSSYK model with a one-particle insertion, considering a Schrödinger equation for states in the bulk Hilbert space and momentum shift symmetry (\ref{eq:Dirac constraints}); we perform the canonical quantization of the bulk dual theory; and we formulate the double-scaled algebra in bulk terms.

\paragraph{Deriving the holographic dictionary from Krylov complexity}Recent works have found a precise match between spread complexity in the DSSYK model with a minimal geodesic length in an (effective) AdS$_2$ black hole background without matter \cite{Rabinovici:2023yex,Heller:2024ldz}. Previous literature also found that Krylov operator complexity in the finite N SYK model is closely related to the minimal volume in an AdS$_2$ black hole geometry in JT gravity \cite{Jian:2020qpp}. This motivates a closer inspection of Krylov complexity as a realization of the CV proposal by \cite{Susskind:2014rva,Stanford:2014jda}. A natural expectation is that our results from Sec. \ref{sec:deriving Krylov} should have a bulk interpretation in JT gravity or sine dilaton gravity with matter. 
Indeed, ``wormhole distances'', i.e. minimal geodesic lengths between the asymptotic boundaries in an AdS$_2$ black hole, match the Krylov operator and spread complexity found in Sec. \ref{sec:deriving Krylov}. In particular, Krylov operator complexity has promising properties expected of holographic complexity proposals \cite{Belin:2021bga,Belin:2022xmt}, as we explain in \cite{Aguilar-Gutierrez:2025mxf}.

\paragraph{On the role of chord diagrams}One can also interpret the equality between the semiclassical Krylov operator complexity and the geodesic distance $L_{\rm AdS}$ from the holographic dictionary relating crossed four-point correlation functions in the DSSYK model (calculated from chord diagrams in Sec. \ref{app:twopoint oneparticle}) with two-sided two-point correlation functions in the bulk, as previously realized in \cite{Berkooz:2022fso} (albeit under different considerations). This means
\begin{equation}\label{eq:equality propagators}
\expval{q^{\Delta \hat{N}}}\eqlambda\expval{\rme^{-\Delta L_{\rm AdS}}}~.
\end{equation}
The expectation values above are evaluated in the two-sided HH state in the boundary and bulk sides, and they are normalized {{with respect to}} the partition function of the thermal ensemble. From (\ref{eq:equality propagators}) when we identify $\lambda~\mathcal{C}=L_{\rm AdS}$ (in the regime $\lambda\rightarrow0$).

Thus, we see that there are different ways of filling in the holographic dictionary. Namely, chord diagrams, path integrals and Krylov complexity are all closely related, since the chord diagram Hamiltonian is the input for path integral, and the Lanczos algorithm. For this reason, it might be unsurprising they all yield equivalent results for the dual bulk geodesics in the semiclassical limit.

\paragraph{Outline}We tackle this problem by matching our previous results on saddle point solutions of the DSSYK path integral as semiclassical Krylov complexity in terms of minimal length geodesics in the bulk. In Sec. \ref{ssec:Krylov operator bulk picture}, we consider an AdS$_2$ black hole with a minimally coupled massive scalar field generating a small backreaction. We match the {{corresponding bulk length}} to the Krylov operator complexity of the matter chord operator generating the two-sided HH state (\ref{eq:HH state tL tR}). We derive its holographic dictionary. Later, in Sec. \ref{ssec:towardsquantization} we study a shockwave geometry in an AdS$_2$ black hole. We match the minimal geodesic length connecting the asymptotic boundaries and Krylov operator and state complexity. The {{different}} types of Krylov complexity for the two-sided HH state represent different relative boundary time evolution for the minimal length geodesics. In Sec. \ref{ssec:canonical quantization} we {{explore consequences of our holographic dictionary findings for the canonical}} quantization {{of}} sine dilaton gravity {{with constraints, and to formulate the double-scaled algebra in bulk terms}}. In Sec. \ref{ssec:proper momentum} we conclude exploring the relation between our results on the geometric interpretation of Krylov complexity and the proper radial momentum of a probe particle in the bulk, motivated by recent developments \cite{Caputa:2024sux}.

\subsection{Minimal length geodesic with a massive particle in AdS\texorpdfstring{$_2$}{}}\label{ssec:Krylov operator bulk picture}
As a warm-up, we will discuss about the bulk interpretation for Krylov operator complexity of the two-sided HH state (\ref{eq:HH state tL tR}) in terms of minimal geodesic lengths in AdS$_2$ with a massive particle insertion. We explore this first as this calculation {{has limitations regarding the backreaction of the matter field which are alleviated in the shockwave geometry}}. However, since the shockwave case is a particular type of backreaction in the bulk, the goal of this subsection is to see if there is a more generic type of backreaction that reproduces the same observable, after which we will specialize in shockwave case (Sec. \ref{ssec:towardsquantization}).

Thus, following our discussion on the holographic dynamical properties captured by Krylov complexity, we search for a bulk interpretation of the results so far, and in particular, to confirm if (\ref{eq:womrhole semiclassical answer many}) describes the dynamics of a wormhole distance corresponding to Krylov complexity of the two-sided HH state (\ref{eq:HH state tL tR}). By wormhole distance, we refer to a geodesic distance between boundary particles with different {{two-sided boundary}} times; namely $t_R=\eta t_L$ with $\eta=\pm1$ correspond to Krylov complexity in the semiclassical limit shown in Fig. \ref{fig:one_particle_Euclidean}. More general saddle point solutions to the path integral (\ref{eq:Hamilton PI}) describe other {{two-sided boundary times}}. For this reason, we will study AdS$_2$ black hole geometries with matter operator insertions in this subsection and match the saddle points with the corresponding wormhole distance. We remark that the computations are valid for a general backreacted hyperbolic AdS$_2$ geometry by a minimally coupled massive field, which is precisely that for the massive scalar field in sine dilaton gravity (\ref{eq:non-minimal scalar}). 
We will find that Krylov operator complexity (\ref{eq:Krylov complexity Operator many}) indeed reproduces the geodesic distance depicted in Fig. \ref{fig:one_particle_Euclidean} (b) (albeit the bulk side of the calculation is perturbative); while spread complexity (\ref{eq:Krylov complexity State many}), does not precisely match with the corresponding geodesic distance ($t_L=t_R$) beyond the $\lambda\rightarrow0$ limit.

A possible description of the double-scaled PETS in bulk terms, similar to the finite $N$ SYK case \cite{Goel:2018ubv} is that the two-sided HH state (\ref{eq:HH state tL tR}) are constructed by inserting an operator in the Euclidean boundary of the geometry that bisects the interior, as shown in Fig. \ref{fig:one_particle_Euclidean}. 
\begin{figure}
    \centering
    \subfloat[]{{\includegraphics[width=0.52\textwidth]{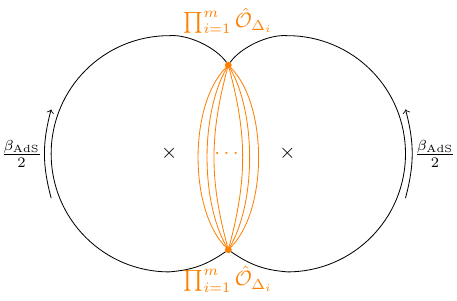}}}\hfill\subfloat[]{\includegraphics[width=0.47\textwidth]{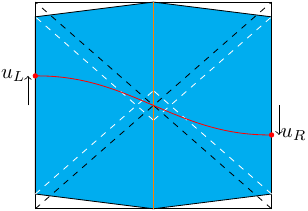}}
    \caption{Bulk interpretation of the double-scaled PETS (\ref{eq:HH state tL tR}) described by the dual quantum system (\ref{eq:H LR many particles1}, \ref{eq:H LR many particles2}). (a) Representation of the thermal partition function $Z_{\tilde{\Delta}}(-\beta,\beta)=\bra{\Psi_{\tilde{\Delta}}(-\frac{\beta}{2},\frac{\beta}{2})}\ket{\Psi_{\tilde{\Delta}}(-\frac{\beta}{2},\frac{\beta}{2})}$, where $\prod_i\hat{\mathcal{O}}_{\Delta_i}$ are inserted at the Euclidean boundary (orange dots), causing backreaction in the putative Euclidean bulk geometry, and dividing it into left and right sectors of the thermal circle with fake periodicity $\beta_{\rm fake}^{R}=\beta_{\rm fake}^L=\beta_{\rm AdS}$. The crosses, $\times$, represent the location of the conical singularities. (b) Lorentzian continuation. The operator insertions are collapsed to a single line (orange), and the curve joining the asymptotic boundaries (in red) represents the wormhole distance $(Z_{\tilde{\Delta}}(-\beta,\beta))^{-1}\bra{\Psi_{\tilde{\Delta}}(-\tau,\tau)}\hat{\ell}\ket{\Psi_{\tilde{\Delta}}(-\tau,\tau)}$ (\ref{eq:womrhole semiclassical answer many}), where $\tau=\frac{\beta}{2}+\rmi t$. Dashed white lines denote the location of the black hole horizons of the AdS geometries without backreaction from the massive particle, and the black dashed line the location of the event horizon with backreaction.}
    \label{fig:one_particle_Euclidean}
\end{figure}
In Lorentzian signature, we consider AdS$_2$ black holes
\begin{equation}\label{eq:effective geometry}
\begin{aligned}
    &\rmd s_{\rm AdS}^2=-f_{L/R}(r)\rmd u_{L/R}^2+\frac{\rmd r^2}{f_{L/R}(r)}~,\\
    &f_{L/R}(r)=r^2-\qty(\Phi_h^{L/R})^2~,
\end{aligned}
\end{equation}
where $r=\Phi_h^{L/R}={2\pi}/{\beta_{\rm AdS}^{L/R}}$ is the location of the event horizon, and $\beta_{\rm AdS}^{L/R}$ the inverse temperature. We allow for $\beta_{\rm AdS}^{L}$ and $\beta_{\rm AdS}^{R}$ to be independent of each other. Meanwhile, from the boundary side, we require $\beta_L=-\beta_R$ (Sec. \ref{ssec:total chord as Krylov}) in order for the Krylov operator complexity (computed from the algorithm (\ref{eq:Lanczos eta})) to have a semiclassical description in terms of an expectation value of the total chord number. However, why should there be negative temperatures\footnote{\label{fnt:negative temp}{{Note that negative absolute temperatures are known in several areas in physics (see e.g. \cite{onsager1949statistical,Gauthier_2019,purcell1951nuclear,hsu1992statistics,braun2013negative}). This occurs when the system has a bounded energy spectrum (such as (\ref{eq:energy spectrum}) for the DSSYK), and their thermodynamic entropy become bounded, as in (\ref{eq:entropy temp}); so that their microcanonical temperature ($\partial E/\partial S$) can take either sign.}}} in the bulk?
 We interpret this in terms of fake vs real temperature in sine dilaton gravity \cite{Blommaert:2024ymv}, where the inverse temperature in the effective AdS$_2$ geometry (\ref{eq:effective geometry}) corresponds to the decay rate of thermal two-point correlation functions, instead of the real DSSYK inverse temperature, which we derived in (\ref{eq:saddles 1 particle beta}) (with $n=0$ for the thermodynamically stable saddle point). Comparing these quantities from (\ref{eq:fake temperature result}),
\begin{equation}\label{eq:fake temperature}
\beta_{\rm AdS}^{L/R}=\beta_{\rm fake}(\theta_{L/R}):=\frac{\pi}{J\sin\theta_{L/R}}~,\quad\text{vs}\quad \beta_{L/R}:=\beta(\theta_{L/R})=\frac{2\theta_{L/R}-\pi}{J\sin\theta_{L/R}}~,
\end{equation}
we notice that $\beta_{\rm fake}(\theta_{L/R})\geq0$ for $\theta_{L/R}\in[0,\pi]$. To make sense of the $\beta_{L}=-\beta_{R}$ condition for Krylov operator complexity in the AdS$_2$ dual geometry, the notion of physical and fake temperature sine dilaton gravity action (\ref{eq:sine dilaton gravity}) is {crucial} \cite{Blommaert:2024ymv}. In contrast, this condition would have no interpretation in JT gravity, unless we take the infinite temperature limit.

Previously, the Krylov operator complexity corresponding to a two-sided PETS in the finite $N$ SYK model was correlated with the CV conjecture in JT gravity with matter by \cite{Jian:2020qpp}. However, the derived relations did not yield a precise match. Essentially, it is possible to write a holographic dictionary relating bulk and boundary parameters that match the wormhole distance with Krylov operator complexity, but the entries in the early and late-time dictionary do not agree. However, as we show below, there is a precise agreement between the two sides once we consider the DSSYK model instead of its finite $N$ version.

Importantly, the chord number and its conjugate momenta are part of the canonical variables in the DSSYK Hamiltonian with or without matter (the others being their conjugate momenta, Sec. \ref{ssec:many chords}). By matching bulk distances with the chord number in the semiclassical limit, one may match the boundary and gravitational Hamiltonians at the quantum level \cite{Lin:2022rbf,Rabinovici:2023yex,Blommaert:2024ymv}, to show they are equivalent. There are some limitations since we can only match the $t_L=-t_R$ bulk/boundary side. Regardless, we will test our formulation of spread and Krylov operator complexity using the two-sided HH state.

The calculation of the wormhole geodesic distance in an AdS$_2$ black hole (i.e. the distance between the asymptotic boundaries) in the bulk with matter fields has been previously carried out by \cite{Jian:2020qpp}. To be explicit, we first consider JT gravity (in units where the AdS$_2$ length scale is $1$) with a minimally coupled (mc) scalar field
\begin{align}
    I_{\rm total}=&I_{\rm JT}[g,\Phi]+I_{\rm mc\,scalar}[g,\xi]~,\label{eq:total JT gravity action}\\
    I_{\rm JT}=&-\frac{\Phi_0}{2\kappa^2}\qty(\int_{\mathcal{M}}\sqrt{g}\mathcal{R}+2\int_{\partial\mathcal{M}}\sqrt{h}K)\label{eq:JT gravity}\\
    &-\frac{1}{2\kappa^2}\qty(\int_{\mathcal{M}}\rmd^2x\sqrt{g}\Phi(\mathcal{R}+2)+\int_{\partial\mathcal{M}}\sqrt{h}\Phi_b(K-1))~,\nonumber\\
    I_{\rm mc\,scalar}=&\int_{\mathcal{M}}\rmd^2x\sqrt{g}\qty(g^{ab}\partial_a\xi\partial_b\xi+m_\xi^2\xi^2)~,\label{eq:matter action}
\end{align}
where $\kappa^2=8\pi G_N$, $\Phi_0$ is a topological term, $\mathcal{M}$ denotes the spacetime manifold, $h_{\mu\nu}$ the boundary metric, $\Phi_b=\eval{\Phi}_{\partial\mathcal{M}}$, and $\xi$ is a free massive scalar field.

The same geodesic distances in AdS$_2$ can be recovered from sine dilaton gravity with a non-minimally coupled (nmc) scalar \cite{Blommaert:2024ymv},\footnote{Notice that one recovers JT gravity by rescaling $\Phi\rightarrow \lambda\Phi$.}
\begin{align}\label{eq:total sdg matter}
&I_{\rm total}=I_{\rm SDG}[g,\Phi]+I_{\rm nmc\,scalar}[g,\xi,\Phi]~,\\
\label{eq:sine dilaton gravity}
    &I_{\rm SDG}=-\frac{\Phi_0}{2\kappa^2}\qty(\int_{\mathcal{M}}\sqrt{g}\mathcal{R}+2\int_{\partial\mathcal{M}}\sqrt{{h}}K)\\
    &\qquad~~-\frac{1}{2\kappa^2}\qty(\int_{\mathcal{M}}\rmd^2x\sqrt{g}\qty(\Phi\mathcal{R}+2\sin\Phi)+2\int_{\partial\mathcal{M}}\rmd x\sqrt{h}\qty(\Phi_{b} K-\rmi\rme^{-\rmi\Phi_b}))~,\nonumber\\
    &I_{\rm nmc\,scalar}=\int_{\mathcal{M}} \rmd^2 x \sqrt{g}\,\left(g^{\mu\nu}\partial_\mu \xi \partial_\nu \xi +m_\xi^2 \rme^{-\lambda\rmi\Phi}\xi^2\right)~.\label{eq:non-minimal scalar}
\end{align}
Note that (\ref{eq:sine dilaton gravity}) has a shift symmetry in $\Phi\rightarrow\Phi+2\pi$ when we consider a cylinder surface.\footnote{It turns out that the disk partition function can also be expressed in terms of a cylinder partition function \cite{Okuyama:2025fhi}, so that the shift symmetry in $\Phi$ can still implemented as in \cite{Blommaert:2024whf}.} It has been argued in \cite{Blommaert:2024ymv,Blommaert:2024whf} (see also \cite{Blommaert:2023opb,Blommaert:2023wad}) that the bulk holographic dual to the DSSYK model (at the disk topology level) is (\ref{eq:sine dilaton gravity}). The DSSYK model is located {{at}} the asymptotic boundary of an effective AdS$_2$ black hole geometry found by Weyl rescaling the metric, $\rmd s^2_{\rm eff}=\rme^{-\rmi \Phi}\rmd s^2$ (with $\Phi^{L/R}_h$ in (\ref{eq:effective geometry}) being the energy parametrization $\theta_{L/R}$ in (\ref{eq:conserved energies})). The non-minimally coupled scalar (\ref{eq:non-minimal scalar}) becomes (\ref{eq:matter action}) with $g_{ab}\rightarrow g_{ab}^{\rm eff}$. According to the proposal in \cite{Blommaert:2024whf,Blommaert:2024ymv,Bossi:2024ffa}, the dual description of these fields correspond to $\hat{\mathcal{O}}_\Delta$ in the DSSYK model, where $m_\xi=\Delta$ (when the AdS$_2$ scale is set to one).

The wormhole distance with $u_R=-u_L=u$, and equal bulk temperature $\beta_{\rm AdS}=\frac{2\pi}{\Phi_h}$ between the left/right sides of the particle insertion (Fig. \ref{fig:one_particle_Euclidean} (b)) is given by \cite{Jian:2020qpp}\footnote{See their (3.25) and (3.27). The $t\ll \Phi_h^{-1}$ entry in (\ref{eq:AdS2PETS}) also follows from their expression (3.25). Note that $u\gg u_{\rm sc}$ we keep the additive $-\frac{\kappa^2m_\xi L_{b}}{\pi^2 \Phi_b}$ constant term.  \cite{Jian:2020qpp} ignores it in the late time regime (3.25). We thank Zhuo-Yu Xian for sharing a notebook to evaluate geodesic distances in JT gravity with matter in \cite{Jian:2020qpp} to confirm their results.}
\begin{equation}\label{eq:AdS2PETS}
L_{\rm AdS}(-u,u)\simeq\begin{cases}
L_{\rm reg}+\frac{\kappa^2m_\xi L_{b}(\pi-4)}{4\pi^2 \Phi_b}+\frac{\kappa^2m_\xi L_{b}}{8\pi \Phi_b}(\Phi_h u)^2~,&u\ll\Phi_h^{-1}~,\\
L_{\rm reg}-\frac{\kappa^2m_\xi L_{b}}{\pi^2 \Phi_b}+\frac{\kappa^2m_\xi L_{b}}{4\pi \Phi_b}\rme^{\Phi_h u}~,&\Phi_h^{-1}\ll u\ll u_{\rm sc}~,\\
L_{\rm reg}-\frac{\kappa^2m_\xi L_{b}}{\pi^2 \Phi_b}+2\log(\frac{\kappa^2m_\xi L_b}{8\pi \Phi_b})+2\Phi_h u~,&u\gg u_{\rm sc}~,
\end{cases}
\end{equation}
where $L_{b}$ is the asymptotic boundary length, and $r=\Phi_h$ the location of the event horizon (in thermal equilibrium) in the coordinates (\ref{eq:effective geometry}), and $L_{\rm reg}$ is a regulator, and the bulk scrambling time is
\begin{equation}\label{eq:scrambling AdS}
   \Phi_h u_{\rm sc}=\log\frac{8\pi \Phi_b}{\kappa^2m_\xi L_{b}}~.
\end{equation}
Note that the overall ${\kappa^4m^2_{\xi} L_{\rm b}^2}/{\Phi^2_b}$ coefficient is a perturbative backreaction parameter.

We remark that the above relation is valid for both JT and sine dilaton gravity in the $\lambda\rightarrow0$ limit and with the effective metric, $\rmd s^2_{\rm eff}$. The reason is that the expressions (3.15, 18, 19) in \cite{Jian:2020qpp} only rely on the Gauss law (for $SL(2,\mathbb{R})$ charges) and hyperbolic geometry.

Let us now compare with the results in Sec. \ref{ssec:special}. We found that Krylov operator complexity for the two-sided HH state (\ref{eq:HH state tL tR}) as reference state before the scrambling time (\ref{eq:intermediate time COp}) and at late times (\ref{eq:scrambling time}) is given by
\begin{equation}\label{eq:KrylovOpPETS}
\mathcal{C}^{(\eta=-1)}(t)\simeq\begin{cases}
\frac{\ell^{(-)}_*}{\lambda}+\frac{(2J\sin\theta~t)^2}{\lambda}\frac{\sin^2\theta-q^{\Delta_{\rm S}}\rme^{-\ell_*^{(-)}}}{2\sin^2\theta}~,&t\ll\tfrac{1}{J\sin\theta}~,\\
{\tfrac{\ell_*^{(-)}}{\lambda}+\tfrac{2}{\lambda}\log(\tfrac{\sin^2\theta+q^{ \Delta_{\rm S}}\rme^{-\ell_*^{(-)}}}{2\sin^2\theta})+\tfrac{1}{\lambda}\tfrac{\sin^2\theta-{q}^{\Delta_{\rm S}}\rme^{-\ell_*^{(-)}}}{\sin^2\theta+{q}^{\Delta_{\rm S}}\rme^{-\ell_*^{(-)}}}\rme^{2J\sin\theta~t}~,}&\tfrac{1}{J\sin\theta}\ll t\ll t_{\rm sc}~,\\
\frac{\ell_*^{(-)}}{\lambda}+\frac{2}{\lambda}\log(\frac{\sin^2\theta-q^{\Delta_{\rm S}}\rme^{-\ell_*^{(-)}}}{4\sin^2\theta})+\frac{4J\sin\theta~t}{\lambda}~,&t\gg t_{\rm sc}~,
\end{cases}
\end{equation}
where $\ell^{(-)}_*(\theta)$ is given by (\ref{eq:ell * operator}), and the scrambling time $t_{\rm sc}$ by (\ref{eq:scrambling time}). Thus, we can match (\ref{eq:AdS2PETS}) and (\ref{eq:KrylovOpPETS}) 
\begin{equation}
    L_{\rm AdS}(-u,u)=\lambda ~\mathcal{C}^{(\eta=-1)}(t)~,
\end{equation}
provided that we identify the holographic dictionary:
\begin{subequations}\label{eq:holographic dictionary massive particle}
    \begin{align}
&u=t~,\quad \Phi_h=2J\sin\theta~,\label{eq:u t scalar}\\
&\frac{m_\xi \kappa^2L_b}{\Phi_b}=4\pi\qty(\frac{\sin^2\theta-q^{\Delta_{\rm S}}\rme^{-\ell_*^{(-)}(\theta)}}{\sin^2\theta+q^{\Delta_{\rm S}}\rme^{-\ell_*^{(-)}(\theta)}})~,\label{eq:holographic u dictionary Krylov operator}\\
&L_{\rm reg}=\ell_*^{(-)}(\theta)+\frac{4}{\pi}\qty(\frac{\sin^2\theta-q^{\Delta_{\rm S}}\rme^{-\ell_*^{(-)}(\theta)}}{\sin^2\theta+q^{\Delta_{\rm S}}\rme^{-\ell_*^{(-)}(\theta)}})+2\log(\frac{\sin^2\theta+q^{\Delta_{\rm S}}\rme^{-\ell_*^{(-)}(\theta)}}{2\sin^2\theta})~.\label{eq:match Phib}
\end{align}
\end{subequations}
Note that the fake temperature of the DSSYK model (\ref{eq:fake temperature}) matches the bulk one (\ref{eq:holographic u dictionary Krylov operator}), i.e. $\beta_{\rm fake}=\beta_{\rm AdS}$ just as in the zero-particle space of the DSSYK model \cite{Blommaert:2024ymv}.

We stress that we require $\frac{\kappa^2L_b m_\xi}{\Phi_b}\ll1$ in the derivation of (\ref{eq:AdS2PETS}) since we do not specify the exact backreaction profile generated by the scalar field. We recover the dictionary when we specify the bulk backreaction as in the shockwave geometries (next section) for any $t_L$, $t_R$.

As a sanity check, we can use the above dictionary to express the DSSYK scrambling time defined as the transition from exponential to linear growth in Krylov operator complexity (\ref{eq:scrambling time}) in bulk terms. Indeed, it matches the bulk scrambling time in (\ref{eq:holographic u dictionary Krylov operator}). Thus, at least at the perturbative level in the backreaction of the scalar field, we can indeed match Krylov operator complexity with the wormhole distance displayed in Fig. \ref{fig:one_particle_Euclidean} (b).

\subsection{Minimal length geodesic with a shockwave}\label{ssec:towardsquantization}
Next, we will show that our boundary calculation (\ref{eq:shockwave approx}) has a bulk interpretation without the restrictions of the previous subsection. We focus on single shockwave geometries, while \cite{Aguilar-Gutierrez:2025mxf} has a more comprehensive analysis for building multiple shockwave geometries from the boundary perspective.

For this reason, we would like to compare the semiclassical total chord number with the double-scaled precursor operator insertion (\ref{eq:double scaled precursor shockwave}), which we derived in  (\ref{eq:shockwave approx}), and the geodesic length of an AdS$_2$ black hole backreacted by a shockwave insertion, which follows from the analysis of \cite{Shenker:2013pqa} (where the geodesic lengths correspond to those in a BTZ black hole in the s-wave sector). In bulk terms, this corresponds to an infalling shell of matter with very small mass at very early times in the past \cite{Shenker:2013pqa,Shenker:2013yza,Berkooz:2022fso}.

To be explicit, we now consider the shockwave limit for the minimally coupled scalar field $\xi$ in the JT gravity action (\ref{eq:total JT gravity action}) (or as well for the scalar in sine dilaton (\ref{eq:sine dilaton gravity})), where the stress tensor is given by \cite{Shenker:2013pqa,Goto:2018iay}
\begin{equation}\label{eq:shockwave stress tensor}
    T_{ab}=\frac{-2}{\sqrt{g}}\frac{\delta I_{\rm mc\,scalar}}{\delta g^{ab}}\rightarrow \frac{\alpha_{\rm sw}}{4\pi G_N}\delta(x^-)\delta_{a}^{x^-}\delta_{b}^{x^-}~,\quad\qty(\alpha_{\rm sw}=\frac{\kappa^2E_{\rm sw}}{4\pi\Phi_h^2}\rme^{\Phi_hu_w})~,
\end{equation}
where $\alpha_{\rm sw}$ is called the shockwave shift parameter, which is fixed as the insertion time $u_w\rightarrow\infty$, $E_{\rm ADM}$ the ADM energy (which we discuss further later in this section), and the asymptotic energy of the perturbation $E_{\rm p}\rightarrow0$ \cite{Shenker:2013pqa}. The coordinates being used in (\ref{eq:shockwave stress tensor}) appear in the backreacted metric (e.g. \cite{Mertens:2022irh})
\begin{equation}\label{eq:shockwave metric}
    \rmd s^2=\frac{-4\rmd x^+~\rmd x^-}{(1+x^-(x^++\alpha_{\rm sw})H(x^-))^2}~,
\end{equation}
where $H(x)$ is the Heaviside step function. (\ref{eq:shockwave metric}) also applies analogously for sine dilaton gravity in the Weyl-rescaled frame (\ref{eq:effective geometry}). The wormhole distance can be calculated as \cite{Shenker:2013pqa}
\begin{equation}\label{eq:length AdS sw}
    L_{\rm AdS}(u_L,u_R)=2\log(\cosh(\frac{\Phi_h}{2} (u_L-u_R))+\frac{\alpha_{\rm sw}}{2}\rme^{-\frac{\Phi_h}{2}(u_L+u_R)})+L_{\rm reg}~,
\end{equation}
where $L_{\rm reg}$ is a regularization scheme dependent constant (as in other sections). We represent the bulk preparation of state in Euclidean signature and its evolution in Fig. \ref{fig:shockwave_picture}.
\begin{figure}
    \centering
    \subfloat[]{\includegraphics[width=0.28\textwidth]{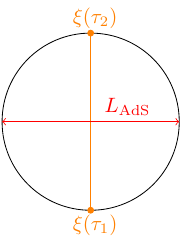}}\hspace{1.5cm}\subfloat[]{\includegraphics[width=0.48\textwidth]{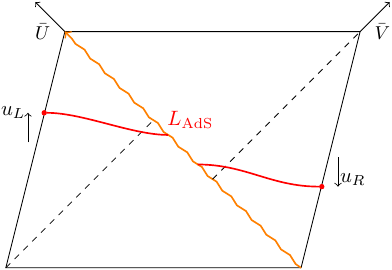}}
    \caption{Shockwave geometry in (a) Euclidean signature, where we prepare the state at $t_L=t_R=-t_{\rm sw}$ by a small perturbation to the HH state using the scalar field $\xi(\tau)$ (orange), which is massless in the limit where $t_{\rm sw}\rightarrow\infty$ \cite{Shenker:2013pqa}. Here $\tau_1=\frac{\beta(\theta)}{2}+\rmi t_{\rm sw}$, $\tau_2=\beta(\theta)-\tau_1$. (b) We display the Penrose diagram of the shockwave geometry, with the wormhole length (\ref{eq:length AdS sw}) in coordinates where $\overline{U}=\arctan((x^-)^{\Phi_h})$, $\overline{V}=-\arctan((x^+)^{\Phi_h})$ (see e.g. \cite{Aguilar-Gutierrez:2023pnn}).}
    \label{fig:shockwave_picture}
\end{figure}
\paragraph{Holographic dictionary}By equating the expectation value of the chord number in the two-sided HH state (\ref{eq:shockwave approx}), and the wormhole minimal geodesic length (\ref{eq:length AdS sw}) in the shockwave geometry,
\begin{equation}\label{eq:holographic dictionary N}
    L_{\rm AdS}(u_L,u_R)=\lambda ~\expval{\hat{N}}(t_L,t_R)~,
\end{equation}
we arrive at the following holographic dictionary
\begin{subequations}\label{eq:holographic dictionary shockwave}
    \begin{align}
        t_L&=u_L~,\quad t_R=-u_R~,\label{eq:u t shockwave}\\
        \alpha_{\rm sw}&=\qty(\frac{1}{2}-\frac{q^{\Delta}\rme^{-\ell^{(+)}_*(\theta)}}{2\sin^2\theta})\rme^{2J\sin\theta t_w}~,\label{eq:shockwave no approx}\\
        \Phi_h&=2J\sin\theta~,\quad L_{\rm reg}=\ell^{(+)}_*(\theta)~.\label{eq:dictionary Phih shockwave}
\end{align}
\end{subequations}
where $\ell_*^{(+)}$ is shown in (\ref{eq:ell * state}). While we do not need to make further approximations at this point, it is an useful consistency check to compare with the triple-scaling limit previously studied in \cite{Berkooz:2022fso}. In the strict shockwave limit $\alpha_{\rm sw}$ is a fixed parameter, while $t_w\rightarrow\infty$. This means that the term in parenthesis in (\ref{eq:shockwave no approx}) must be small. This can be recovered for light operators ($\Delta\sim\mathcal{O}(1)$) when 
\begin{equation}\label{eq:alpha sw2}
\alpha_{\rm sw}\eqlambda\Delta_{\rm S}\frac{\lambda~\rme^{2J\sin\theta~t_w}}{2\sin\theta}\quad\text{fixed}~,
\end{equation}
which is in agreement with a chord diagram calculation in the triple-scaling result considered by \cite{Berkooz:2022fso}.\footnote{The authors in \cite{Berkooz:2022fso} considered a different scaling of the energy spectrum, $E=\frac{2}{\sqrt{1-q}}\cos\theta$, compared to our (\ref{eq:energy spectrum}), and they work in the canonical ensemble. Nevertheless, (5.48) in \cite{Berkooz:2022fso} agrees with our (\ref{eq:alpha sw2}) when $T$ takes the role of the fake temperature (\ref{eq:fake temperature result}) (without $J$ in $\alpha_{\rm sw}$ since it is a dimensionless quantity) and after rescaling $\lambda\rightarrow\lambda^{3/2}$ due to the difference in conventions.} Thus, remarkably, \emph{all the saddle point solutions of the total chord number in }(\ref{eq:shockwave approx}) \emph{have a geometric interpretation} in the shockwave limit, which are related to spread and operator complexity in the semiclassical limit. We expect that the most general form for the semiclassical total chord number that we derived (\ref{eq:almost shockwave}) for arbitrary $\theta_{L/R}\in[0,~\pi]$ still describes an AdS$_2$ bulk geodesic with a shockwave insertion, but it no longer describes an early perturbation of the zero chord state. The insertion of the operator $\hat{\mathcal{O}}_{\Delta}$ may then take the system out of thermal equilibrium. It would be interesting to confirm this from the bulk perspective by solving the corresponding dilaton gravity EOM with a shockwave insertion and two different temperatures. In terms of the original construction by \cite{Shenker:2013pqa} this might be interpreted to taking the TFD out of thermal equilibrium in the early past by a heavy operator insertion; or by inserting a light operator in the later evolution.

\paragraph{Comparison with the massive particle case}
Let us compare (\ref{eq:holographic dictionary massive particle}) with (\ref{eq:holographic dictionary shockwave}). When considering the shockwave as perturbations of the TFD state, we demand that $\beta_L=\beta_R$ and so we need to replace $\ell^{(-)}_*\rightarrow \ell^{(+)}_*$ in (\ref{eq:holographic dictionary massive particle}), since we instead matched the bulk distance to the Krylov operator complexity of the HH state derived in (\ref{eq:Krylov complexity Operator many}) when $\beta_L=-\beta_R$. 

Next, in the semiclassical limit for light operators, it follows from (\ref{eq:ell * state}) that $\ell^{(+)}_*\rightarrow\sin^2\theta+2\lambda\Delta_{\rm S}\sin\theta(1-\sin\theta)+\mathcal{O}(\lambda^2)$. This means that to leading order (\ref{eq:holographic u dictionary Krylov operator}, \ref{eq:match Phib}) can be expressed as: 
\begin{subequations}
    \begin{align}
    \frac{m_\xi\kappa^2 L_b}{\Phi_b}&=\frac{2\pi\lambda\Delta}{\sin\theta}(1-\sin\theta)+\mathcal{O}(\lambda^2)~,\label{eq:backreaction scalar to shockwave}\\
    L_{\rm reg}&=\ell^{(+)}_*(\theta)+\mathcal{O}(\lambda)~.\label{eq:regulator scalar}
\end{align}
\end{subequations}
We note that the backreaction parameter (\ref{eq:backreaction scalar to shockwave}) is then proportional to shift parameter $\alpha_{\rm sw}$ in (\ref{eq:alpha sw2}); however, they are not equal since they describe different backreaction profiles; the scalar case is a stationary matter particle according to the frames in the asymptotic boundary, while the shockwave is a massless shell inserted at $t_w\rightarrow\infty$.

It can be seen that (\ref{eq:u t scalar}), (\ref{eq:regulator scalar}) (at leading order) are also consistent with $\Phi_h$ and $L_{\rm reg}$ in (\ref{eq:dictionary Phih shockwave}). Note that the overall sign difference between the exponentially growing part of (\ref{eq:AdS2PETS}) and the exponentially decreasing (\ref{eq:scrambling shockwave}). This occurs since the massive scalar is inserted at the $t=0$ slice in a HH preparation of state, instead of the shockwave perturbation which is send at $t=-t_w$($\rightarrow-\infty$).

\subsection{{{Predictions for sine dilaton gravity:}} Canonical quantization {{and}} the dual of the double-scaled algebra}\label{ssec:canonical quantization}
{{As an application of the holographic dictionary that we have developed earlier this section, we look for implications of our results to make predictions for the putative dual of a single DSSYK model, such as sine dilaton gravity \cite{Blommaert:2023opb,Blommaert:2024whf,Blommaert:2024ymv,Blommaert:2025avl} (in contrast to a pair of decoupled DSSYK models \cite{Narovlansky:2023lfz,Verlinde:2024znh,Verlinde:2024zrh,Gaiotto:2024kze,Tietto:2025oxn}). We focus on the parameters one should find for the canonical quantization of the bulk theory and its algebra of observables, interpreted as the bulk dual of the double-scaled algebra. We leave a bulk derivation of the one-sided lengths involved in these arguments for future directions.}}
We {{first discuss about}} the canonical quantization of the bulk theory \eqref{eq:total sdg matter} as an application of our results from previous sections. Given the relationship between geodesic lengths in the bulk and the discreteness of the chord number (\ref{eq:holographic dictionary N}), the bulk is in fact a constrained system (as in higher dimensional settings, e.g. \cite{Brown:1994py,Giesel:2007wi,Kuchar:1993ne,Lapchinsky:1979fd}). For this reason, to study the {{canonical}} quantization of the bulk theory, one must apply constraints (e.g. by reduced phase space quantization \cite{faddeev1969feynman}, Dirac quantization \cite{Dirac:1931kp,dirac2013lectures}, or alternative approaches; see e.g. \cite{Kunstatter:1990hz}). Previous works in JT gravity and sine dilaton gravity have taken similar steps \cite{Harlow:2018tqv,Blommaert:2024whf,Blommaert:2024ymv}. 

{{Later, we conjecture what would be}} the bulk dual of the double-scaled algebra {based on the holographic dictionary we developed in Sec. \ref{ssec:towardsquantization}}. To achieve this, we first need to find the corresponding dual operators with respect to the double-scaled algebra (\ref{eq:operators DS algebra}). We know that the dual theory needs to reproduce the AdS$_2$ geodesic lengths that we found in Sec. \ref{ssec:Krylov operator bulk picture}, \ref{ssec:towardsquantization}. The main candidates are then JT and sine dilaton gravity. Given that JT gravity is only the low energy (low temperature) description of the DSSYK (in the triple-scaling limit), we will focus on sine dilaton gravity with matter. As we just saw in Sec. \ref{ssec:Krylov operator bulk picture}, Krylov operator complexity with finite temperature distinguishes between physical and fake temperatures. The latter is the temperature in the effective AdS$_2$ spacetime, which is different from the physical temperature of the DSSYK model with matter. Moreover, the bulk theory needs to have a finite energy spectrum range (\ref{eq:energy spectrum}). We will {{relate}} the shockwave geometry in sine dilaton gravity {{to}} the DSSYK model with a matter chord.

\paragraph{The bulk Hamiltonian}{{Based on our results for $\ell_{L/R}:=\ell_{0/m}$ in \eqref{eq:ell0m}, we can lay out how to canonically quantize the bulk Hamiltonian corresponding to \eqref{eq:total sdg matter} dual}} to the DSSYK model. {{We focus on the case with shockwave backreaction, rather than the original derivation without introducing precursor operators \eqref{eq:solutions HH}, since in this limiting case}} we know that the total length matches the total chord number of the DSSYK model under the identification of the holographic dictionary (\ref{eq:holographic dictionary shockwave}).

When $\theta_L=\theta_R\equiv \theta$, sine dilaton gravity has the following ADM energy \cite{Blommaert:2024ymv},
\begin{equation}\label{eq:constrained HADM}
    H^{(\rm ADM)}_{L/R}=E_{\rm ADM}(\theta)\equiv\frac{1}{2\kappa^2}\cos\theta~,
\end{equation}
which is not modified by the insertion of shockwaves at this level since they do not reach the asymptotic boundary at finite boundary times \cite{Shenker:2013pqa}. 

{{Let us emphasize there is no reason a priori for the bulk and boundary theories to have the same energy spectrum; however, this allows us to match the bulk wormhole length with the total chord number $\lambda\hat{N}$, which is the canonical variable allowing for the bulk Hamiltonian to match the one particle DSSYK Hamiltonian (\ref{eq:pair DSSYK Hamiltonians 1 particle}) if the holographic dictionary in (\ref{eq:holographic dictionary shockwave}) is satisfied. Curiously, this type of constraint is also imposed in the canonical quantization of JT gravity at the disk topology level to match to the triple-scaled SYK (see e.g. \cite{Rabinovici:2023yex}).}}

Furthermore, the overall constant matches the one of the DSSYK model (\ref{eq:energy spectrum}) when\footnote{{{Note that $\kappa^2$ is dimensionless in \eqref{eq:total sdg matter}, so in \eqref{eq:constants relation} we have identified $1/J$ as the AdS length scale in the holographic dictionary, which we had set to 1 in (\ref{eq:sine dilaton gravity}).}}}
\begin{equation}\label{eq:constants relation}
    \frac{1}{2\kappa^2}=\frac{2}{\sqrt{\lambda(1-q)}}~.
\end{equation}
We can then construct the symplectic form{{, $\omega_{ab}$}}, in the covariant phase space. {{One can define it in terms of its inverse through the relation,}}
\begin{equation}\label{eq:phase space}
    \dv{x^a}{t_{L/R}}=(\omega^{-1})^{ba}\partial_b H_{L/R}~,\quad\text{where}\quad x^a=\qty{\ell_i,~P_i}~,
\end{equation}
{{where $\dv{x^a}{t_{L/R}}$ obeys the Hamilton equations, e.g. \eqref{eq:more general EOM}.}}

The {{symplectic form in}} (\ref{eq:phase space}) can either be expressed in \emph{boundary} terms, using the classical phase space solutions of (\ref{eq:evol many}) of the form \cite{Harlow:2018tqv}
\begin{equation}\label{eq:xsympletic form}
    \omega=\rmd t_L\wedge \rmd E_L+\rmd t_R\wedge \rmd E_R=\frac{2}{\sqrt{\lambda(1-q)}}\sum_{i}\rmd \ell_i\wedge\rmd P_i~,
\end{equation}
or in \emph{bulk} terms, as
\begin{equation}\label{eq:sympletic LR}
    \omega=\rmd u_L\wedge\rmd H^{(\rm ADM)}_L+\rmd u_R\wedge\rmd H^{(\rm ADM)}_R=\frac{1}{2\kappa^2}\sum_i \rmd L_{\rm AdS}^{i}\wedge\rmd P_{\rm AdS}^{i}~.
\end{equation}
We know the length variable in the effective AdS$_2$ shockwave geometry is given by (\ref{eq:length AdS sw}), which matches the DSSYK total chord number {{(\ref{eq:shockwave approx}) provided that the ADM energy \eqref{eq:constrained HADM} matches the energy spectrum \eqref{eq:conserved energies} (for $\theta_L=\theta_R\equiv\theta$), and that the relation between Newton's constant and $\lambda$ in \eqref{eq:constants relation} holds. However, if one were to use known results in JT gravity for the one-sided minimal geodesic lengths spanning between the asymptotic boundary to the shockwave insertion \cite{Shenker:2013pqa}, given by
\begin{equation}\label{eq:JT one sided lengths}
    L_{\rm AdS}^{L/R}=\frac{L_{\rm reg}}{2}+\log(\frac{1}{2}\qty(1+\alpha_{\rm sw}\rme^{\Phi_h u_{L/R}}+\rme^{\Phi_h(u_{R/L}-u_{L/R})}))~,
\end{equation}
then there would be no match with the DSSYK since the lengths \eqref{eq:JT one sided lengths} are real in contrast to $\ell_{L/R}$ in \eqref{eq:ell0m} which is complex when using the total length and the momenta in (\ref{eq:shockwave approx}, \ref{eq:Pom shockwave}), as we explain in the next paragraph. This indicates that the JT gravity one-sided lengths is not always a good description for the bulk dual theory of the DSSYK model. One should verify how this changes this explicitly in sine dilaton gravity \cite{Blommaert:2023opb,Blommaert:2024ymv,Blommaert:2024whf,Blommaert:2025avl} with matter \eqref{eq:sine dilaton gravity}, or other dual proposal.}}

{{More explicitly, based on the semiclassical chord number \eqref{eq:ell0m} and the holographic dictionary \eqref{eq:holographic dictionary shockwave} the corresponding bulk one-sided lengths that we \emph{expect} to find in the dual bulk theory are
\begin{equation}\label{eq:Length AdS}
\rme^{-L_{\rm AdS}^{L/R}}=\frac{2\cos P_{\rm AdS}^{L/R}-2\cos\theta-\rme^{-\frac{\kappa^2}{16\pi}\frac{E_{\rm sw}}{\Phi_h}}\rme^{-L_{\rm AdS}+\rmi P_{\rm AdS}^{L/R}}}{\rme^{\rmi P_{\rm AdS}^{L/R}}-\rme^{-\frac{\kappa^2}{16\pi}\frac{E_{\rm sw}}{\Phi_h}}\rme^{\rmi P_{\rm AdS}^{R/L}}}~,
\end{equation}
where $L_{\rm AdS}$ appears in \eqref{eq:length AdS sw}, and the constant factor in the exponential follows from (\ref{eq:shockwave stress tensor}) and the dictionary \eqref{eq:holographic dictionary shockwave}
\begin{equation}
    \lambda\Delta_w=\frac{\kappa^2}{16\pi}\frac{E_{\rm sw}}{\Phi_h}~.
\end{equation}
Here the factor $E_{\rm sw}/\Phi_h$ is proportional to the proper energy of the perturbation released in the past \cite{Shenker:2013pqa}.\footnote{{{One can translate our notation to the one in \cite{Shenker:2013pqa} by $\Phi_h\rightarrow R$, $E_{\rm sw}\rightarrow E$ while setting the AdS length scale $=1$.}}} while the conjugate momenta (\ref{eq:canonical momenta0 many}, \ref{eq:canonical momentam many}) with respect to the ADM energy (\ref{eq:constrained HADM}) are:
\begin{equation}\label{eq:Momenta AdS}
P_{\rm AdS}^{L/R}=\rmi\log(\cos\theta+\rmi\sin\theta\frac{\tanh(\delta u_{L/R})+\rme^{-\frac{\kappa^2}{16\pi}\frac{E_{\rm sw}}{\Phi_h}}\tanh(\delta u_{R/L})}{1+\rme^{-\frac{\kappa^2}{16\pi}\frac{E_{\rm sw}}{\Phi_h}}\tanh(\delta u_{L})\tanh(\delta u_{R})})~,
\end{equation}
and}} we introduced the shorthand
\begin{equation}
    \delta u_{L}=\frac{\Phi_h}{2}(u_L-u_w)~,\quad \delta u_{R}=-\frac{\Phi_h}{2}(u_R+u_w)~,
\end{equation}
with $u_w$ being the insertion time in the bulk theory \eqref{eq:shockwave stress tensor}. 

{{We stress that $\theta$ in \eqref{eq:Length AdS} is the energy spectrum with the one-particle insertion when $\theta_L=\theta_R\equiv\theta$ since we consider the shockwave limit as in \cite{Shenker:2013pqa}, where the perturbation at very early times is very weak, and it does not take the system out of equilibrium, as we discussed in Sec. \ref{ssec:shockwaves}. While the following arguments are based on the shockwave limit, where the holographic dictionary \eqref{eq:holographic dictionary shockwave} is easier to interpret; however, we expect the implications for the bulk theory can be derived under more general conditions.}}

{{Then, by then inverting (\ref{eq:Length AdS}, \ref{eq:Momenta AdS}) we can express the conserved ADM Hamiltonian (\ref{eq:constrained HADM}) in terms of canonical coordinates as
\begin{equation}\label{eq:classical Hamiltonian}
H_{L/R}^{(\rm ADM)}=\frac{1}{\kappa^2}\qty(\rme^{-\rmi P_{\rm AdS}^{L/R}}+\rme^{\rmi P_{\rm AdS}^{L/R}}\qty(1-\rme^{-L_{\rm AdS}^{L/R}})+\rme^{-\frac{\kappa^2}{16\pi}\frac{E_{\rm sw}}{\Phi_h}}\rme^{\rmi P_{\rm AdS}^{R/L}}\rme^{-L_{\rm AdS}^{L/R}}\qty(1-\rme^{-L_{\rm AdS}^{R/L}}))~.
\end{equation}}}
To construct the bulk Hamiltonian at the quantum level from the classical one (\ref{eq:classical Hamiltonian}) we promote classical variables to quantum ones
\begin{equation}
    \qty{L_{\rm AdS}^{L/R},~P_{\rm AdS}^{L/R}}\rightarrow \qty{\hat{L}_{\rm AdS}^{L/R},~\hat{P}_{\rm AdS}^{L/R}}~.
\end{equation}
This approach is equivalent to the path integral in Sec. \ref{ssec:many chords} {{where}} the canonical coordinates in the gravity theory $L_{\rm AdS}^{L/R}$ are promoted to the Heisenberg operators which the gravitational path integral (dual to (\ref{eq:Hamilton PI})) prepares.
 While there is the usual operator ambiguity in canonical quantization {{of \eqref{eq:classical Hamiltonian}}}; we choose the following natural ordering
\begin{equation}\label{eq:ADM SDG Hamiltonian}
    \hH_{L/R}^{(\rm ADM)}=\frac{1}{\kappa^2}\qty(\rme^{-\rmi \hat{P}_{\rm AdS}^{L/R}}+\rme^{\rmi \hat{P}_{\rm AdS}^{L/R}}\qty(1-\rme^{-\hat{L}_{\rm AdS}^{L/R}})+\rme^{-\frac{\kappa^2}{16\pi}\frac{E_{\rm sw}}{\Phi_h}}\rme^{\rmi \hat{P}_{\rm AdS}^{R/L}}\rme^{-\hat{L}_{\rm AdS}^{L/R}}\qty(1-\rme^{-\hat{L}_{\rm AdS}^{R/L}}))~.
\end{equation}
{{One could then identify $\lambda~\hat{n}_{L/R}$ and $\hat{P}$ \eqref{eq:pair DSSYK Hamiltonians 1 particle} with $\hat{L}_{\rm AdS}^{L/R}$ and $\hat{P}_{\rm AdS}^{L/R}$ in the bulk, respectively.}} 

Before moving on, there are several remarks regarding \eqref{eq:ADM SDG Hamiltonian}.
\begin{itemize}
\item Note that the operator ordering in (\ref{eq:ADM SDG Hamiltonian}) is not very important. For instance, a different natural ordering choice 
\begin{equation*}
\qty(1-\rme^{-\hat{L}_{\rm AdS}^{L/R}})\rme^{\rmi \hat{P}_{\rm AdS}^{L/R}} \quad \text{instead of}\quad ~\rme^{\rmi \hat{P}_{\rm AdS}^{L/R}}\qty(1-\rme^{-\hat{L}_{\rm AdS}^{L/R}})~,
\end{equation*}
and similarly for the last term in (\ref{eq:ADM SDG Hamiltonian}), this alternative ordering only amounts to an overall shift in $\hat{L}_{\rm AdS}^{L/R}$ by a factor $\lambda\mathbb{1}$. 

\item {{The Hermicity of the bulk Hamiltonian \eqref{eq:ADM SDG Hamiltonian} depends on the set of states where it acts and the inner product they satisfy. The bulk Hamiltonian below, based on the one-particle DSSYK Hamiltonian \eqref{eq:pair DSSYK Hamiltonians 1 particle}, is non-Hermitian in a length basis which we present later, see \eqref{eq:bulk chord basis}. This feature has also been observed in (42) \cite{Lin:2022rbf} (the triple-scaling version of the DSSYK Hamiltonian $\hH_{L/R}$ in \eqref{eq:Hamiltonians LR}) which is non-Hermitian with respect to the chord basis $\ket{\tilde{\Delta},n_0,\dots,n_m}$ representation. Nevertheless, $\hH_{L/R}$ can be expressed in a Hermitian form by working in a different basis, see \cite{Lin:2023trc} (18), \cite{Ambrosini:2024sre} (2.53, 54).}}

\item {{At last, we emphasize that \eqref{eq:ADM SDG Hamiltonian} is not a derivation of the bulk Hamiltonian, which instead would require to derive \eqref{eq:Length AdS} for instance from a sine dilaton gravity calculation. Instead, we are stating the steps that one should take to complete the derivation after \eqref{eq:pair DSSYK Hamiltonians 1 particle} is established from a bulk derivation.}}
\end{itemize}

\paragraph{Dirac quantization}
Now, we examine the quantization of the bulk theory {{defined in \eqref{eq:ADM SDG Hamiltonian}}} with constraints {{using Dirac's approach \cite{dirac2013lectures,Dirac:1931kp}}}. 
We know from \cite{Lin:2022rbf} that the chord Hilbert space of the DSSYK has a bulk interpretation where the chord number is related to a geodesic length in the bulk. It then natural to examine the Dirac quantization method, since it acts directly in the Hilbert space of the theory that we seek to quantize. Based on the previous discussion, we need to implement:
\begin{subequations}\label{eq:Dirac constraints}
    \begin{align}\label{eq:ADM Hamiltonian}
        \qty(\hH^{\rm ADM}_{L/R}-E(\theta_{L/R}))\ket{\psi(\theta_{L},\theta_{R})}&=0~,\\ \sum_{m=-\infty}^\infty\qty(\rme^{2m\pi \rmi~\hat{n}_{\rm AdS}^{L/R}}-\mathbb{1})\ket{\psi(\theta_{L},\theta_{R})}&=0~,\label{eq:momentum shift}
    \end{align}
\end{subequations}
$\forall\ket{\psi(\theta_{L},\theta_{R})}\in\mathcal{H}_{\rm phys}$ (the physical bulk Hilbert space), and where $\hat{n}_{\rm AdS}^{L/R}\equiv\hat{L}_{\rm AdS}^{L/R}/\lambda$ (which can be expressed in bulk terms from (\ref{eq:constants relation})), while $\hH^{\rm ADM}_{L/R}$ appears in (\ref{eq:ADM SDG Hamiltonian}). 

The first constraint in (\ref{eq:Dirac constraints}) equates the energy spectrum of the DSSYK Hamiltonian with the bulk one, {{corresponding to the ADM energies at the classical level \cite{Blommaert:2024whf}, such as in \eqref{eq:constrained HADM}}}. {{As explained below of \eqref{eq:sympletic LR}, this condition allows to match the corresponding length geodesic with the total chord number.}} Meanwhile, the last constraint in (\ref{eq:Dirac constraints}) is motivated by \cite{Blommaert:2024whf} (requiring that the eigenvalues of the lengths $\hat{L}_{\rm AdS}^{L/R}$ are quantized). This is equivalent to imposing constraints in the \emph{chord Hilbert space} of the DSSYK model from the boundary-to-bulk map in \cite{Lin:2022rbf}. Moreover, note that:
\begin{equation}
[\hH^{\rm ADM}_{L/R},\hat{n}_{\rm AdS}^{L/R}]=\hH^{\rm ADM}_{L/R}-\frac{2J}{\sqrt{\lambda(1-q)}}\rme^{\hat{P}_{\rm AdS}^{L/R}}\neq 0~,
\end{equation}
i.e. the constraints above are second class \cite{Dirac:1931kp,dirac2013lectures}. We can solve the constraints (\ref{eq:Dirac constraints}) using the explicit form of the bulk Hamiltonian (\ref{eq:ADM SDG Hamiltonian}). We work under the assumption that the chord Hilbert space $\in\mathcal{H}_m$ is the Hilbert space of the bulk dual theory based on \cite{Lin:2022rbf}. This means that the one-particle irrep. chord state ${\Delta,n_L,n_R}$ denotes a bulk basis, where
\begin{equation}\label{eq:bulk chord basis}
\hat{n}_{\rm AdS}^{L/R}\ket{\Delta;n_L,n_R}=n_{L/R}\ket{\Delta;n_L,n_R}~,
\end{equation}
which is consistent with (\ref{eq:momentum shift}). Denoting
\begin{equation}\label{eq:notation psi nLnR}
\begin{aligned}
\psi_{n_L,n_R}(\theta_L,\theta_R;q,q^\Delta)&:=\bra{\psi(\theta_{L},\theta_{R})}\ket{\Delta;n_L,n_R}~,\\
x:=2\cos\theta_L~,\quad &y:=2\cos\theta_R~.
\end{aligned}
\end{equation}
We can then express the bulk Schrödinger equation (\ref{eq:ADM Hamiltonian}) as\footnote{Here, we allow unequal left/right-chord sector energies for generality, although the explicit form of the quantum Hamiltonian is based on (\ref{eq:ADM SDG Hamiltonian}), which is based on a limiting case where $\theta_L=\theta_R$.}
\begin{equation}\label{eq:nL nR HADM}
\bra{\psi(\theta_{L},\theta_{R})}\hH^{\rm ADM}_{L/R}\ket{\Delta;n_L,n_R}=E(\theta_{L/R})\psi_{n_L,n_R}(\theta_L,\theta_R;q,q^\Delta)~.
\end{equation}
Now, we perform Dirac quantization by plugging the explicit form of the quantum bulk Hamiltonian (\ref{eq:ADM SDG Hamiltonian}) in the bulk Schrödinger equation with the constraint (\ref{eq:ADM Hamiltonian}), which takes the form of the DSSYK Hamiltonian with creation and annihilation operators (\ref{eq:Hmultiple}) acting on the chord basis $\ket{\Delta;n_L,n_R}$. This results in the following recurrence relation
\begin{equation}
\begin{aligned}
    &2x~\psi_{n_1,n_2}(x,y;q,r)=\psi_{n_1+1,n_2}+(1-q^{n_1})\psi_{n_1-1,n_2}+r~q^{{n_1}}(1-q^{n_2})\psi_{n_1,n_2-1}~,\\
    &2y~\psi_{n_1,n_2}(x,y;q,r)=\psi_{n_1,n_2+1}+(1-q^{n_2})\psi_{n_1,n_2-1}+r~q^{{n_2}}(1-q^{n_1})\psi_{n_1-1,n_2}~,
\end{aligned}
\end{equation}
where {{$x,~y$ appear in (\ref{eq:notation psi nLnR}) and}} the initial conditions are
\begin{equation}
    \psi_{n_1,0}(x,y;q,r)=\psi_{n_1}(x;q)~,\quad \psi_{0,n_2}(x,y;q,r)=\psi_{n_2}(y;q)~.
\end{equation}
Here $r$ is an arbitrary parameter, which in our specific problem is $r=q^\Delta$. The solutions to the problem above are the bivariate $q^\Delta$-weighted q-Hermite polynomial of order $(m,n)$,
\begin{equation}
\psi_{n_1,n_2}(x,y;q,q^\Delta)=H_{n_L,n_R}(\cos\theta_L,\theta_R;q,q^\Delta)~.
\end{equation}
One could also discuss how the partition function is modified once we gauge momentum shift symmetry vs if one does not \cite{Blommaert:2024whf}. We leave this calculation for future research.

\paragraph{{{Multiple particles}}}{{To summarize the results in this subsection so far, we have a prediction for the bulk Hamiltonian \eqref{eq:ADM SDG Hamiltonian}, based on the DSSYK model \eqref{eq:pair DSSYK Hamiltonians 1 particle} with one-particle insertion and the holographic dictionary that we derived in \eqref{eq:holographic dictionary shockwave}.}} What happens with arbitrary many particles? Given that the double-scaled algebra of observables acts on $\mathcal{H}_m$, one can suspect that, since we reproduce shockwave geometries from the boundary side (\ref{eq:shockwave approx}), more particles correspond to geometry with multiple shockwaves in the AdS$_2$ bulk, similar to \cite{Shenker:2013yza} (in the s-wave reduction). We show this explicitly in \cite{Aguilar-Gutierrez:2025mxf}. Based on the results {{matching bulk geodesics in multiple shockwave geometries to the total chord number with more matter insertions}} in \cite{Aguilar-Gutierrez:2025mxf}, one {{could}} use similar arguments {{as in this work to canonically quantize the corresponding}} bulk theory (sine dilaton gravity with matter (\ref{eq:total sdg matter}) for instance) with a bounded energy spectrum (\ref{eq:constrained HADM}).\footnote{We interpret the alternative proposals in dS$_3$ space \cite{Narovlansky:2023lfz} in terms of the same DSSYK model with matter but with different constraints from those above. More details appear in \cite{Aguilar-Gutierrez:2025hty}.}{{ We leave this for future directions; however, we expect}} that the bulk algebra of operators dual to the double-scaled algebra (\ref{eq:operators DS algebra}) then corresponds to (\ref{eq:bulk algebra}), which we repeat for the reader
\begin{equation}\label{eq:new bulk algebra}
    \expval{\hH^{(\rm ADM)}_{L/R},~\hat{\xi}_{\mathfrak{R}}^{L/R}(x)}''~,
\end{equation}
where $m_\xi=\Delta$, and $\hat{\xi}_{\mathfrak{R}}^{L/R}(x)$ is the quantized bulk matter field in (\ref{eq:non-minimal scalar}), {{while}} the label $L/R$ indicates the location in the two-sided AdS$_2$ effective geometry (\ref{eq:effective geometry}) where field is evaluated, {{and $\mathfrak{R}$ indicates dressing with respect to the asymptotic boundary (see e.g. \cite{Donnelly:2015hta}), so that the operator is diffeomorphism invariant. For instance, diffeomorphism invariant observables in JT gravity have been worked out explicitly in different contexts (by shooting light rays \cite{Mertens:2025rpa,Blommaert:2019hjr,Mertens:2019bvy,Blommaert:2020yeo,Blommaert:2020seb}; boundary anchored geodesics \cite{Harlow:2021dfp}, and other types of boundary anchorings \cite{Nitti:2024iyj}).}} We comment more about (\ref{eq:new bulk algebra}) in Sec. \ref{ssec:outlook}. Altogether, the double-scaled algebra serves as a guiding structure for the holographic correspondence of the DSSYK model.

\subsection{Connection with the proper radial momentum of a probe}\label{ssec:proper momentum}
Recently, it has been argued through different examples that that the radial proper momentum of probe particles in asymptotically AdS spacetimes can be matched with spread complexity of a TFD state excited by primary CFT operator \cite{Caputa:2024sux,Caputa:2025dep}, (or for Krylov operator complexity \cite{Fan:2024iop}; see also \cite{He:2024pox}). However, the UV finiteness in holographic CFTs have not been considered in the development of the dictionary relating the proper radial momentum and Krylov complexity. Here, we take first steps in this direction in a one-dimensional model.

As mentioned in our study of the bulk description for the Krylov operator complexity (\ref{eq:Krylov complexity Operator many}) in Sec. \ref{ssec:Krylov operator bulk picture} the physical temperatures on the left and right sides of Fig. \ref{fig:one_particle_Euclidean} need to be opposite signed. For this reason, our bulk interpretation in this section is based on the effective AdS$_2$ geometry of sine dilaton gravity. We start from the bulk geometry (\ref{eq:effective geometry}) (with $\Phi_h^{L}=\Phi_h^{R}\equiv\Phi_h$) as:
\begin{equation}
\begin{aligned}
        \rmd s^2=\rmd\sigma^2-\Phi_{h}^2\sinh^2\sigma~\rmd u^2~,\label{eq:silly metric}
\end{aligned}
\end{equation}
where we performed a change of variables $r=\Phi_{h}\cosh\sigma$ and gauge fixed $u_L=-u_R\equiv u$, {{which is the boundary time determining the evolution of the proper radial momentum described below}}. We seek a holographic dictionary relating the black hole horizon parameter $\Phi_h$ in terms of DSSYK parameters $J$ and $\theta$.

We would like to describe the geodesic of a probe point particle falling from the asymptotic boundary of an AdS$_2$ geometry, which is determined by
\begin{equation}
    I_{\rm point-particle}:=-m_{\rm particle}\int\rmd u\sqrt{\Phi_h^2\sinh^2 \sigma(u)-\dot{\sigma}(u)^2}:=\int\rmd u~\mathcal{I}~,
\end{equation}
{{where $\dot{\sigma}(u):=\dv{\sigma}{u}$}}.
{{The radial momentum of this theory is defined as the conserved charge\footnote{{{Although we use coordinates to define the proper radial momentum, there is a recent work \cite{Li:2025fqz} which defines it in a coordinate invariant matter.}}}
\begin{equation}\label{eq:radial momentum eval}
    \mathcal{P}_\sigma(u):=\dv{\mathcal{I}}{\dot{\sigma}(u)}~,
\end{equation}
and we are interested in describing the proper frame of the particle}} (where $\dot{\sigma}(0)=0$ {{since the particle does not have speed relative to itself) which is then}} given by
\begin{equation}
    \mathcal{P}_\sigma(u)=-m_{\rm particle}\cosh(\sigma_b)\sinh(\Phi_hu)~,
\end{equation}
where we used the geodesic solution $\coth(\sigma(u))=\coth(\sigma_b)\cosh(\Phi_hu)$, with the initial location of the infalling particle $\sigma(u=0)=\sigma_b$ being the asymptotic boundary location. {Note that, since the particle is located at the asymptotic boundary, its mass is related to the conformal dimension of a dual DSSYK operator $\hat{\mathcal{O}}_{\Delta}$ through $m_{\rm particle}=\Delta$.} 

Next, we compare the time dependence of the proper radial momentum (\ref{eq:radial momentum eval}) and the rate of growth of Krylov operator complexity (\ref{eq:Krylov complexity Operator many}); which is given explicitly by
\begin{subequations}\label{eq:early time operator complexity}
\begin{align}
    \dv{t}\mathcal{C}^{(\eta=-1)}(t)=&{2\lambda^{-1}J\sin\theta}\frac{\qty(\sin^2\theta-{\rme^{-\ell^{(-)}_*(\theta)}q^{\Delta_{\rm S}}})\sinh(2J\sin\theta~t)}{\sin^2\theta+\qty(\sin^2\theta-{\rme^{-\ell^{(-)}_*(\theta)}q^{\Delta_{\rm S}}})\sinh^2(J\sin\theta~t)}\label{eq:display 1}\\
\simeq&\frac{2J\sin\theta}{\lambda}\qty(1-\frac{\rme^{-\ell^{(-)}_*(\theta)}q^{\Delta_{\rm S}}}{\sin^2\theta})\sinh(2J\sin\theta~t)~,\quad \abs{t}\ll  t_{\rm max}~.\label{eq:display 2}
\end{align}
\end{subequations}
In the last line, we consider the early $\sinh(2J\sin\theta~t)$ growth of Krylov operator complexity (\ref{eq:classical phase space OTOC}) {{that occurs}} below the time scale {{$ t_{\rm max}$ when the denominator in the right-hand-side of (\ref{eq:display 1}) is much greater than the constant term $\sin^2\theta$, i.e.}} 
\begin{equation}\label{eq:relation new}
\qty(1-{\csc^2\theta}~{\rme^{-\ell^{(-)}_*}}q^{\Delta_{\rm S}})\sinh^2(J\sin\theta~t_{\rm max})\approx1~,    
\end{equation}
where for late enough times, {{we may approximate $\sinh^2(J\sin\theta~t)\simeq\tfrac{1}{4}\qty(\rme^{2J\sin\theta~t}-2)$ in (\ref{eq:relation new}), and thus recover}}
\begin{equation}\label{eq:tprm}
t_{\rm max}=-\frac{1}{2J\sin\theta}\log(\frac{6\sin^2\theta-2\rme^{-\ell_*^{(-)}}q^{\Delta_{\rm S}}}{\sin^2\theta-\rme^{-\ell_*^{(-)}}q^{\Delta_{\rm S}}})~.
\end{equation}
Note this time scale is smaller than the scrambling time (\ref{eq:scrambling time}) since we study the regime where the Krylov operator complexity growth behaves as $\sinh(2J\sin\theta t)$ instead of just $\rme^{2J\sin\theta t}$. Thus, \eqref{eq:display 2} relates to the proper radial momentum (\ref{eq:radial momentum eval}) as:
\begin{subequations}
    \begin{align}\label{eq:match proper radial momentum}
    \frac{1}{J}\dv{t}\mathcal{C}^{(\eta=-1)}(t)&\simeq-\frac{\qty(\Phi_h/J)}{\lambda~ m_{\rm particle}~\epsilon_{\rm reg}}\qty(1-4\frac{\rme^{-\ell^{(-)}_*}q^{\Delta_{\rm S}}}{(\Phi_h/J)^2})~\mathcal{P}_\sigma(u)~,\quad \abs{t}\ll  t_{\rm max}~,\\
    &\text{provided}\quad \Phi_h u=2J\sin(\theta)~t~,\label{eq:dictionary prm}
\end{align}
\end{subequations}
where we introduced a regulator $\epsilon_{\rm reg}=\sech(\sigma_b)$ associated with the location of asymptotic boundary location $\sigma_b$. Note that (\ref{eq:dictionary prm}) agrees with (\ref{eq:holographic dictionary massive particle}) and (\ref{eq:dictionary Phih shockwave}).

From (\ref{eq:dictionary prm}) we may express the constant $\ell_*^{(-)}$ (\ref{eq:ell * operator}) entirely in terms of the parameters $\Phi_h/J$, and $q^{\Delta_{\rm S}}$. Moreover, one can also express the right hand side of (\ref{eq:dictionary prm}) purely in bulk terms by (i) identifying $\lambda\equiv\lq=8\pi G_N$ as in e.g. sine dilaton gravity \cite{Blommaert:2024ymv}, (ii) expressing the conformal dimension $\Delta_S$ in terms of the mass $m_\xi$ (\ref{eq:non-minimal scalar}) of the dual minimally coupled scalar field in the AdS$_2$ black hole, and (iii) rescaling $J$ in the time variable $t$ and of $\Phi_h$. Thus, we confirm that the proposal in \cite{Caputa:2024sux} has a realization in the DSSYK model, at least below the time scale (\ref{eq:tprm}). We comment on related future directions in Sec. \ref{ssec:outlook}

On the other hand, the spread complexity of the two-sided HH state (\ref{eq:Krylov complexity State many}) does not display the transition to exponential growth and thus it cannot be matched unless one considers only the parabolic growth regime. This is in contrast with the CFT$_2$ case considered in \cite{Caputa:2024sux}. They showed that the spread complexity of states of the form $\hat{\mathcal{O}}_\Delta\ket{\rm TFD}$, with $\hat{\mathcal{O}}_\Delta$ being a primary CFT$_2$ operator, does indeed display exponential growth that can be matched with proper radial momentum.

\paragraph{Relation with operator size}
As initially defined in the finite $N$ SYK model, operator size measures the length of the operator string of a given orthonormal operator basis. It was shown in \cite{Lin:2022rbf} that the operator size in the DSSYK model is related to the total chord number through
\begin{equation}\label{eq:size}
    {\rm Size}=\frac{1}{p}\qty(\expval{\hat{N}}+\Delta_{\rm S})~,
\end{equation}
where $p$ is the number of all-to-all interactions in the SYK Hamiltonian (\ref{eq:DDSSYK Hamiltonian}) and the expectation value depend on the state under consideration. On the other hand, Krylov operator complexity corresponds to the expectation value (with respect to the two-sided HH state (\ref{eq:HH state tL tR})) of the total chord number in the semiclassical limit, in (\ref{eq:womrhole semiclassical answer many}). Thus, they are related to the radial proper momentum of a probe particle in an AdS$_2$ geometry as
\begin{equation}\label{eq:relation operator size}
    \dv{{~\rm Size}}{t}=\frac{1}{p}\dv{t}\mathcal{C}^{(\eta=-1)}(t)\propto \mathcal{P}_\sigma~.
\end{equation}
The specific proportionality factor appears in (\ref{eq:match proper radial momentum}). On the other hand, the relation between operator size, radial momentum, and Krylov complexity growth in the SYK model and JT gravity was also explored by \cite{Jian:2020qpp}. In contrast to (\ref{eq:relation operator size}), \cite{Jian:2020qpp} found (in particular examples) that operator size is proportional to the proper radial momentum of a probe particle in AdS$_2$ space. It is curious that the relation between the proper momentum, operator size and Krylov complexity is modified in the DSSYK model in contrast to the finite N SYK \cite{Jian:2020qpp}.

\section{Discussion}\label{sec:dis}
\paragraph{Summary}{{In this work, we have derived the holographic dictionary of the DSSYK model with a one-particle insertion, and we discussed several of its features.}} We have studied general properties and observables in the two-sided HH state of the DSSYK model with matter chords (\ref{eq:HH state tL tR}), to elucidate the bulk description of its double-scaled algebra of observables. In the semiclassical limit, we found that there are three type of approaches to recover the same dual bulk minimal geodesic length. \emph{The path integral} of the DSSYK allowed us to prepare the HH state, and to derive the classical phase space of the DSSYK model. This method is very convenient to derive two-sided two-point correlation functions by solving EOM, and without taking a triple-scaling limit. Meanwhile, the \emph{chord diagrams} allowed our derivation of the semiclassical thermodynamics of the one-particle DSSYK, and to verify our path integral results on the two-sided two-point functions with light and heavy operators in App. \ref{app:twopoint oneparticle}.
At last, the \emph{Krylov complexity for states and operators} plays a central role in deriving the holographic dictionary of the DSSYK model, and it allows for a boundary interpretation of the results closely related to quantum chaos, as discussed in Sec. \ref{ssec:special}. To derive it, we formulated a Lanczos algorithm (\ref{eq:Lanczos eta}) that evaluates Krylov operator and spread complexity simultaneously (which depends on the {{relations between the boundary times}} of the two-sided Hamiltonian evolution) and including finite temperature effects. This version of the Lanczos algorithm is also valid for generic two-sided Hamiltonian systems with/without a bounded spectrum. The Krylov operator and spread complexity in the two-sided HH state correspond to geodesic lengths anchored to the asymptotic boundaries with different boundary time {{relations}}. We showed that the generating function of Krylov complexity is a two-sided two-point function. An interesting special case is when the boundary times evolve as $t_L=-t_R$. In this case, we observed a transition from exponential to linear growth in Krylov operator complexity; exponential decrease to a power law decay in an OTOC (resulting from the two-sided two-point correlation). We found that the scrambling time of Krylov operator complexity corresponds to that of the OTOC, and to the one in a shockwave geometry, all at finite temperature. The explicit OTOC obeys bounds found in the literature \cite{Milekhin:2024vbb}. It also features sub-maximal chaos, and its scrambling time is independent of the system size, which is associated to hyperfast growth \cite{Susskind:2021dfc}.

As for the bulk description, we showed that the backreaction of a massive particle at rest in AdS$_2$, or by a shockwave profile can be matched to the saddle point solutions of the total chord number evaluated in the two-sided HH state. This allowed us to derive the holographic dictionary in both cases, and then {{make predictions regarding the canonically quantization}} of the bulk theory {{and its algebra of observables}}. Thus, our results suggest that these three techniques take a \emph{significant role} in understanding holography in the DSSYK model; and particularly, on the dynamics of its bulk dual in the presence of matter fields. For the convenience of the reader, we collected the main outcomes of the paper in App. \ref{app:results}.

\subsection{Outlook}\label{ssec:outlook}
Besides the points below, additional future directions can be found in footnotes \ref{fnt:generalized Ehrenfest}, \ref{fnt:MX bound}, \ref{fnt:TTbar}, \ref{fnt:matrix models}, and \ref{fnt:nonperturbative Okuyama}.

\paragraph{A bulk description of the double-scaled algebra {{and its observables}}}
{{We formulated}} (\ref{eq:bulk algebra}) based on the holographic dictionary of the DSSYK (Tab. \ref{tab:holographic_dictionary}). {{Dressing the bulk fields with respect to the asymptotic boundary allows one to evaluate diffeomorphism invariant correlation functions, such as (see Fig. \ref{fig:dressed_fields} for an illustration)
\begin{figure}
    \centering
    \includegraphics[width=0.45\textwidth]{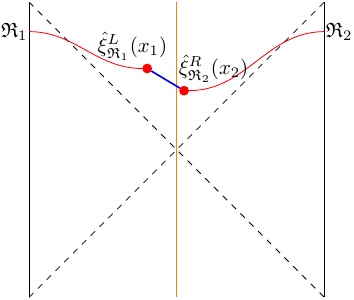}
    \caption{{{Representation of the scalar field $\hat{\xi}$ dressed through geodesics lines (red curves) with respect to points $\mathfrak{R}_{1/2}$ in the asymptotic boundary of an effective AdS$_2$ black hole background, which we denote $\hat{\xi}^{L/R}_{\mathfrak{R}_{1/2}}$ (represented by red dots). The dashed black lines represent the black hole horizon; while orange solid line represents the particle insertion in the bulk (similar to Fig. \ref{fig:one_particle_Euclidean}); and the blue solid line represents the evaluation of \eqref{eq:relational correlator}.}}}
    \label{fig:dressed_fields}
\end{figure}
\begin{equation}\label{eq:relational correlator}
    \bra{\Psi}\hat{\xi}^{L/R}_{\mathfrak{R}_1}(x_1)\hat{\xi}^{L/R}_{\mathfrak{R}_2}(x_2)\ket{\Psi}~,
\end{equation}
where $\ket{\Psi}\in\mathcal{H}_{\rm phys}$ (the latter being the physical bulk Hilbert space) is defined in a Cauchy slice that intercepts the spacetime points $x_1$ and $x_2$. Thus, the algebra of observables in the bulk is relational in the sense that it is defined by dressing the operators to the asymptotic boundary, resulting in non-local observables (such as \eqref{eq:relational correlator}). It would be a very interesting future development to evaluate \eqref{eq:relational correlator} explicitly in the putative bulk dual theory to confirm whether there is a divergence when $x_1\rightarrow x_2$ (expected in UV divergent models; but possibly not present in this model). Furthermore, assuming that the holographic correspondence with the DSSYK model holds, \eqref{eq:relational correlator} should also correspond to the expectation value of an operator in the boundary theory; which should be confirmed.}}

One might also try to study (\ref{eq:bulk algebra}) {{in sine dilaton gravity}} independently of the boundary theory, deduce that it is type II$_1$ (which in principle involves arbitrary more particle insertions). One should carefully check that it has the same properties as the double scaled algebra. Moreover, given that sine dilaton gravity has multiple black hole and cosmological horizons, it is also important to have a bulk interpretation of its generalized horizon entropy (see e.g. \cite{Chandrasekaran:2022cip,Chandrasekaran:2022eqq,Penington:2023dql,Witten:2023qsv,Witten:2025xuc,Penington:2024sum,Witten:2023xze,Witten:2021unn,DeVuyst:2024pop,DeVuyst:2024uvd,Jensen:2023yxy,Faulkner:2024gst,Kudler-Flam:2024psh,AliAhmad:2023etg,AliAhmad:2024saq,AliAhmad:2025ukh,Kudler-Flam:2023qfl}).

From the boundary perspective, it was realized in \cite{Tang:2024xgg} that there is a natural notion of entanglement entropy for the DSSYK model that reproduces some expectations for a type II$_1$ algebra, and the JT gravity RT surface in the triple-scaling limit. However, this is  not a satisfactory interpretation of the generalized entropy of the DSSYK and its dual, as there is no good understanding of the entanglement entropy for most states even in $\mathcal{H}_0$. {{Furthermore,}} there is no holographic interpretation of the would-be generalized horizon entropy beyond the triple-scaling limit. It would be interesting to study this in detail by constructing traces and reduced density matrices in chord space (which we associate with the bulk Hilbert space of the DSSYK \cite{Lin:2022rbf}). 

\paragraph{Lessons dS\texorpdfstring{$_3$}{} holography and sine dilaton gravity} There are different proposals on dS$_3$ holography based on the DSSYK model which incorporate other kinds of constraints and gauge fixing conditions as in the series of works by \cite{Verlinde:2024znh,Verlinde:2024zrh,Narovlansky:2023lfz,Gaiotto:2024kze,Tietto:2025oxn}, which we have commented earlier on. Our work found {{shares several similarities with sine dilaton gravity; however, one should check that the one-sided lengths and their canonical momenta in reproduce our predictions based on the holographic dictionary derived in (\ref{eq:Length AdS}) and (\ref{eq:Momenta AdS}) respectively}}. We {{also}} expect that the techniques developed here can also be applied in other settings where there are other type of constraints to reproduce the dS$_3$ holographic model in \cite{Narovlansky:2023lfz}. We discuss this in \cite{Aguilar-Gutierrez:2025hty}. To provide evidence that the original model by \cite{Narovlansky:2023lfz} is consistent with our study, one simply must impose other constraints in the chord Hilbert space. Additionally, the Krylov complexity for physical states and observables in the model by \cite{Narovlansky:2023lfz} have been previously studied by \cite{Aguilar-Gutierrez:2024nau}. It would be worth revisiting some of these results to connect with the findings in this work.

There was also an interesting work \cite{Blommaert:2025avl} on End-Of-The-World (ETW) branes, Euclidean wormholes and matrix models from sine dilaton gravity. In \cite{Aguilar-Gutierrez:2025hty} we address how imposing constraints in the one-particle space leads to their bulk model, which makes manifest that the ETW brane is itself the worldline of a massive particle in the effective AdS$_2$ description. The Krylov space methods in this work also play an important role. It would be interesting to connect \cite{Aguilar-Gutierrez:2025hty} with our discussion in Sec. \ref{ssec:special} for how to relate the Krylov basis in these models with a proper definition of quantum chaos from free probability theory, e.g. \cite{Camargo:2025zxr,Jahnke:2025exd}.

\paragraph{DSSYK correlation functions}
Our evaluation of the crossed four-point function (\ref{eq:generating function explicit DSSYK 1 particle}) from phase space methods can be seen as a prediction of a chord diagram computation (Fig. \ref{fig:OTOC_chord_diagram}) where a light and a composite heavy matter chord operator intercepts a light one. We evaluated the corresponding correlation function near the end of App. \ref{app:twopoint oneparticle}. However, to obtain an explicit answer in the semiclassical limit that indeed matches our prediction for heavy-heavy-light-light operators, we need to make an ansatz relating the energy variables of the particle chords. We can take some inspiration from an analogous diagram in JT gravity \cite{Lam:2018pvp} to carry out this computation. However, this results in a very similar ansatz to having only light operators in the crossed four-point function (which we discuss at the end on App. \ref{app:twopoint oneparticle}). Thus, this does not quite work for heavy operators. We hope to find an appropriate ansatz to carry out this evaluation. An alternative way to carry out these calculations is to express the diagrams in energy basis and work out the integrals numerically, as \cite{Xu:2024gfm}.

It is also important to learn how to compute more general correlation functions involving several heavy or light operators (Fig. \ref{fig:composite_op} (b)). We believe that a better understanding of the classical phase space methods for these more general cases would help us to derive similar predictions that can help to predict the answer for chord diagram correlation functions.

Another technical point to add in our chord diagram calculations is to include quantum corrections to the semiclassical results. We discuss this next.

\paragraph{{{Quantum corrections}}}
First, we comment on the {{quantization methods in this work}}. {{Our definition of the quantum gravity theory \eqref{eq:ADM SDG Hamiltonian} followed from the canonical quantization of \eqref{eq:total sdg matter}. It is manifestly of the same form as the DSSYK Hamiltonian with a one-particle insertion (\ref{eq:pair DSSYK Hamiltonians 1 particle}) provided the validity of the holographic dictionary (Tab. \ref{tab:holographic_dictionary}), which we deduced at the classical level and it is not altered by the canonical quantization (as we only promote degrees of freedom in the Hamiltonian to operators). One can use our definition of the quantum gravity theory to evaluate partition functions or correlation functions, in terms of a power series in $\kappa^2$ (i.e. the gravitational coupling \eqref{eq:ADM SDG Hamiltonian}) for the bulk theory, and check that they match those of the DSSYK in the corresponding $\lambda$ power series. This evaluation would be similar to \cite{Bossi:2024ffa}, which considers a similar problem for sine dilaton gravity and the DSSYK model without matter at the first order quantum correction to the semiclassical results.}}

{{Also, while discussing the quantization of the bulk theory, we mentioned that}}, there is an additional treatment regarding the partition functions after or without implementing constraints in the bulk Hilbert space that have been explored in sine dilaton gravity without matter \cite{Blommaert:2024whf}. Perhaps exploring a similar treatment could be a useful check of the boundary-to-bulk map in \cite{Lin:2022rbf}, and to {{provide an alternative method to}} confirm our arguments in Sec. \ref{ssec:canonical quantization}.

It would also be interesting to consider quantum corrections in our saddle point solutions of the path integral, {{such as (\ref{eq:chord number as Krylov complexity op})}}. {The path integral offers some {{practical}} advantages over computing the chord diagram correlation functions {{in the sense that}} we can skip tedious calculations {{involving R-matrices, instead by}} solving EOM, {{which nevertheless leads to the}} semiclassical answer in the double-scaling limit. {{In contrast, this}} is often limited to {{a}} triple-scaling {{evaluation}} in the literature (e.g. \cite{Berkooz:2022fso}). Nevertheless, an important \emph{limitation} {{of the path integral method}} is that if we want to derive quantum corrections, we should work instead with an effective action to a given loop order, which is not well developed in the DSSYK model. {{On the other hand,}} loop corrections in chord diagram correlation functions are much better studied, e.g. \cite{Goel:2023svz,Okuyama:2023bch,Xu:2024hoc}, and {{as well as}} from a bulk perspective in \cite{Bossi:2024ffa}. {{Furthermore, chord diagrams are advantageous in expressing observables (such as partition functions; correlation functions) in terms of integrals $\forall q\in[0,1)$, that may be solved in terms of a saddle point approximation or with numerical methods (see e.g. \cite{Xu:2024gfm,Heller:2024ldz}). Nevertheless, no technique is better than the other, each one has different practical uses, and in principle one might need to exploit both techniques in complex problems, as for instance, when we studying multiple shockwave configurations \cite{Aguilar-Gutierrez:2025mxf}}. It would be interesting to develop effective actions in the DSSYK model to incorporate quantum effects in the path integral evaluations of the observables studied in this work.}} A first step in this direction would be to derive the first order quantum corrections in the two-point correlation function of the DSSYK without matter \cite{Okuyama:2023bch,Goel:2023svz}. This could help us predict quantum corrections in the wormhole distance dual to the total chord number in the DSSYK model with matter (which would no longer be related to a Krylov complexity operator). If we consider quantum corrections to the semiclassical limit, one should account for backreaction of matter effects inside the bulk. This might be interesting to connect with bulk matter contributions in the CV in JT gravity that has been previously mentioned in \cite{Carrasco:2023fcj}. 

{{Similarly, we discussed in Sec. \ref{ssec:special} that one should incorporate quantum corrections in \eqref{eq:womrhole semiclassical answer many} to verify that the expression remains well-approximated past the scrambling time \eqref{eq:plus case} when considering light operators (i.e. $\Delta_{\rm S}$ is fixed as $\lambda\rightarrow0$). There is numerical evidence supporting that this approximation holds when $\beta_L=\beta_R=0$ \cite{Ambrosini:2024sre}; but it should be verified also for other values of $\beta_L=\beta_R$. Note however that this does not play a role in our discussion of scrambling dynamics for $\beta_L=-\beta_R$, as discussed in more detail in Sec. \ref{ssec:special}.}}

Another interesting limit is $q\rightarrow 0$, where we recover a random matrix model behavior for the DSSYK model without matter. In this limit the matterless DSSYK density of states in $\mathcal{H}_0$ is controlled by the Wigner surmise. This might be a useful test ground to {{extend our study}}; for instance, to study the thermodynamics of the system in the quantum regime (i.e. without $\lambda\rightarrow0$).
 
\paragraph{More on the holographic dictionary of the two-sided DSSYK Hamiltonian}
It would be interesting to relate our results with other proposals of complexity. For instance, the non-perturbative overlaps in JT gravity put forward by \cite{Miyaji:2024ity,Miyaji:2025yvm} have been used to introduce generating functions of the type $q^{\Delta\hat{N}}=\rme^{-\Delta\hat{\ell}}$. Thus, we expect that our results can be seen as a q-deformation of theirs (albeit at the semiclassical limit $\lambda\rightarrow0$, with $q^\Delta$ finite). A way to connect our results would be to study higher genus contributions to the correlation function (which we will discuss in future work). One might then compare with those in \cite{Miyaji:2024ity,Miyaji:2025yvm} in the low-energy limit. It would also be fascinating to see what aspects of the fake disk (introduced by \cite{Lin:2023trc} in the semiclassical limit of the one-particle chord algebra) are captured in the semiclassical wormhole length (\ref{eq:womrhole semiclassical answer many}). For instance, the difference between physical and fake temperature in the DSSYK model was associated with a geometric description in sine dilaton gravity by \cite{Blommaert:2024ymv}, which appears to be captured by the exponential growth of Krylov operator complexity below the scrambling time (\ref{eq:intermediate time COp}).

\paragraph{Rate of growth of Krylov complexity vs the wormhole distance}
We would like to interpret our results from Sec. \ref{sec:bulk} which seems to relate the growth of a wormhole geodesic length with the proper momentum of a point particle in the geometry. Both quantities are related to Krylov operator complexity (and its rate of growth) provided (\ref{eq:match proper radial momentum}). It might be worth to find a purely bulk derivation confirming this relation between seemly unrelated dynamical parameters. We also remark that we could only match the early time behavior of the rate of growth of Krylov operator complexity with the radial proper momentum (\ref{eq:match proper radial momentum}). It could be interesting to explore if this can be extended to the late time regime where the Krylov operator complexity grows linearly instead.

\section*{Acknowledgments}
I am happy to thank Hugo Camargo, Rathinda Nath Das, Johanna Erdmenger, Viktor Janke, Keun-Young Kim, Mitsuhiro Nishida, and Zhuo-Yu Xian for collaboration on related projects; especially Jiuci Xu for several discussions, and comments on an earlier draft. I am also grateful for useful feedback by Stefano Baiguera, Andreas Blommaert, Kuntal Pal, and Pratik Nandy; and to Gonçalo Araujo-Regado, Masamichi Miyaji, Klaas Parmentier, Adrián Sánchez-Garrido and Masataka Watanabe for illuminating discussions. I am indebted to the organizers of ``New Avenues in Quantum Many-body Physics and Holography'' in APCTP, where this work started; ``Recent Developments in Black Holes and Quantum Gravity'' in the YITP at Kyoto University; ``Strings 2025'' in NYU Abu Dhabi; ``Workshop on Low-dimensional Gravity and SYK Model'' in Shinshu University; ``QISS 2025 Conference'' in Vienna University, the YITP-I-25-01 on ``Black Holes, Quantum Chaos, and Quantum Information'', where this work was completed; and especially to Don Marolf in UC Santa Barbara for hosting me during the development of this manuscript, and for travel support from the APCTP, The Scholarships Office, the QISS consortium, and the YITP. SEAG is supported by the Okinawa Institute of Science and Technology Graduate University. This project/publication was also made possible through the support of the ID\#62312 grant from the John Templeton Foundation, as part of the ‘The Quantum Information Structure of Spacetime’ Project (QISS), as well as Grant ID\# 62423 from the John Templeton Foundation. The opinions expressed in this project/publication are those of the author(s) and do not necessarily reflect the views of the John Templeton Foundation.

\appendix

\section{Notation}\label{app:notation}
\begin{itemize}
\item $\mathcal{A}_{\rm DS}$ (\ref{eq:operators DS algebra}): Double-scaled algebra.
\item $J$ (\ref{eq:DDSSYK Hamiltonian}), $N$, and $p$ (\ref{eq:double scaling}): coupling constant in the DSSYK Hamiltonian; total number of fermions; and number of all-to-all interactions, respectively.

\item $\lambda=2N/p^2\equiv-\log q$ (\ref{eq:double scaling}): Ratio related to the penalty factor for DSSYK Hamiltonian interceptions $q=\rme^{-\lambda}\in[0,1)$.

\item $[n]_q=\frac{1-q^n}{1-q}$: q-deformed integer.

\item $(a;q)_n=\prod_{k=0}^{n-1}(1-aq^k)$ (\ref{eq:q-Pochhammer symbol}): q-Pochhammer symbol.

\item $(a_1,a_2,\dots a_m;q)_n=\prod_{i=1}^N(a_i;~q)_n$ (\ref{eq:combine Pochhammer}).

\item $\hat{n}$ (\ref{eq: Oscillators 1}): chord number operator in the zero-particle sector $\mathcal{H}_0$.

\item $\ket{0}$: Tracial state of the DSSYK model.

\item $\hat{n}_i^>=\sum_{j={i+1}}^{m}\qty(\hat{n}_j+\Delta_{j})$ and $\hat{n}_i^<=\sum_{j=0}^{i-1}\qty(\hat{n}_j+\Delta_{j+1})$ in (\ref{eq:Hmultiple}). Similarly, for $\hat{\ell}_i^{>}=\lambda ~\hat{n}_i^{>}$, and $\hat{\ell}_i^{<}=\lambda ~\hat{n}_i^{<}$.

\item $Z(\beta)=\bra{0}\rme^{-\beta\hH}\ket{0}$ (\ref{eq:partition function zero particle chord}): Partition function for the DSSYK Hamiltonian $\hH$ (\ref{eq:Transfer matrix}) without matter.

\item $E(\theta)=\frac{2J~\cos(\theta)}{\sqrt{\lambda(1-q)}}$ (\ref{eq:energy spectrum}): Energy spectrum of the DSSYK model.

\item $S(\theta)$ (\ref{eq:entropy temp}): Thermodynamic entropy.

\item $\beta(\theta)$ (\ref{eq:entropy temp}): Inverse temperature.

\item $\beta_{\rm fake}=\frac{\pi}{J\sin\theta}$ (\ref{eq:fake temperature result}): Fake temperature.

\item $H_n(x|q)$ (\ref{eq:H_n def}): q-Hermite polynomials.

\item $\rmd\mu(\theta)$ (\ref{eq:norm theta}): Integration measure in the energy parametrization basis $\theta$.

\item $G_\beta^{(\Delta)}(\tau)$ (\ref{eq:correlator DSSYK}): Thermal two-point correlation function/Krylov generating function in $\mathcal{H}_{0}$.

\item $\hat{\mathcal{O}}_\Delta$ (\ref{eq:correlator DSSYK}): Matter chord operator with conformal dimension $\Delta$.

\item $\hat{\mathcal{O}}_\Delta(t)\equiv\rme^{-\rmi t\hH}\hat{\mathcal{O}}_\Delta\rme^{\rmi t\hH}$.

\item $\mathcal{H}_{m}$ (\ref{eq:Fock space with matter}): $m$-particle chord space.

\item $\hat{N}$ (\ref{eq:total chord number basis}): Total chord number with one or more particle chords.

\item $\ket{\tilde{\Delta};0,\dots,0}\equiv\hat{\mathcal{O}}_{\Delta_1}\dots\hat{\mathcal{O}}_{\Delta_m}\ket{0}$, where $\tilde{\Delta}=\qty{\Delta_1,\dots,\Delta_m}$ is a collective index.

    \item $\chi_{k, m_L, m_R}$ (\ref{eq:LinStanfordIdentity}): Clebsch-Gordan coefficients.

    \item $E_{L/R}$ (\ref{eq:conserved energies}): Energy spectrum of the left/right sectors in $\mathcal{H}_{m}$.

\item $Z_{\tilde{\Delta}}(\beta_L,\beta_R)$ (\ref{eq:overall partition function}): Partition function in the two-sided HH state.

    \item $C_\Delta$ (\ref{eq:heat capacity Delta}): Heat capacity in the DSSYK model (at fixed $\lambda$, $J$, and $\Delta$).
    
    \item $G^{(\Delta_w)}_{\tilde{\Delta}}(\tau_L,\tau_R)$ (\ref{eq:generating function 1 particle}): Thermal two-point correlation function in chord matter space, corresponding to a four-point function in $\mathcal{H}_0\otimes\mathcal{H}_0$.

\item $\ket{\Psi_{\tilde{\Delta}}(\tau_L,\tau_R)}$ (\ref{eq:HH state tL tR}): Double-scaled PETS, a generalization of the HH state with matter chord insertions, and complex times $\tau_{L/R}=\frac{\beta_{L/R}}{2}+\rmi t_{L/R}$.

    \item $\hat{\mathcal{L}}_\eta=\hH_R+\eta\hH_L$ (\ref{eq:Lanczos eta}): Liouvillian operator.

    \item $\eta=\tau_L/\tau_R$: Ratio between the two-side complex time variables, which we take to be constant in order to use the Lanczos algorithm (\ref{eq:Lanczos eta}).

\item $\mathcal{C}^{(\eta)}$ \eqref{eq:Krylov complexity eta}: Krylov complexity for two-sided Hamiltonian systems . $\eta=-1$ corresponds to Krylov operator complexity for $\mathcal{O}_\Delta$, and $\eta=1$ to spread complexity.

\item $a^{(\eta)}_n$ and $b^{(\eta)}_n$ (\ref{eq:Lanczos eta}): Lanczos coefficients.

    \item $\ket{K_n^{(\eta)}}$ (\ref{eq:normal ordered total chord number}): Krylov basis in the Lanczos algorithm (\ref{eq:Lanczos many}).

\item $\Psi_{n}^{(\eta)}(\tau)=\bra{K_n^{(\eta)}}\ket{\Psi_{\tilde{\Delta}}(\tau_L=\eta\tau,\tau_R=\tau)}$ (\ref{eq:decomposition Krylov}): Amplitudes solving the Lanczos algorithm (\ref{eq:Lanczos algorithm eta}).
    
    \item $\hat{a}^\dagger_{i}$ and $\hat{\alpha}_{i}$ (\ref{eq:non-conjugate ops many particles}): Non-Hermitian conjugate creation and annihilation operators in $\mathcal{H}_{m}$.

    \item $\hat{\ell}(t_L,t_R)$ (\ref{eq:evol many}), $\hat{P}_{0/m}$ (\ref{eq:evol many}): Canonical distance and momenta in $\mathcal{H}_{m}$, with a two-sided Hamiltonian $\hH_{L/R}$ (\ref{eq:H LR many particles1}, \ref{eq:H LR many particles2}). The corresponding quantities without hat ${\ell}(t_L,t_R)$ (\ref{eq:evol many}) and ${P}_{0/m}$ (\ref{eq:evol many}) indicate classical variables.

    \item ${\Delta}_{\rm S}=\sum_{i=1}^m\Delta_i$: Sum of conformal dimensions for operators $\hat{\mathcal{O}}_{\Delta_1},\dots\hat{\mathcal{O}}_{\Delta_m}$.

    \item $\ell_*$ (\ref{eq:in cond many}): Initial condition in the semiclassical wormhole distance, determined by the operator dimension and the energy spectrum of the system.

\item $\ell_*^{(+)}$ (\ref{eq:ell * state}), $\ell_*^{(-)}$ (\ref{eq:ell * operator}): Initial values Krylov operator and spread complexity (at finite temperature) respectively.

    \item $\gamma_{L/R}=J\sin\theta_{L/R}$ (\ref{eq:gamma parameter}): Frequency parameters in classical phase space solutions.

\item $I_{\rm JT}[g,\Phi]$ (\ref{eq:JT gravity}), $I_{\rm SDG}[g,\Phi]$ (\ref{eq:sine dilaton gravity}), $I_{\rm (n)mc-scalar}[g,\xi,\Phi]$ (\ref{eq:matter action}): JT gravity, sine dilaton gravity and (non-)minimally coupled scalar field ($\xi$) actions respectively. 

\item $\kappa^2=8\pi G_N$ (\ref{eq:JT gravity}): Gravitational coupling.

    \item $\rmd s^2_{\rm eff}$ (\ref{eq:effective geometry}): Effective AdS$_2$ geometry in sine dilaton gravity.
    
    \item $L_{\rm AdS}(u_L,u_R)$ (\ref{eq:length AdS sw}): Geodesic distance in an AdS$_2$ black hole background with a minimally coupled scalar field. The asymptotic boundaries are located at $r=\Phi_b$ and boundary times $u_L$ and $u_R$, in the coordinates (\ref{eq:effective geometry}).

    \item $t_{\rm sc}$ (\ref{eq:scrambling time}), $u_{\rm sc}$ (\ref{eq:scrambling AdS}): Scrambling time in the boundary and bulk theory respectively.

     \item $H_{L/R}^{(\rm ADM)}$ (\ref{eq:sympletic LR}): two-sided ADM Hamiltonian.
     
     \item $\omega$: Symplectic form in the boundary (\ref{eq:xsympletic form}), and in the bulk (\ref{eq:sympletic LR}).


\item $\ket{\psi(\theta_{L},\theta_{R})}$ (\ref {eq:ADM Hamiltonian}): Bulk states with fixed ADM energy $E_{\rm ADM}(\theta_{L/R})$ (\ref{eq:constrained HADM}).

\item $ \psi_{n_L,n_R}(\theta_L,\theta_R;q,r)$ \eqref{eq:notation psi nLnR}: Bulk wavefunction with left/right energies $E(\theta_{L/R})$. 

   \item $\hat{\mathcal{F}}$ (\ref{eq:isometric map}): Isometric map $\mathcal{H}_1\rightarrow\mathcal{H}_0\otimes\mathcal{H}_0$.

\item $\hat{a}_L=\hat{a}\otimes\mathbb{1}$, $\hat{a}_R=\mathbb{1}\otimes\hat{a}$: Creation and annihilation operators in $\mathcal{H}_1$.

\item $|m,n)\equiv\ket{m}\otimes\ket{n}$ \eqref{eq:prod 1}; $|\theta_1,\theta_2)\equiv\ket{\theta_1}\otimes\ket{\theta_2}$ \eqref{eq:prod 2}.

    \item $\gamma^{(q)}_\Delta(\theta_L,\theta_R)$ (\ref{eq:gammaq}): factor in crossed four-point correlation function (\ref{eq:Gdelta}).
    
    \item $R_{\theta_2\theta_4}
^{(q)}\begin{pmatrix}
\theta_1&\Delta_1\\
\theta_2&\Delta_2
\end{pmatrix}$ (\ref{eq:Rmatrix}): R-matrix in the crossed four-point correlation function (\ref{eq:Gdelta}).

\item $_8W_7(a_1;a_4,\ldots,a_{r+1};q,z)$ is a very-well-poised basic hypergeometric series (\ref{eq:very-well-poised basic hypergeometric series}).
\end{itemize}

\paragraph{Acronyms}
\begin{itemize}[noitemsep]
\item ADM: Arnowitt–Deser–Misner
\item (A)dS: (Anti-)de Sitter
\item CA: Complexity equals action
    \item CFT: Conformal field theory
    \item CV: Complexity equals volume
    \item (DS)SYK: (Double-scaled) Sachdev–Ye–Kitaev
    \item EOM: Equations of motion
    \item ETW: End-Of-The-World
    \item GFF: Generalized free fields
    \item HH: Hartle-Hawking
    \item Irrep: Irreducible representation
    \item JT: Jackiw–Teitelboim
    \item OTOC: Out-of-time-ordered correlator
    \item PETS: Partially entangled thermal state
    \item QRF: Quantum reference frame
    \item TFD: Thermofield double state
    \item UOGH: Universal operator growth hypothesis
    \item WDW: Wheeler-DeWitt
\end{itemize}

\section{List of the main results}\label{app:results}
We refer the reader to App. (\ref{app:notation}) (or the main text) for details about the notation used below.
\paragraph{DSSYK model with $m$-particles in canonical variables (\ref{eq:Hamiltonians LR})}
\begin{align*}
&\tfrac{\sqrt{\lambda(1-q)}}{J}\hH_{L}=\rme^{-\rmi \hat{P}_0}+\sum_{i=0}^m\rme^{\rmi \hat{P}_i}(1-\rme^{-\hat{\ell}_i})\rme^{-\hat{\ell}_i^<}\quad\text{where}\quad\hat{\ell}_i^<=\sum_{j=0}^{i-1}\qty(\hat{\ell}_j+\lambda\Delta_{j+1})~,\\
&\tfrac{\sqrt{\lambda(1-q)}}{J}\hH_{R}=\rme^{-\rmi \hat{P}_m}+\sum_{i=0}^m\rme^{\rmi \hat{P}_i}(1-\rme^{-\hat{\ell}_i})\rme^{-\hat{\ell}_i^>}\quad\text{where}\quad\hat{\ell}_i^>=\sum_{j={i+1}}^{m}\qty(\hat{\ell}_j+\lambda\Delta_{j})~.
\end{align*}
\paragraph{Two-sided HH state (\ref{eq:HH state tL tR})}
\begin{equation*}
\ket{\Psi_{\tilde{\Delta}}(\tau_L,\tau_R)}=\rme^{-\tau_L\hH_L-\tau_R\hH_R}\prod_{i=1}^m\hat{\mathcal{O}}_{\Delta_i}\ket{0}=\rme^{-\tau_L\hH_L-\tau_R\hH_R}\ket{\Delta_{\rm S};0,0}~,
\end{equation*}
where $\Delta_{\rm S}\equiv\sum_i\Delta_i$.
\paragraph{Partition function (\ref{eq:overall partition function})}
\begin{equation*}
    Z_{\tilde{\Delta}}(\beta_L,\beta_R)= \bra{\Delta_{\rm S};0,0}\rme^{-\beta_L\hH_L-\beta_R\hH_R}\ket{\Delta_{\rm S};0,0}~.
\end{equation*}
The one-loop corrected thermodynamics of the \emph{(thermally) stable }saddle point solutions for both light and heavy particles 
(\ref{eq:saddles 1 particle S}, \ref{eq:saddles 1 particle beta}, \ref{eq:heat capacity Delta}) (with $n=0$) correspond to
\begin{align*}
    &S(\theta_{L/R})\eqlambda-\frac{2}{\lambda}\qty(\theta_{L/R}-\frac{\pi}{2})^2~,\quad\beta_{L/R}\eqlambda\pdv{S(\theta_{L/R})}{E_{L/R}}\frac{2\theta_{L/R}-\pi}{J\sin\theta_{L/R}}~,\\
    &C_{\Delta_{\rm S}}(\theta_{L/R})=-\beta_{L/R}^2\pdv{E_{L/R}}{\beta_{L/R}}\eqlambda\frac{J^2}{\lambda}\frac{\beta_{L/R}^2\sin^2\theta_{L/R}}{1+\qty(\qty(\frac{1}{2})\pi-\theta_{L/R})\cot\theta_{L/R}}~.
\end{align*}
\paragraph{Path integral of the DSSYK (\ref{eq:Hamilton PI})}
\begin{equation*}
    \int\prod_{i=0}^m[\rmd \ell_i][\rmd P_i]\exp\qty[\int\rmd\tau_L\rmd\tau_R\qty(\frac{\rmi}{\lambda}\sum_{i=0}^m\qty(P_i(\partial_{\tau_L}+\partial_{\tau_R})\ell_i)+H_L+H_R)]~.
\end{equation*}
The saddle point solutions for $\ell(t_L=t_R=0)=\ell_*$ and $\eval{\partial_{t_{L/R}}\ell}_{t_L=t_R=0}=0$ are (\ref{eq:solutions HH})
\begin{align*}
&\ell=2\log \qty(\cosh(\gamma_L t_L)\cosh(\gamma_R t_R)+\frac{J^2q^{\Delta_{\rm S}}\rme^{-\ell_*}}{\gamma_L\gamma_R}\sinh(\gamma_L t_L)\sinh(\gamma_R t_R))+\ell_*~,\\
&P_{0}=\rmi\log(\cos\theta_{L}+\rmi\sin\theta_{L}\frac{\rme^{\ell_*}\sin\theta_L\sin\theta_R\tanh(\gamma_{L}t_{L})+q^{\Delta_{\rm S}}\tanh(\gamma_{R}t_{R})}{\rme^{\ell_*}\sin\theta_L\sin\theta_R+q^{{\Delta_{\rm S}}}\tanh(\gamma_Rt_R)\tanh(\gamma_Lt_L)})~,\\
&P_{m}=\rmi\log(\cos\theta_{R}+\rmi\sin\theta_{R}\frac{\rme^{\ell_*}\sin\theta_L\sin\theta_R\tanh(\gamma_{R}t_{R})+q^{\Delta_{\rm S}}\tanh(\gamma_{L}t_{L})}{\rme^{\ell_*}\sin\theta_L\sin\theta_R+q^{{\Delta_{\rm S}}}\tanh(\gamma_Rt_R)\tanh(\gamma_Lt_L)})~.
\end{align*}
where $\gamma_{L/R}\equiv J\sin\theta_{L/R}$. Here $\ell_*$ is determined by energy conservation \eqref{eq:ell cond thetaLR}:
\begin{align*}
    &\rme^{-\ell_*(\theta_L,\theta_R)}=\frac{q^{-2\Delta_{\rm S}}}{2}\biggl(1+q^{2\Delta_{\rm S}}-2q^{\Delta_{\rm S}}\cos\theta_L\cos\theta_R\\
    &-\sqrt{1+q^{4\Delta_{\rm S}}+2 q^{2\Delta_{\rm S}}(1+\cos(2\theta_L)+\cos(2\theta_R))-4q^{\Delta_{\rm S}}(1+q^{2\Delta_{\rm S}})\cos\theta_L\cos\theta_R}\biggr)~.\nonumber
\end{align*}
We also denote
\begin{equation*}
\ell_*^{(+)}(\theta)\equiv\ell_*(\theta_L=\theta,\theta_R=\theta) ~,\quad \ell_*^{(-)}(\theta)\equiv\ell_*(\theta_L=\pi-\theta,\theta_R=\theta) ~.
\end{equation*}

\paragraph{Two-sided two-point correlation function (\ref{eq:generating function explicit DSSYK 1 particle})}
\begin{equation*}
\begin{aligned}
    &G^{(\Delta_w)}_{\tilde{\Delta}}(\tau_L,\tau_R)\equiv\frac{\bra{\Psi_{\tilde{\Delta}}(\tau_L,\tau_R)}q^{\Delta_w \hat{N}}\ket{\Psi_{\tilde{\Delta}}(\tau_L,\tau_R)}}{Z_{\tilde{\Delta}}(\beta_L,\beta_R)}\\
&\eqlambda\qty(\frac{\rme^{-\ell_*/2}}{\cosh(\gamma_Lt_L)\cosh(\gamma_Rt_R)+\frac{J^2\rme^{-\ell_*(\theta_L,\theta_R)}q^{{\Delta_{\rm S}}}}{\gamma_L\gamma_R}\sinh(\gamma_Lt_L)\sinh(\gamma_Rt_R)})^{2\Delta_w}~.
\end{aligned}    
\end{equation*}
\paragraph{Double-scaled precursors \eqref{eq:double scaled precursor shockwave}}
These are operators in the Schrödinger picture
\begin{equation*}
    \hat{\mathcal{O}}^{L}_\Delta(t_w)\equiv\rme^{\rmi t_w \hH_{L}}\hat{\mathcal{O}}^{L}_\Delta\rme^{-\rmi t_w\hH_{L}}~,\quad \hat{\mathcal{O}}^{R}_\Delta(t_w)\equiv\rme^{-\rmi t_w \hH_{R}}\hat{\mathcal{O}}^{R}_\Delta\rme^{\rmi t_w\hH_{R}}~,
\end{equation*}
where $t_{w}\in\mathbb{R}$ is the insertion time. When $t_w\gg\abs{t_{L/R}}$ and $\theta_L=\theta_R=\theta$, the saddle point solution for the rescaled total chord number is (\ref{eq:shockwave approx})
\begin{equation*}
\begin{aligned}
\ell(t_L,t_R)\simeq&2\log \biggl(\cosh(J\sin\theta(t_L+t_R))+\qty(\frac{1}{4}-\frac{q^{\Delta_{\rm S}}\rme^{-\ell_*^{(+)}(\theta)}}{4\sin^2\theta})\rme^{-J\sin\theta (t_L-t_R-2t_w)}\biggr)+\ell^{(+)}_*(\theta)~.
\end{aligned}
\end{equation*}
\paragraph{Lanczos algorithm (\ref{eq:Lanczos eta}) for the two-sided DSSYK}
Consider a reference state
\begin{equation}
    \ket{\Psi_{\tilde{\Delta}}(\tau_L=\eta\tau,\tau_R=\tau)}=\rme^{-\mathcal{L}_\eta\tau}\ket{\Delta_{\rm S};0,0}=\sum_n\Psi_n^{(\eta)}(\tau)\ket{K^{(\eta)}_n}~,
\end{equation}
where $\hat{\mathcal{L}}_\eta=\hH_R+\eta\hH_L$ is the generalized Liouvillian operator (considering $\eta=\pm1$), and the \emph{Krylov basis} $\qty{\ket{K_n^{(\eta)}}}$ obeys a Lanczos algorithm
\begin{equation*}
    \hat{\mathcal{L}}_\eta\ket{K^{(\eta)}_n}=b^{(\eta)}_{n+1}\ket{K_{n+1}^{(\eta)}}+b^{(\eta)}_{n}\ket{K_{n-1}^{(\eta)}}~,
\end{equation*}
where (\ref{eq:normal ordered total chord number}) gives
\begin{equation*}
    \ket{K^{(\eta)}_n}=c^{(\eta)}_n\sum_{l=0}^n\eta^k\begin{pmatrix}
        n\\
        l
    \end{pmatrix}\ket{\Delta_{\rm S};l,n-l}+\text{subleading}~,
\end{equation*}
with $$c^{(\eta)}_n=\sqrt{\frac{\lambda^n(1-q)^n(1+\eta)}{2^n(q^{1/2};q^{1/2})_n(-\eta;q^{1/2})_{n+1}}}~,$$and the norm of other terms vanishes as $\lambda\rightarrow0$ when $\eta=\pm$. The \emph{Lanczos coefficients} (\ref{eq:Lanczos many}) are
\begin{equation*}
b^{(\eta)}_n\eqlambda\frac{2J}{\sqrt{\lambda(1-q)}}\sqrt{(1-q^{n/2})(1+\eta q^{n/2+\Delta_{\rm S}})}~.    
\end{equation*}
\emph{Krylov operator} ($\eta=-1$) and \emph{spread} ($\eta=+1$) \emph{complexity} correspond to (\ref{eq:Krylov complexity Operator many}) and (\ref{eq:Krylov complexity State many}) (from (\ref{eq:womrhole semiclassical answer many})) respectively
    \begin{align*}
        &\mathcal{C}^{(\eta=\pm1)}(t)\eqlambda\frac{2}{\lambda}\log(A(\theta, \pm q^{\Delta_{\rm S}})+B(\theta, \pm q^{\Delta_{\rm S}})\cosh(2J\sin\theta t))~,\\
&A(\theta,q^\Delta)=\frac{(\cos \theta
   -\cos 2 \theta ) q^{\Delta
   }+\left(1-q^{\Delta }\right) \left(q^{\Delta
   }+\sqrt{q^{2 \Delta }-2 \cos \theta 
   q^{\Delta }+1}-1\right)}{2
   \sin ^2\theta \left(q^{\Delta }+\sqrt{q^{2 \Delta }-2 \cos
   \theta  q^{\Delta }+1}-1\right)}~,\\
&B(\theta,q^\Delta)=\frac{q^{2 \Delta
   }-\left((\cos \theta +\cos 2 \theta )
   q^{\Delta }\right)-\left(1-q^{\Delta
   }\right) \sqrt{q^{2 \Delta }-2 \cos \theta
    q^{\Delta }+1}+1}{2
   \sin ^2\theta \left(q^{\Delta }+\sqrt{q^{2 \Delta }-2 \cos
   \theta  q^{\Delta }+1}-1\right)}~.
    \end{align*}
The $\eta=-1$ case has an associated scrambling time (i.e. a transition from exponential to linear growth) (\ref{eq:scrambling time}),
\begin{equation*}
   t_{\rm sc}=\frac{1}{2J\sin\theta}\log\frac{{\sin^2\theta}-{\rme^{-\ell^{(-)}_*(\theta)}}q^{\Delta_{\rm S}}}{{\sin^2\theta}+{\rme^{-\ell^{(-)}_*(\theta)}}q^{\Delta_{\rm S}}}~.
\end{equation*}

\paragraph{Holographic dictionary}The reader is referred to Tab. \ref{tab:holographic_dictionary} for a brief summary.

\paragraph{Bulk quantization}
The bulk variables take the form
\begin{align*}
&{P_{\rm AdS}^{L/R}=\rmi\log(\cos\theta+\rmi\sin\theta\frac{\tanh(\delta u_{L/R})+\rme^{-\frac{\kappa^2E_{\rm sw}}{16\pi\Phi_h}}\tanh(\delta u_{R/L})}{1+\rme^{-\frac{\kappa^2E_{\rm sw}}{16\pi\Phi_h}}\tanh(\delta u_{L})\tanh(\delta u_{R})})~,}\\
&\rme^{-L_{\rm AdS}^{L/R}}=\frac{2\cos P_{\rm AdS}^{L/R}-2\cos\theta-\rme^{-\frac{\kappa^2E_{\rm sw}}{16\pi\Phi_h}}\rme^{-L_{\rm AdS}+\rmi P_{\rm AdS}^{L/R}}}{\rme^{\rmi P_{\rm AdS}^{L/R}}-\rme^{-\frac{\kappa^2E_{\rm sw}}{16\pi\Phi_h}}\rme^{\rmi P_{\rm AdS}^{R/L}}}~,
\end{align*}
with $\delta u_{L}=\frac{\Phi_h}{2}(u_L-u_w)$, $\delta u_{R}=-\frac{\Phi_h}{2}(u_R+u_w)$, and {{$E_{\rm sw}/\Phi_h$ is proportional to the proper energy of the shockwave \cite{Shenker:2013pqa}; $\kappa^2$ is Newton's constant in two-dimensions}}. Canonical quantization leads to the bulk theory (\ref{eq:ADM SDG Hamiltonian})
\begin{equation*}
    \hH_{L/R}^{(\rm ADM)}=\frac{1}{\kappa^2}\qty(\rme^{-\rmi \hat{P}_{\rm AdS}^{L/R}}+\rme^{\rmi \hat{P}_{\rm AdS}^{L/R}}\qty(1-\rme^{-\hat{L}_{\rm AdS}^{L/R}})+\rme^{-\frac{\kappa^2E_{\rm sw}}{16\pi\Phi_h}}\rme^{\rmi \hat{P}_{\rm AdS}^{R/L}}\rme^{-\hat{L}_{\rm AdS}^{L/R}}\qty(1-\rme^{-\hat{L}_{\rm AdS}^{R/L}}))~.
\end{equation*}

\section{Background material}\label{sec:background material}
In this appendix, we briefly review general results on the DSSYK model (\ref{sapp:DSSYK review}) and Krylov complexity for operators and states (\ref{sapp:Krylov review}) which we use throughout the manuscript. This is \emph{not} intended to be a pedagogical review. The reader can find excellent reviews on the DSSYK model \cite{Berkooz:2024lgq} and Krylov complexity in \cite{Baiguera:2025dkc,Nandy:2024htc}. Other useful references with some review material can be found in \cite{Lin:2022rbf,Rabinovici:2023yex,Verlinde:2024znh,Aguilar-Gutierrez:2024yzu}, besides the original works on the DSSYK \cite{Berkooz:2018jqr,Berkooz:2018qkz}.

\subsection{Review of the DSSYK model}\label{sapp:DSSYK review}
\paragraph{DSSYK model without matter chords}
The SYK model (see \cite{Sachdev:1992fk,kitaevTalks,Cotler:2016fpe,Maldacena:2016upp,Saad:2018bqo,Maldacena:2018lmt,Jensen:2016pah,Polchinski:2016xgd,Chowdhury:2021qpy} among many others) consists on $N$ Majorana fermions in $(0+1)$-dimensions, and $p$-body all-to-all interactions,
\begin{equation}\label{eq:DDSSYK Hamiltonian}
    \hat{H}_{\rm SYK}=\rmi^{p/2}\sum_{I}J_{I}\hat{\psi}_{I}~,\quad J_{I}=0~,\quad\expval{J_{I}J_{I'}}=\begin{pmatrix}
        N\\
        p
    \end{pmatrix}^{-1}\frac{J^2N}{2p^2}\delta_{II'}~,
\end{equation}
where $I=i_1,\dots,~i_p$ is a collective index, $1\leq i_1<\dots<i_p\leq N$; $J_I=J_{i_1\dots i_p}$ are Gaussian distributed coupling constants; and  $\hat{\psi}_I\equiv\hat{\psi}_{i_1}\dots\hat{\psi}_{i_p}$ is a string of Majorana fermions. They obey a Clifford algebra $\qty{\hat{\psi}_{i},~\hat{\psi}_{j}}=2\delta_{ij}\mathbbm{1}$. We define the double-scaling limit as
\begin{equation}\label{eq:double scaling}
    N,~ p \rightarrow \infty~,\quad q = \rme^{-\lambda}~,\quad\lambda\equiv\frac{2p^2}{N}~ \text{fixed}~.
\end{equation}
The DSSYK model (and other models in the same universality class \cite{Erdos:2014zgc,Berkooz:2024lgq,Parisi_1994,Berkooz:2024evs,Berkooz:2024ofm,Almheiri:2024xtw,Gao:2024lem}) becomes solvable through combinatorial methods \cite{Berkooz:2020fvm,Berkooz:2018jqr,Berkooz:2018qkz}, even away of the low energy regime. The ensemble-averaged description of (\ref{eq:DDSSYK Hamiltonian}) can be formulated in an auxiliary (H-chord) Hilbert space
\begin{equation}\label{eq:Fock H0}
    \mathcal{H}=\bigoplus_{n=0}^\infty\mathbb{C}\ket{n}~,
\end{equation}
where $n$ denotes the number of open chords at a given time slice in the chord diagram. We pick an orthonormal chord number basis $\qty{\ket{n}}$, where the Hamiltonian in the ensemble theory is given by \cite{Berkooz:2018jqr,Berkooz:2018qkz}
\begin{align}\label{eq:Transfer matrix}
    &\hH=\frac{J}{\sqrt{\lambda}}\qty(\hat{a}^\dagger+\hat{a})~,\\
    &\hat{a}^\dagger\ket{n}=\sqrt{[n+1]_q}\ket{n+1}~,\quad \hat{a}^\dagger\ket{n}=\sqrt{[n]_q}\ket{n-1}~,\label{eq:a adagger DSSYK}
\end{align}
where $[n]_q:= \frac{1-q^n}{1-q}$, $\hat{a}^\dagger$ and $\hat{a}$ are the creation and annihilation operators of a q-deformed harmonic oscillator. 
We may also make a redefinition $\hat{a}^\dagger=\sqrt{[n]_q}\rme^{-\rmi \hat{p}}$, and $\hat{a}=\rme^{\rmi \hat{p}}\sqrt{[n]_q}$ to formulate the theory in terms of the chord number operator $\hat{n}$ and its conjugate momentum, $\hat{p}$:
\begin{equation}\label{eq: Oscillators 1}
    \hat{n}\ket{n}=n\ket{n}~,\quad \rme^{\pm\rmi \hat{p}}\ket{n}=\ket{n\mp1}~.
\end{equation}
The commutation relations become:
\begin{equation}\label{eq:chord numb op}
\begin{aligned}
    &\qty[\hat{n},~\rme^{\pm\rmi \hat{p}}]=\mp\rme^{\pm\rmi \hat{p}}~.
    \end{aligned}
\end{equation}
It can be seen eigenvalues of the Hamiltonian have the following form
\begin{equation}\label{eq:energy spectrum}
    \hH\ket{\theta}=E(\theta)\ket{\theta}~,\quad E(\theta)=\frac{2J}{\sqrt{\lambda(1-q)}}{\cos\theta}~.
\end{equation}
The solutions to the eigenvalue problem are
\begin{equation}\label{eq:qHermite recurrence}
E(\theta)\bra{\theta}\ket{n}=\frac{J}{\sqrt{\lambda}}\qty(\sqrt{[n]_q}\bra{\theta}\ket{n-1}+\sqrt{[n+1]_q}\bra{\theta}\ket{n+1})~,
\end{equation}
with $\bra{\theta}\ket{0}=1$ are given by
\begin{equation}
    \label{eq: proj E0 theta}
    \bra{\theta}\ket{n}={\frac{H_n(\cos\theta|q)}{\sqrt{(q;q)_n}}}~,
\end{equation}
with $(a;~q)_n$ is the q-Pochhammer symbol and related product:
\begin{align}
        (a;~q)_n\equiv\prod_{k=0}^{n-1}(1-aq^k)~,\label{eq:q-Pochhammer symbol}\\
    \label{eq:combine Pochhammer}
    (a_0,\dots, a_N;q)_n=\prod_{i=1}^N(a_i;~q)_n~.
    \end{align}
Meanwhile, $H_n(x|q)$ is the q-Hermite polynomial:
\begin{equation}\label{eq:H_n def}
    H_n(\cos\theta|q)=\sum_{k=0}^n\begin{bmatrix}
        n\\
        k
    \end{bmatrix}_q\rme^{\rmi(n-2k)\theta}~,\quad
    \begin{bmatrix}
        n\\
        k
    \end{bmatrix}_{q}\equiv\frac{(q;~q)_{n}}{(q;~q)_{n-k}(q;~q)_{k}}~.
\end{equation}
The $\ket{\theta}$ basis is normalized such that
\begin{align}
&\mathbb{1}=\int_{\theta=0}^{\theta=\pi}\rmd\mu(\theta)\ket{\theta}\bra{\theta}~,\quad \rmd\mu(\theta)\equiv\frac{\rmd\theta}{2\pi}(q,~\rme^{\pm 2 \rmi \theta};q)_\infty~,\quad \label{eq:norm theta}\\
\label{eq:identity theta}
&\text{where}\quad (x^{\pm a_1\pm a_2};q)_n\equiv(x^{a_1+ a_2};q)_n(x^{- a_1+a_2};q)_n(x^{-a_1+ a_2};q)_n(x^{- -a_1-a_2};q)_n~.
\end{align}
One can then evaluate the partition function of the model as
\begin{equation}\label{eq:partition function zero particle chord}
Z(\beta)=\bra{0}\rme^{-\beta\hH}\ket{0}=\int_{\theta=0}^{\theta=\pi}\rmd\mu(\theta)\rme^{-\beta E(\theta)}~,
\end{equation}
which in the semiclassical limit reduces to 
\begin{align}\label{eq:partition H0}
Z(\beta)&\eqlambda\int{\rmd E(\theta)}\rme^{S(\theta)-\beta E(\theta)}~,
\end{align}
where we have used the following approximation for (\ref{eq:identity theta})
\begin{equation}
\begin{aligned}\label{eq:approx (x;q)infty}
    (x;q)_\infty&=\exp\qty(-\frac{{\rm Li}_2(x)}{\lambda}+\frac{1}{2}\log(1-x)+\mathcal{O}(\lambda))~,\\
    (q;q)_\infty&=\sqrt{\frac{2\pi}{\lambda}}\exp\qty(-\frac{\pi^2}{6\lambda}+\mathcal{O}(\lambda))~,\\
    &{\rm Li}_2(z)+{\rm Li}_2(z^{-1})=-\frac{\pi^2}{6}-\frac{1}{2}\qty(\log(-z))^2~.
\end{aligned}
\end{equation}
with ${\rm Li}_2(x)$ the dilogarithm function. The thermodynamic entropy and the microcanonical temperature in (\ref{eq:partition H0}) are:
\begin{equation}\label{eq:entropy temp}
    S(\theta)=-\frac{2(\frac{\pi}{2}-\theta)^2}{\lambda}~,\quad \beta(\theta)=\dv{S}{E}=2\frac{\theta-\pi/2}{J\sin\theta}~,
\end{equation}
respectively. {{Note that $\beta$ can be positive or negative since $\theta\in[0,\pi]$; which we comment further in footnote \ref{fnt:negative temp}.}} We can also introduce matter chords operators (also called cords), which have the form:
\begin{align}\label{eq:matter ops DSSYK}
\hat{\mathcal{O}}_\Delta&=\rmi^{\frac{p'}{2}}\sum_{I'}M_{I'}\psi_{I'}~,\quad \expval{M_{I}}=0~,\quad \expval{M_{I}J_{I'}}=0~,\quad \expval{M_{I}M_{I'}}=\begin{pmatrix}
        N\\
        p'
    \end{pmatrix}^{-1}\delta_{II'}~.
    \end{align}
    Here $\Delta\equiv p'/p$, $I'=i_1,\dots,i_{p'}$; and $M_{I'}$ are Gaussian random couplings that are independent of $J_{I}$, and we have chosen the normalization of $\expval{M_{I}M_{I'}}$ by requiring {$\expval{\tr(\hat{\mathcal{O}}_{\Delta}^2)}_{M}=1$, following the convention in \cite{Berkooz:2018jqr,Berkooz:2018qkz}.

One can then evaluate correlation functions in the auxiliary Hilbert space as \cite{Berkooz:2018jqr,Berkooz:2018qkz}
\begin{equation}\label{eq:correlator DSSYK}
    \begin{aligned}
G_\beta^{(\Delta)}(\tau)&=\frac{\bra{0}\rme^{-\qty(\beta-\tau)\hH}q^{\Delta\hat{n}}\rme^{-\tau\hH}\ket{0}}{Z(\beta)}\\
    &=\frac{1}{Z(\beta)}\prod_{j=1,~2}\int_{\theta_j=0}^{\theta_j=\pi}\rmd\mu(\theta_j)\rme^{-(\beta-\tau) E(\theta_1)-\tau E(\theta_2)}\bra{\theta_1}q^{\Delta\hat{n}}\ket{\theta_2}~,
    \end{aligned}
\end{equation}
where \cite{Berkooz:2018jqr,Goel:2023svz} 
\begin{equation}\label{eq:expand qDelta}
    \bra{\theta_i}q^{\Delta\hat{n}}\ket{\theta_j}=\frac{(q^{2\Delta};q)_\infty}{(q^{\Delta\pm\rmi\theta_i\pm\rmi\theta_j};q)_\infty}~.
\end{equation}
Using saddle point methods, one can evaluate the integral (\ref{eq:correlator DSSYK}) {{in the same way as, e.g. \cite{Goel:2023svz},
\begin{equation}\label{eq:two point semiclassical saddle}
    G_\beta^{(\Delta)}(\tau)\eqlambda\qty(\frac{\sin\theta}{\sin(\theta+\frac{\pi-2\theta}{\beta(\theta)}\tau)})^{2\Delta}~,
\end{equation}
where $\beta(\theta)$ denotes the function in \eqref{eq:entropy temp}.}}

\paragraph{On the chord space with particle insertions}\label{sapp:DSSYK one particle}
As originally discussed by \cite{Lin:2022rbf}, the Fock space of q-deformed harmonic oscillators (\ref{eq:Fock H0}) is modified under the addition of chord particles in the chord diagrams (e.g. Fig. \ref{fig:composite_op}). Once we introduce $m\in\mathbb{Z}_{\geq0}$ number of chord particles into the chord diagrams, the new auxiliary Hilbert space takes the form \cite{Lin:2022rbf}
\begin{equation}\label{eq:Fock space with matter}
    \mathcal{H}_{m}=\bigoplus_{n_0,n_1,\cdots,n_m=0}^\infty\mathbb{C}\ket{\tilde{\Delta},n_0,n_1,\cdots,n_m}~,
\end{equation}
where the states in $\mathcal{H}_m$ are represented with the same notation as (\ref{eq:states notation matter}). The inner product for the Hilbert space in (\ref{eq:Fock space with matter}) was worked out in \cite{Lin:2023trc}.
To describe this system as a q-deformed harmonic oscillator, one needs to introduce creation and annihilation operators (\ref{eq:a adagger DSSYK}) acting on the different chord number sectors in (\ref{eq:Fock space with matter}),
\begin{subequations}\label{eq:Fock Hm}
    \begin{align}\label{eq:Fock Hm 1}
    \hat{a}^\dagger_{i}\ket{\tilde{\Delta};n_0,\dots n_i,\dots, n_m}&=\ket{\tilde{\Delta};n_0,\dots, n_i+1,\dots n_m}~,\\\label{eq:Fock Hm 2}
    \hat{\alpha}_{i}\ket{\tilde{\Delta};n_0,\dots n_i,\dots, n_m}&=\ket{\tilde{\Delta};n_0,\dots, n_i-1,\dots n_m}~,
\end{align}
\end{subequations}
We also introduce the total chord number operator,
\begin{equation}\label{eq:total chord number basis}
    \hat{N}\ket{\tilde{\Delta},n_0,\cdots,n_m}=\sum_{i=0}^mn_i\ket{\tilde{\Delta},n_0,\cdots,n_m}~.
\end{equation}
The two-sided Hamiltonian acting on the Fock space with matter (\ref{eq:Fock space with matter}) takes the form  (\ref{eq:Hamiltonians LR}) \cite{Lin:2022rbf} 
\begin{align}\label{eq:two-sided new}
    &\hH_{L/R}=\frac{J}{\sqrt{\lambda}}\qty(\hat{a}_{L/R}+\hat{a}^\dagger_{L/R})~,\quad \text{where}\\
&\hat{a}^\dagger_L=\hat{a}^\dagger_0~,\quad \hat{a}_L=\sum_{i=0}^m\hat{\alpha}_i[\hat{n}_i]_qq^{\hat{n}_i^<}~,\quad\text{with}\quad\hat{n}_i^<=\sum_{j=0}^{i-1}\qty(\hat{n}_j+\Delta_{j+1})~,\label{eq:aLdagger,aL}\\
&\hat{a}^\dagger_R=\hat{a}^\dagger_m~,\quad\hat{a}_R= \sum_{i=0}^m\hat{\alpha}_i[\hat{n}_i]_qq^{\hat{n}_i^>}~,\quad\text{with}\quad\hat{n}_i^>=\sum_{j={i+1}}^{m}\qty(\hat{n}_j+\Delta_{j})~.\label{eq:aRdagger,aR}
\end{align}
This is discussed in more detail in Sec. \ref{ssec:many chords}.

For generality, we also discuss how the operators $\hat{\mathcal{O}}_\Delta$ act on the Fock space (\ref{eq:Fock space with matter}). Since $\hat{\mathcal{O}}_\Delta$ can be seen as a double-scaled auxiliary Hamiltonian, (\ref{eq:matter ops DSSYK}), one can decompose it into creation an annihilation operators acting on chord matter space, i.e. $\hat{\mathcal{O}}_\Delta=\hat{b}_\Delta+\hat{b}_\Delta^\dagger$ (in a similar way as (\ref{eq:two-sided new})), where \cite{Xu:2024hoc}
\begin{align}\label{eq:property operator}
&\hat{b}_{\Delta}^{\dagger}\ket{\tilde{\Delta};n_{0},\cdots,n_{m}}=\ket{\Delta,\tilde{\Delta};0,n_{0},\cdots,n_{m}}~.
\end{align}
Meanwhile, assuming that $\Delta_1=\dots=\Delta_m=\Delta$, we also have 
\begin{equation}\label{eq:property operator2}
    \begin{aligned}
        &\hat{b}_{\Delta}\ket{\Delta^{(m)};n_{0},\cdots,n_{m}}	\\
        &=\sum_{j=1}^{m}q^{(j-1)\Delta^2}q^{\Delta\sum_{l<j}n_{l}}\ket{\Delta^{(m-1)};n_{0},\cdots,n_{j-2},n_{j-1}+n_{j},\cdots,n_{m}}~.
    \end{aligned}
\end{equation}
Here $\Delta^{(m)}$ indicates $m$ times the same $\Delta$; $q^{\Delta^2}$ is the penalty factor for a crossing between two matter chords, and $q^\Delta$ for a crossing between Hamiltonian and matter chords. If instead we considered some $\Delta_{1\leq i\leq m}$ to be different from each other, then we only need to account for the chords with the same conformal weight as $\hat{\mathcal{O}}_\Delta$, so that they can be annihilated by $\hat{b}_\Delta$.

\subsection{Krylov complexity}\label{sapp:Krylov review}
Quantum complexity (see reviews in \cite{Chapman:2021jbh,Chen:2021lnq,Baiguera:2025dkc}) are among several measures quantum chaos\footnote{Other common measures include level spacing spectral statistics \cite{Dyson:1962es,Dyson:1962oir,wigner1993characteristic}, out-of-time-ordered correlators \cite{Rozenbaum:2016mmv}, the spectral form factor \cite{Cotler:2016fpe}. See also recent progress from algebraic methods in \cite{Camargo:2025zxr,Gesteau:2024rpt,Gesteau:2024rpt,Ouseph:2023juq}.} which is expected to be one of the distinguishing features of holographic systems \cite{Hayden:2007cs,Sekino:2008he,Shenker:2013pqa,Shenker:2013yza,Maldacena:2015waa}. Generically, it is expected that quantum complexity measures in holographic systems probe regions of the black hole interior that are not accessible to entanglement entropy \cite{Susskind:2014moa}. This has motivated several conjectures for geometrical observables dual to quantum complexity (called \emph{holographic complexity}); most commonly capturing the properties expected for circuit complexity \cite{Susskind:2014rva}. This has given rise to several holographic complexity proposals, which include the CV proposal \cite{Susskind:2014rva,Stanford:2014jda}, among many others \cite{Brown:2015bva, Brown:2015lvg,Couch:2016exn}. In fact, it has been realized that there is an infinite number of gravitational probes exhibiting the same universal behavior as circuit complexity \cite{Belin:2021bga, Belin:2022xmt}. The defining properties in holographic complexity include a late time linear growth (which is supposed to eventually saturate,\footnote{See recent discussion of this aspect in the DSSYK model by \cite{Balasubramanian:2024lqk}} and undergo a Poincare recurrence in finite dimensional systems \cite{Chapman:2021jbh,Susskind:2014rva}), as well as the so-called switchback effect \cite{Susskind:2014rva}, a decrease in the growth of complexity due to alternating thermal scale operator insertions, also called timefolds \cite{Susskind:2014rva}, which translate to shockwave pulses in an alternating order in the bulk \cite{Stanford:2014jda}. If the late time evolution decreases by an overall multiple of the scrambling time for the system, then, a notion of quantum complexity is said to experience a switchback effect.

The ample number of bulk proposals for holographic complexity \cite{Belin:2021bga,Belin:2022xmt,Myers:2024vve} is reminiscent of the ambiguity in defining circuit complexity. However, despite enormous progress, holographic complexity remains, arguably, not a well-developed entry in the holographic dictionary, with a few exemptions \cite{Erdmenger:2024xmj,Erdmenger:2022lov,Rabinovici:2023yex,Heller:2024ldz}. 

Arguably, the least ambiguous measure of quantum complexity is Krylov complexity. This is constructed by decomposing a state or operator in an ordered basis, called the Krylov basis, by sequential applications of a Hamiltonian or Liouvillian operator, and summing over all amplitudes resulting from this decomposition, with an appropriate weight. 

Both measures have seen several applications. For instance, as originally observed in \cite{Parker:2018yvk}, there is a strong connection between the late time exponential behavior of Krylov operator complexity with the chaos bound of OTOCs \cite{Maldacena:2015waa} at least in several examples (while there are some limitation, e.g. \cite{Bhattacharjee:2022vlt,Dymarsky:2021bjq,Camargo:2022rnt,Chapman:2024pdw}). There has also been a lot of recent interest in spread complexity (and higher moments of its generating function \cite{Camargo:2024rrj}), as it exhibits a peak value in chaotic quantum systems, and other non-integrable (except for saddle-point dominated scramblers \cite{Huh:2023jxt}) \cite{Balasubramanian:2022tpr,Erdmenger:2023wjg,Baggioli:2024wbz,Jeong:2024oao,Huh:2023jxt,Fu:2024fdm,Camargo:2024rrj,Camargo:2023eev,Balasubramanian:2024ghv,Alishahiha:2024vbf}. The evolution of spread complexity can be directly associated with the level spacing statistics \cite{Alishahiha:2024vbf}. Moreover, it has been found that the time derivative of spread complexity for CFT excited states can be sharply related with the proper momentum of a point particle in asymptotically AdS spacetimes \cite{Caputa:2024sux,Caputa:2025dep}. This provides evidence that Krylov complexity is indeed an entry in the holographic dictionary, at least in specific examples.

In the case of two-sided Hamiltonians one can also derive concrete relations between Krylov operator complexity and spread complexity of an excited state; as we develop in the main text (Sec. \ref{ssec:total chord as Krylov}). The reader is also referred to \cite{Sanchez-Garrido:2024pcy} for a rigorous procedure to map the spread complexity of the TFD state with an operator insertion $\hat{\mathcal{O}}$ into the corresponding operator Krylov complexity of $\hat{\mathcal{O}}$, and App. F.2 in \cite{Ambrosini:2024sre} for a discussion about this.

\paragraph{For states (spread complexity)}In the original work by \cite{Balasubramanian:2022tpr}, the authors showed that the basis $\qty{\ket{B_n}}$ minimizing $\sum_n c_n\abs{\bra{\varphi(t)}\ket{B_n}}^2$, with $\ket{\varphi(t)}$ being a general evolving state in the Schrödinger picture, and c$_n$ being arbitrary monotonically increasing real coefficients, is given by the Krylov basis, $\ket{K_n}$, defined through a Gram-Schmidt orthogonalization of $\qty{H^k\ket{\varphi(t=0)}}$. The formalism was later extended by \cite{Erdmenger:2023wjg}, who shows that one can threat this problem following the same Lanczos algorithm (which we introduce shortly) while allowing for complex time evolution
\begin{equation}\label{eq:phi Schrödinger}
\begin{aligned}
-\partial_\tau\ket{\varphi(\tau)}&=\hH\ket{\varphi(\tau)}~,\\
    	\ket{\varphi(\tau)}&=\rme^{-\tau\hH }\ket{\varphi(0)}~,\quad \tau\in\mathbb{C}~.
\end{aligned}
\end{equation}
The Lanczos algorithm, used to construct the Krylov basis, $\qty{\ket{K_n}}$, is then given by
\begin{equation}\label{eq:lanczos alg}
    \begin{aligned}
	&\ket{A_{n+1}}\equiv (\hH - a_n)\ket{K_n} - b_n \ket{K_{n-1}}~,\\
	&\text{If }b_n=0\text{ stop; otherwise } \ket{K_n} \equiv b_n^{-1}\ket{A_n}~,
\end{aligned}
\end{equation}
which is initialized by $\ket{K_0}\equiv\ket{\varphi(0)}$. The Lanczos coefficients are given by
\begin{equation}\label{eq:spread lanczos}
	a_{n} \equiv \bra{K_n} \hH \ket{K_n}, \qquad b_{n} \equiv (\braket{A_n})^{1/2}~.
\end{equation}
We can now express (\ref{eq:phi Schrödinger}) as
\begin{equation}\label{eq: phi (t) spread}
	\ket{\varphi(\tau)} = \sum^{\mathcal{K}}_{n=0} \varphi_n (\tau) \ket{K_n}~,
\end{equation}
where $\mathcal{K}$, the Krylov space dimension, is bounded by the total Hilbert space dimension. Using the Schrödinger equation (\ref{eq:phi Schrödinger}), we can see that the Hamiltonian is tridiagonal in the Krylov basis:
\begin{equation} \label{eq:Schro}
	-\partial_\tau \varphi_n(\tau) = a_n \varphi_n (\tau) + b_{n+1}\varphi_{n+1}(\tau) + b_n \varphi_{n-1}(\tau)~,
\end{equation}
and $\sum_n|\varphi_n(\tau)|^2=\bra{\varphi_0}\rme^{-\beta\hH}\ket{\varphi_0}$ is a conserved probability in real time $t:={\rm Im}(\tau)$; and we denote $\beta:=2{\rm Re}(\tau)$. The \emph{spread complexity} of evolving states in complex time is then defined as
\begin{equation} \label{eq:spread Complexity}
	\mathcal{C}(t)\equiv\eval{\frac{\bra{\varphi(\tau)}\hat{\mathcal{C}}\ket{\varphi(\tau)}}{\bra{\varphi(\tau)}\ket{\varphi(\tau)}}}_{\tau=\frac{\beta}{2}+\rmi t} =\eval{\frac{\sum_n n |\varphi_n(\tau)|^2}{\sum_n\abs{\varphi_n(\tau)}^2}}_{\tau=\frac{\beta}{2}+\rmi t} ~,
\end{equation}
where the Krylov complexity operator is identified with
\begin{equation}\label{eq:operator spread complexity}
    \hat{\mathcal{C}}=\sum_nn\ket{K_n}\bra{K_n}~.
\end{equation}
(\ref{eq:spread Complexity}) provides the average position of a particle propagating through a one-dimensional ordered lattice generated by the Krylov basis, where each step requires more applications of the Hamiltonian $\hH$, which heuristically provides a measure of quantum chaos.\footnote{There have been more formal arguments that spread complexity can be used to distinguish integrable from chaotic quantum systems, including \cite{Baggioli:2024wbz,Alishahiha:2024vbf}.}

In particular, in the DSSYK model without matter chords, one can see from the symmetric Hamiltonian (\ref{eq:qHermite recurrence}) that the Krylov basis for a given seed state $\ket{K_0}=\ket{0}$ is simply the $\ket{K_n}=\ket{n}$, and the Lanczos coefficients are
\begin{equation}\label{eq:Lanczos coeff spread H0}
a_n=0~,\quad b_n=\frac{J}{\sqrt{\lambda}}\sqrt{[n]_q}~.    
\end{equation}
The spread complexity for $\rme^{-\hH\tau}\ket{0}$ in the $\lambda\rightarrow0$ limit is then (\ref{eq:womrhole semiclassical answer many}) with $t_L=t_R=t/2$ and $\sum_i\Delta_i=0$ (corresponding to no particle chord since the conformal dimensions for matter chords are strictly non-negative, as seen from (\ref{eq:matter ops DSSYK})).

\paragraph{For operators}
We can to express the evolution of $\hat{O}$ in the Heisenberg picture in terms of a complete basis $\qty{\ket{x_n}}$ in operator space as
\begin{equation}\label{eq:Op as state Krylov}
	\begin{aligned}
		|\hat{O})&\equiv\sum_{m,\,n}O_{nm}|x_m,~x_n)~,
	\end{aligned}
\end{equation}
with $O_{nm}\equiv\bra{x_m}\hat{O}\ket{x_n}$, and $|x_m,~x_n)=\ket{x_m}\otimes\ket{x_n}$. To define the Hilbert space associated with these states, one needs to specify the inner product. There are different choices when considering finite temperature ensembles \cite{Parker:2018yvk,Barbon:2019wsy,Anegawa:2024yia}, see \cite{Sanchez-Garrido:2024pcy} for a comprehensive review. For instance, in the context of thermal correlation functions in boundary theories, it is often convenient to use regulated thermal inner product, where  Boltzmann factors are inserted between every matter operator $\hat{O}$. The so-called Wightman inner product, introduced by \cite{Parker:2018yvk} for Krylov operator complexity, is then given
\begin{equation}\label{eq:inner prod}
	(\hat{X}|\hat{Y})=\frac{1}{Z(\beta)}\Tr(\rme^{-\frac{\beta}{2}\hH}\hat{X}^\dagger\rme^{-\frac{\beta}{2}\hH} \hat{Y})~.
\end{equation}
The Heisenberg equation can be then expressed as
\begin{align}\label{eq: Heisenberg time evol}
	\partial_t|\hat{O}(t))&=\rmi\hat{\mathcal{L}}|\hat{O}(t))~,
\end{align}
with $\hat{\mathcal{L}}$ the Liouvillian super-operator, defined as
\begin{equation}
	\hat{\mathcal{L}}=\qty[\hH,~\cdot~],\quad \hat{O}(t)=\rme^{\hat{\rmi\mathcal{L}}t}\hat{O}=\rme^{\rmi \hH t}\hat{O}\rme^{-\rmi \hH t}~.
\end{equation}
Note that when $\hH=\hH^\dagger$ (which we assume throughout our discussion), then the inner product (\ref{eq:inner prod}) is equivalent to evolving the operator in complex time,
\begin{equation}
    (\hat{X}(t_x)|\hat{Y}(t_y))=Z(\beta)^{-1}\eval{\Tr\qty(\rme^{-\tau_x\hH}\hat{X}^\dagger\rme^{-\tau^*_x\hH}\rme^{-\tau_y^*\hH}\hat{Y}\rme^{-\tau_y\hH})}_{\tau_{x/y}=\frac{\beta}{4}+\rmi t_{x/y}}~.
\end{equation}
We can therefore associate this choice of inner product with complex time evolution, similar to the Schrödinger picture in (\ref{eq:phi Schrödinger}). 
We would like to use the so-called Krylov basis, $\qty{|\hat{K}_n)}$, defined through the recursion relation
\begin{equation}\label{eq:Lanczos alg Operator comp}
    |\hat{A}_{n+1})=\mathcal{L}|\hat{K}_n)-b_{n}|\hat{K}_{n-1})~,\quad (\hat{K}_m|\hat{K}_n)=\delta_{mn}~,
\end{equation}
with $|\hat{K}_0)=|\hat{O}(t=0))$. In writing (\ref{eq:Lanczos alg Operator comp}), we have assumed that the system is Hermitian; otherwise, we would need to determine the Lanczos coefficients $a_n=(\hat{K}_n|\mathcal{L}|\hat{K}_n)$ \cite{Parker:2018yvk}. We may then express the solution $|\hat{O}(t))$ of (\ref{eq: Heisenberg time evol}) as
\begin{equation}\label{eq:amplitudes}
		|\hat{O} (t))=\sum_{n=0}^{\mathcal{K}-1} \rmi^n \tilde{\varphi}_n(t)~ |\hat{K}_n)~.
\end{equation}
where $\mathcal{K}$ refers to the dimension of the Krylov space (i.e. the operator subspace that is spanned by $\qty{|\hat{K}_n)}$), and
\begin{equation}
\tilde{\varphi}_n(\tau)=(\hat{K}_n|\rme^{\rmi\hat{\mathcal{L}}t}|\hat{K}_0)=\frac{\Tr(\hat{K}_n\rme^{-\tau^*\hH}\hat{K}_0\rme^{-\tau\hH})}{Z(\beta)}~,
\end{equation}
where $\tau=\beta/2+\rmi t$. One can then define the Krylov complexity of $\hat{O}(t)$ \cite{Parker:2018yvk}\footnote{Note that $(\hat{O}(t)|\hat{O}(t))=(\hat{O}(0)|\hat{O}(0))$.}
\begin{equation}\label{eq:Krylov complexity}
	\mathcal{C}(t)=\frac{(\hat{O}(t)|\hat{\mathcal{C}}|\hat{O}(t))}{(\hat{O}(t)|\hat{O}(t))}=\eval{\frac{\sum_{n}n\abs{\tilde{\varphi}_n(\tau)}^2}{\sum_n\abs{\tilde{\varphi}_n(\tau)}^2}}_{\tau=\beta/2+\rmi t}~,
\end{equation}
where we have introduced the Krylov complexity operator 
\begin{equation}
    \hat{\mathcal{C}}\equiv\sum_nn|\hat{K}_n)(\hat{K}_n|~,
\end{equation}
in a similar way as (\ref{eq:operator spread complexity}). Krylov complexity for operators serves as a measure of the operator growth due to Hamiltonian evolution, in the sense of representing the mean wavepacket amplitude width in the Krylov space. It has found several applications. For instance, one can gain knowledge of the OTOC Lyapunov exponent of a chaotic quantum system without having to compute it directly, as conjectured in \cite{Parker:2018yvk} through several non-trivial examples.\footnote{There have been several counterexamples in both quantum mechanical systems \cite{Chapman:2024pdw,Huh:2023jxt} and QFTs \cite{Avdoshkin:2022xuw,Camargo:2022rnt}.}

\paragraph{Two-sided Hamiltonians} Let us now study how to generalize the arguments above when the total Hamiltonian of the system is $\hH=\hH_L\otimes\mathbb{1}+\mathbb{1}\otimes\hH_R$, one may represent the Liovillian acting on an operator $\hat{\mathcal{O}}_\Delta\otimes\mathbb{1}$ perturbing the TFD state, as
\begin{equation}
    \hat{\mathcal{L}}~\hat{\mathcal{O}}_\Delta\ket{\rm TFD}=[\hH_L,\hat{\mathcal{O}}_\Delta]\ket{\rm TFD}=(\hH_L-\hH_R)\hat{\mathcal{O}}_\Delta\ket{\rm TFD}~,
\end{equation}
where we used the property: $(\hH_L-\hH_R)\ket{\rm TFD}=0$. Thus, we may represent $\hat{\mathcal{L}}=\hH_L-\hH_R$ when analyzing the double-scaled algebra of operators, as we do in the main text.

Using the eigenstate basis of the TFD state to perform the Choi–Jamiołkowski isomorphism in (\ref{eq:Op as state Krylov}), it can be seen that we threat Krylov operator complexity of $|\hat{\mathcal{O}}_\Delta(t))$ in a thermal ensemble in the same formalism for as the spread complexity (\ref{eq:spread Complexity}) of the perturbed TFD state in a thermal state
\begin{equation}\label{eq:thermal state Krylov}
\begin{aligned}
    \rme^{-\frac{\beta_L}{2}\hH_L-\frac{\beta_R}{2}\hH_R}|\hat{\mathcal{O}}_\Delta(t))&=\rme^{-\frac{\beta_L}{2}\hH_L-\frac{\beta_R}{2}\hH_R+\rmi\hat{\mathcal{L}}t}\hat{\mathcal{O}}_\Delta\ket{\rm TFD}\\
    &=\rme^{-\frac{\beta_L}{2}\hH_L-\frac{\beta_R}{2}\hH_R+\rmi(\hH_R-\hH_L)t}\hat{\mathcal{O}}_\Delta\ket{\rm TFD}~.
\end{aligned}
\end{equation}
Meanwhile, the Wightman inner product (\ref{eq:inner prod}) is not uniquely generalizable in two-sided Hamiltonian systems for a thermal ensemble where $\beta_L\neq \beta_R$. In principle, one might use other equally valid definitions as well, which are expected to provide similar dynamical behavior (e.g. \cite{Parker:2018yvk,Sanchez-Garrido:2024pcy}) at least in the one-sided Hamiltonian systems. For these reasons, we pick a particular choice that leads to simplifications in the Lanczos algorithm (\ref{eq:Lanczos algorithm eta}) for technical convenience in our discussion of Sec. \ref{ssec:many chords}.

\section{Details about the isometric map and proof of (\ref{eq:initial velocity})}\label{app:isometric map}
In order to do the evaluations in the following sections, we study the isometric map between Hilbert spaces $\hat{\mathcal{F}}:~\mathcal{H}_{1}\rightarrow\mathcal{H}_{0}\otimes\mathcal{H}_{0}$. We first explain its definition; how to invoke it to evaluate partition functions or correlation function in the energy basis; and lastly, we prove the initial conditions for the expectation value of total chord number in the two-sided HH state (\ref{eq:initial velocity}).

The isometric map is defined \cite{Xu:2024hoc,Xu:2024gfm,Okuyama:2024yya,Okuyama:2024gsn}
\begin{equation}\label{eq:isometric map}
    \hat{\mathcal{F}}\ket{\Delta;n_l,n_R}=(q^\Delta\hat{a}_L\hat{a}_R;q)_\infty|m,n)~,
\end{equation}
 where $\hat{a}_L=\hat{a}\otimes\mathbb{1}$, and $\hat{a}_R=\mathbb{1}\otimes\hat{a}$ (with $\hat{a}$ being the endomorphism in $\mathcal{H}_0$ (\ref{eq:a adagger DSSYK})), and 
 \begin{equation}\label{eq:prod 1}
 |m,n):=\ket{m}\otimes\ket{n}~.    
 \end{equation}
 Using this map, we can simplify functionals of the Hamiltonians of the form
 \begin{equation}\label{eq:evaluation 0 isometric}
     (\theta_1,\theta_2|h(\hH_L.\hH_R)\ket{\Delta;0,0}=h(E(\theta_1),E(\theta_2))~,
 \end{equation}
 where $h(\cdot,\cdot)$ is an arbitrary bi-variable function {{and
 \begin{equation}\label{eq:prod 2}
     {{|\theta_1,\theta_2)}}:=\ket{\theta_L}\otimes\ket{\theta_L}~.
 \end{equation}}}
Furthermore, we can evaluate the inner product of generic states in $\mathcal{H}_1$ in energy basis \cite{Xu:2024gfm}
\begin{equation}\label{eq:inner product energy}
    \bra{\Psi_1}\ket{\Psi_2}=\int_{\theta_{1}=0}^{\theta_{1}=\pi}\int_{\theta_{2}=0}^{\theta_{2}=\pi}\rmd\mu(\theta_1)\rmd\mu(\theta_2)\bra{\theta_1}q^{\Delta\hat{n}}\ket{\theta_2}\bra{\Psi_1}\hF^\dagger|\theta_1,\theta_2)(\theta_1,\theta_2|\hF\ket{\Psi_2}~.
\end{equation}
To exemplify the use of the isometric map, we prove the initial condition (\ref{eq:initial velocity}). We begin with
\begin{align}\label{eq:next step eenergy}
\bra{\Delta_{\rm S};0,0}\rme^{-\frac{\beta_L}{2}\hH_L-\frac{\beta_R}{2}\hH_R}&[\hH_{L/R},~\hat{N}]\rme^{-\frac{\beta_L}{2}\hH_L-\frac{\beta_R}{2}\hH_R}\ket{\Delta_{\rm S};0,0}\\
=\int\rmd\mu(\theta_1)\rmd\mu(\theta_2)&\bra{\theta_1}q^{\Delta\hat{n}}\ket{\theta_2}\rme^{-\frac{\beta_L}{2}E_1-\frac{\beta_R}{2}E_2)}E_{1/2}\cdot\\
&\cdot\qty[(\theta_1,\theta_2|\hF\hat{N}\rme^{-\frac{\beta_L}{2}\hH_L-\frac{\beta_R}{2}\hH_R}\ket{\Delta_{\rm S};0,0}+{\rm h.c.}]~,\nonumber
\end{align}
where h.c. represents Hermitian conjugate, and we introduced a shorthand $E_i=E(\theta_i)$, with $E(\theta)$ is (\ref{eq:energy spectrum}). We have also used (\ref{eq:evaluation 0 isometric}) to express
\begin{equation}
    (\theta_1,\theta_2|\hF\hH_{L/R}\rme^{-\frac{\beta_L}{2}\hH_L-\frac{\beta_R}{2}\hH_R}\ket{\Delta;0,0}=E_{1/2}\rme^{-\frac{\beta_L}{2}E_1-\frac{\beta_R}{2}E_2}~.
\end{equation}
We can insert the identify element as
\begin{equation}
   \hF^{-1}\hF=\mathbb{1} ~,\quad \int\rmd\mu(\theta_3)\rmd\mu(\theta_4)|\theta_{3},\theta_{4})(\theta_{3},\theta_{4}|=\mathbb{1}~,
\end{equation}
into (\ref{eq:next step eenergy}) so that it can be expressed as
\begin{equation}\label{eq:initial vel results 1}
    \begin{aligned}
        \bra{\Delta_{\rm S};0,0}&{{(\mathcal{F}}^{\dagger})^{-1}}\rme^{-\frac{\beta_L}{2}\hH_L-\frac{\beta_R}{2}\hH_R}[\hH_{L/R},~\hat{N}]\rme^{-\frac{\beta_L}{2}\hH_L-\frac{\beta_R}{2}\hH_R}\hF^\dagger\ket{\Delta_{\rm S};0,0}\\
        =\int&\qty(\prod_{i=1}^4\rmd\mu(\theta_i))\bra{\theta_1}q^{\Delta\hat{n}}\ket{\theta_2}\rme^{-\frac{\beta_L}{2}(E_1+E_3)-\frac{\beta_R}{2}(E_2+E_4)}E_{1/2}\cdot\\
        &\cdot\qty[(\theta_1,\theta_2|\hF\hat{N}\hF^{-1}|\theta_3,\theta_4)-(\theta_3,\theta_4|(\hF^{-1})^\dagger\hat{N}\hF^\dagger|\theta_1,\theta_2)]~.
    \end{aligned}
\end{equation}
Moreover, given that the eigenvalues of the total chord number are just integers, we can express the total chord number in  a diagonal representation through the chord number basis $\qty{\ket{K_n}}$ as
\begin{equation}
\hat{N}=\sum_{n=0}^\infty n\ket{K_n}\bra{K_n}~,
\end{equation}
which means that
\begin{equation}\label{eq:interchange theta}
   (\theta_3,\theta_4|(\hF^{-1})^\dagger\hat{N}\hF^\dagger|\theta_1,\theta_2)=(\theta_1,\theta_2|\hF\hat{N}\hF^{-1}|\theta_3,\theta_4) ~.
\end{equation}
Thus, by combining (\ref{eq:initial vel results 1}) and (\ref{eq:interchange theta}), (\ref{eq:initial velocity}) follows.

\section{Partition function in the two-sided Hartle-Hawking state}\label{app:partition1p}
In this appendix we evaluate the semiclassical partition function and related thermodynamical quantities (entropy, temperature, and heat capacity) by taking expectation values with respect to the two-sided HH state (\ref{eq:HH state tL tR}), illustrated in Fig. \ref{fig:composite_op}.  From the partition function, we derive the thermodynamic entropy, temperature, and heat capacity for both light (Sec. \ref{sapp:light partition function}), and heavy (Sec. \ref{sapp:heavy partition function}) operators. This allows us to determine the conditions for thermal stability of the saddle point solutions.

We begin deducing the form of the partition in $\mathcal{H}_m$:
\begin{equation}\label{eq:Z from H1}
\begin{aligned}
    Z_{\tilde{\Delta}}(\beta_L,\beta_R)\equiv&\bra{\tilde{\Delta};0,\dots0}\rme^{-\beta_L\hH_L-\beta_R\hH_R}\ket{\tilde{\Delta};0,\dots0}\\
    =&\bra{{\Delta_{\rm S}};0,0}\rme^{-\beta_L\hH_L-\beta_R\hH_R}\ket{{\Delta_{\rm S}};0,0}
\end{aligned}
\end{equation}
where we used the $\mathcal{H}_1$ irrep. in (\ref{eq:LinStanfordIdentity}) with $\Delta_{\rm S}=\sum_{i=0}^m\Delta_i$, and $\tilde{\Delta}=\qty{\Delta_1,\hdots,\Delta_m}$.

Using the isometric map (\ref{eq:isometric map}), we can evaluate the partition function (\ref{eq:Z from H1}) using the inner product (\ref{eq:inner product energy}) in energy basis (with the property (\ref{eq:evaluation 0 isometric})) as
\begin{equation}\label{eq:partition function Delta}
    \begin{aligned}
        Z_{\tilde{\Delta}}(\beta_L,\beta_R)&=
\int_{\theta_{L/R}=0}^{\theta_{L/R}=\pi}\rmd\mu(\theta_L)\rmd\mu(\theta_R)\frac{(q^{2{\Delta_{\rm S}}};q)_\infty}{(q^{{\Delta_{\rm S}}}\rme^{\rmi(\pm\theta_L\pm\theta_R)};q)_\infty}\rme^{-\qty(\beta_L E(\theta_L)+\beta_R E(\theta_R))}~.
    \end{aligned}
\end{equation}
{{Note from \eqref{eq:expand qDelta}}} that (\ref{eq:partition function Delta}) can be expressed as, {{
\begin{equation}\label{eq:explicit partition function}
   \begin{aligned}
Z_{\tilde{\Delta}}(\beta_L,\beta_R)&=\int_{\theta_{L/R}=0}^{\theta_{L/R}=\pi}\rmd\mu(\theta_L)\rmd\mu(\theta_R)\bra{\theta_L}q^{\Delta_{\rm S}\hat{n}}\ket{\theta_R}\rme^{-\qty(\beta_L E(\theta_L)+\beta_R E(\theta_R))}~,
   \end{aligned}
\end{equation}
where we can insert $\bra{0}\ket{\theta_L}=\bra{\theta_R}\ket{0}=1$ and use the identity (\ref{eq:identity theta}) to see that
\begin{equation}\label{eq:new partition result}
    Z_{\tilde{\Delta}}(\beta_L,\beta_R)=\bra{0}\rme^{-\beta_L\hH}q^{\Delta_{\rm S}\hat{n}}\rme^{-\beta_R\hH}\ket{0}~,
\end{equation}
where $\hH$ corresponds to the auxiliary chord Hamiltonian acting in $\mathcal{H}_0$, \eqref{eq:Transfer matrix}. Thus, \eqref{eq:new partition result} is manifestly equivalent to the thermal two-point function in $\mathcal{H}_0$ \eqref{eq:correlator DSSYK} with the replacement $(\beta-\tau)\rightarrow\beta_L$ and $\tau\rightarrow\beta_R$, and the corresponding normalization given by the thermal partition function in $\mathcal{H}_0$ \eqref{eq:partition function zero particle chord} where $\beta\rightarrow\beta_L+\beta_R$, i.e.}}
\begin{equation}\label{eq:isometric map partition function}
    Z_{\tilde{\Delta}}(\beta_L,\beta_R)=Z(\beta_L+\beta_R)G_{\beta_L+\beta_R}^{(\Delta_{\rm S})}({\beta_R})~.
\end{equation}
To evaluate the partition function, {{of the form \eqref{eq:Z from H1}, or \eqref{eq:isometric map partition function}}} we distinguish between two cases.

\subsection{Light particles}\label{sapp:light partition function}
Let us examine the semiclassical partition function \eqref{eq:isometric map partition function} {{when $\lambda\rightarrow0$ and $\Delta_{\rm S}\sim\mathcal{O}(1)$}}. Since we identified that the partition function reduces to the two-point correlation function for light operators in $\mathcal{H}_0$, we can use {{the known result}} \cite{Goel:2023svz,Blommaert:2024ymv} in (\ref{eq:two point semiclassical saddle}){{, which simplifies to}}
\begin{equation}\label{eq:Partition function light fields final}
\begin{aligned}
Z_{\tilde{\Delta}}(\beta_L,\beta_R)\eqlambda ~\qty({\sin\theta})^{2\Delta_{\rm S}}Z(\beta(\theta))~.
\end{aligned}
\end{equation}
{{Here we have used (\ref{eq:two point semiclassical saddle}) where the saddle point evaluation with $\Delta_{\rm S}\sim\mathcal{O}(1)$ fixes $\beta_L=\beta_R=\beta(\theta)/2$, where $\beta(\theta)$ appears in (\ref{eq:entropy temp}), i.e. the operator does not take the system our of thermal equilibrium (as we also find below \eqref{eq:almost done correlator2})}}.

Moreover, the first order quantum correction to the on-shell partition function {{(\ref{eq:explicit partition function})}}
\begin{equation}
    \ln Z_{\tilde{\Delta}}(\beta_L,\beta_R)=S(\theta_L)+S(\theta_R)+\bra{\theta_R}q^{\Delta_{\rm S}\hat{N}}\ket{\theta_L}-\beta_LE_L-\beta_RE_L~,
\end{equation}
follow immediately from (3.9) in \cite{Goel:2023svz} by replacing $\tau\rightarrow\beta_L$ and $\beta-\tau\rightarrow\beta_R$, and $\tilde{\lambda}\rightarrow\Delta_{\rm S}\lambda$.

\subsection{Heavy particles}\label{sapp:heavy partition function}
Consider now \emph{heavy operators}.
We evaluate the thermodynamical properties of the DSSYK model while considering thermal equilibrium in the individual left and right sectors. This means that interaction terms (i.e. $\bra{\theta_R}q^{\Delta\hat{n}}\ket{\theta_L}$ in (\ref{eq:partition function Delta})) between the chord sectors do not contribute to the density of states of the individual sectors. We can then express (\ref{eq:partition function Delta}) in the semiclassical limit as\footnote{\label{fnt:nonperturbative Okuyama}
Recently it was realized in \cite{Okuyama:2025fhi} that the disk partition function of the DSSYK model at low temperature can be expressed in terms of JT gravity with non-perturbative corrections to the saddle point approximation. One can perform a similar analysis for the partition function in the two-sided HH state (\ref{eq:partition one particle semiclassical}) which helps us to understand thermodynamics of the DSSYK model with matter operator insertions and its the bulk theory in Sec. \ref{sec:bulk}.}
\begin{equation}\label{eq:partition one particle semiclassical}
\begin{aligned}
Z_{\tilde{\Delta}}(\beta_L,\beta_R)=&(q^{2\Delta_{\rm S}};q)_\infty\\ &\int\rmd E_L\rmd E_R\rme^{S( \theta_L)+S(\theta_R)+\frac{1}{\lambda}{\rm Li}_2\qty(q^{\Delta_{\rm S}}\rme^{\pm \rmi\theta_L\pm \rmi\theta_R})}\rme^{-\qty(\beta_LE(\theta_L)+\beta_RE(\theta_R))}~,
\end{aligned}
\end{equation}
where we introduced the notation ${\rm Li_2}(\rme^{\pm x\pm y})={\rm Li_2}(\rme^{x+y})+{\rm Li_2}(\rme^{x-y})+{\rm Li_2}(\rme^{y-x})+{\rm Li_2}(\rme^{-(x+y)})$, and we expanded the exact density of states as \cite{Berkooz:2018qkz, Berkooz:2024lgq,Verlinde:2024znh}
\begin{equation}
    \label{eq:exact DOS}(q,\rme^{\pm2\rmi\theta};q)_\infty=2q^{-1/8}{\sin\left(\theta\right)}\sqrt{\frac{2\pi}{\lambda}}\sum_{n=-\infty}^{\infty}\left(-1\right)^{n}\rme^{-\frac{2}{\lambda}\left(\theta+\pi\left(n-\frac{1}{2}\right)\right)^{2}}\,,
\end{equation}
in (\ref{eq:saddles 1 particle S}, \ref{eq:saddles 1 particle beta}). The saddle point solutions of the partition function lead to the following thermodynamic relations
\begin{align}\label{eq:saddles 1 particle S}
    &S(\theta_{L/R})=-\frac{2}{\lambda}\qty(\theta_{L/R}+n\pi-\frac{\pi}{2})^2~,\quad n\in\mathbb{Z}~,\\\label{eq:saddles 1 particle beta}
    &\beta_{L/R}=\pdv{S(\theta_{L/R})}{E_{L/R}}=\frac{2\theta_{L/R}+(2n-1)\pi}{J\sin\theta_{L/R}}~.
\end{align}
In the analogous case of $\mathcal{H}_0$ it has been argued by \cite{Blommaert:2024whf} and \cite{Berkooz:2024ifu} that the $n\geq1$ saddles are associated with conical deficits, and they are all required to reproduce the exact DSSYK partition function at the disk level. Our observation is that the additional saddle points are thermodynamically unstable and that they also appear in $\mathcal{H}_{m\geq0}$, as we show below.

The corresponding heat capacity at fixed $\Delta_{\rm S}$ is given by:
\begin{equation}\label{eq:heat capacity Delta}
    C_{\Delta_{\rm S}}(\theta_{L/R})=-\beta_{L/R}^2\pdv{E_{L/R}}{\beta_{L/R}}\eqlambda\frac{J^2}{\lambda}\frac{\beta_{L/R}^2\sin^2\theta_{L/R}}{1+\qty(\qty(\frac{1}{2}-n)\pi-\theta_{L/R})\cot\theta_{L/R}}~.
\end{equation}
Again, this observable does not explicitly depend on the parameter $\Delta_{\rm S}$, which is contained in $\theta_{L/R}$ though (\ref{eq:ell cond thetaLR}). Moreover, one can notice from the denominator in (\ref{eq:heat capacity Delta}) that the heat capacity remains positive for $n=0$ in the range $\theta_{L/R}\in[0,~\pi]$\footnote{{{This is more subtle when one has a topological term in the dilaton gravity action \eqref{eq:total sdg matter}; see \cite{Blommaert:2025avl} for comments about this.}}} (which is unchanged by an operator $\hat{\mathcal{O}}_\Delta$ \cite{Xu:2024hoc}). However, the denominator becomes negative for $n\geq1$ when $\theta_{L/R}\in[0,~\pi]$. This indicates that the additional saddle points are thermodynamically unstable. Yet, since the would-be-saddle points with $n\neq 0$ add to the exact density of states (\ref{eq:exact DOS}); they should be seen as corrections to the saddle point solution for $n=0$. This is in agreement with the bulk analysis by \cite{Blommaert:2024whf}, where the authors explain that one should gauge the shift symmetries in the putative bulk dual description of the DSSYK model, which amounts to having $\theta\in[0,\pi]$. It is interesting to note that the addition of matter is not important in this argument, as it does not modify the saddle point thermodynamics (other than having to consider left/right chord sectors).

\section{Thermal two-sided two-point correlation function from chord diagrams}\label{app:twopoint oneparticle}
In this section, we evaluate the thermal two-point correlation function for the two-sided HH state (\ref{eq:HH state tL tR}). This corresponds to a crossed four-point function when we map it with the isometric map \cite{Xu:2024hoc} to $\mathcal{H}^{L}_0\otimes\mathcal{H}_0^{R}$. The relevant correlation function is illustrated in Fig. \ref{fig:OTOC_chord_diagram}.
Moreover, this observable corresponds to the Krylov complexity moment generating function in the semiclassical limit (i.e. $\lambda\rightarrow0$). The goal of the computation is to provide a consistency check of our answer from phase space methods in (\ref{eq:generating function explicit DSSYK 1 particle}). So, while the classical phase space answer is valid for heavy or light operators, the explicit ansatz we use to evaluate the correlation function assumes only light operators.

First, we derive the general form of this correlation function $\forall q\in[0,1)$, valid when the two-sided HH state (\ref{eq:HH state tL tR}) contains heavy particle (i.e. when $q^{\Delta_{\rm S}}$ is held fixed as $\lambda\rightarrow0$) operators acting on the $\ket{0}$ state. Later, we will evaluate the $\lambda\rightarrow0$ limit of the correlation function while considering light fields $\Delta\sim\mathcal{O}(1)$ to compare with the answer obtained from classical phase space methods in (\ref{eq:generating function explicit DSSYK 1 particle}). The correlation function we calculate in this section agrees with the one we derived for two light and two heavy fields (\ref{eq:generating function explicit DSSYK 1 particle}) in the limit where all particles are light. We hope to report progress on a more general ansatz that works for heavy-heavy-light-light correlation functions in the future. We also comment about the next steps to perform the evaluation with heavy operators.

\subsection{Setting up the correlation function}\label{sapp:setting amplitude}
We start from the thermal two-point function (\ref{eq:generating function 1 particle}), ${G}^{(\Delta_w)}_{\tilde{\Delta}}({\tau_L,\tau_R})=\expval{q^{\Delta_w\hat{N}}}/Z_{\tilde{\Delta}}(\beta_L,\beta_R)$, where
\begin{equation}\label{eq:numerator two point correlator}
\begin{aligned}
    \bra{\Psi_{\tilde{\Delta}}(\tau_L,\tau_R)}&q^{\Delta_w\hat{N}}\ket{{\Psi_{\tilde{\Delta}}(\tau_L,\tau_R)}}=\\
    &\bra{0}\rme^{-\tau_4\hH_R}\hat{\mathcal{O}}_{\Delta_1}\dots\hat{\mathcal{O}}_{\Delta_m}\rme^{-\tau_3\hH_L}q^{\Delta_w\hat{N}}\rme^{-\tau_1\hH_L}\hat{\mathcal{O}}_{\Delta_m}\dots\hat{\mathcal{O}}_{\Delta_1}\rme^{-\tau_2\hH_R}\ket{0}
\end{aligned}
\end{equation}
\begin{align}
Z_{\tilde{\Delta}}(\beta_L,\beta_R)&=\bra{\tilde{\Delta};0,\dots,0}\rme^{-\beta_L\hH_L-\beta_R\hH_R}\ket{\tilde{\Delta};0,\dots,0}~.
\end{align}
and
\begin{equation}\label{eq:tau choices}
    \tau_1=\frac{\beta_L}{2}-\rmi t_L~,\quad \tau_2=\frac{\beta_R}{2}-\rmi t_R~,\quad\tau_3=\frac{\beta_L}{2}+\rmi t_L~,\quad \tau_4=\frac{\beta_R}{2}+\rmi t_R~,
\end{equation}
As usual, we express the two-sided HH state (\ref{eq:HH state tL tR}) in terms of a $\mathcal{H}_1$ irrep (\ref{eq:important product to sum weights}),
\begin{equation}
    \ket{\Psi_{\tilde{\Delta}}(\tau_L,\tau_R)}=\rme^{-\beta_L\hH_L-\beta_R\hH_R}\ket{\Delta_{\rm S};0,0}~,
\end{equation}
with $\Delta_{\rm S}=\sum_{i=1}^m\Delta_i$. Using the isometric map (\ref{eq:isometric map}) we can now express (\ref{eq:numerator two point correlator}) (see also (B.19) and (B.22) in \cite{Xu:2024gfm})
\begin{equation}\label{eq:set up GbetaLbetaR}
    \begin{aligned}
        \bra{\Psi_{\tilde{\Delta}}(\tau_L,\tau_R)}q^{\Delta_w\hat{N}}\ket{{\Psi_{\tilde{\Delta}}(\tau_L,\tau_R)}}&=\int\qty(\prod_{i=1}^4\rmd\mu(\theta_i)\rme^{-\tau_iE_i})\bra{\theta_1}q^{\Delta_{\rm S}}\ket{\theta_2} \cdot\\
        \cdot&(\theta_1,\theta_2|q^{\Delta_w(\hat{n}_L+\hat{n}_R)}\sum_{l=0}^\infty \frac{(q^{2\Delta_w};q)_l}{(q;q)_l}q^{{\Delta_{\rm S}} l}\hat{a}_L^l\hat{a}_R^l|\theta_3,\theta_4)~,
    \end{aligned}
\end{equation}
where $\hat{n}_L=\hat{n}\otimes\mathbb{1}$, and $\hat{n}_R=\mathbb{1}\otimes\hat{n}$  with $\hat{n}$ in (\ref{eq: Oscillators 1}).

Moreover, we can perform the sum above from (A.7) in \cite{Okuyama:2024yya} (also in \cite{szablowski2013q}):
\begin{equation}\label{eq:Fourier kernel Al Salam Chihara}
    \begin{aligned}
        &(\theta_1,\theta_2|q^{\Delta_w(\hat{n}_L+\hat{n}_R)}\sum_{l=0}^\infty \frac{(q^{2\Delta_w};q)_l}{(q;q)_l}q^{{\Delta_{\rm S}} l}\hat{a}_L^l\hat{a}_R^l|\theta_3,\theta_4)=\\
    &\bra{\theta_1}q^{\Delta_w\hat{n}_L}\ket{\theta_3}\bra{\theta_2}q^{\Delta_w\hat{n}_R}\ket{\theta_4}\sum_{l=0}^\infty q^{{\Delta_{\rm S}}l}\frac{Q_l(\cos\theta_1|q^{\Delta_w}\rme^{\pm\rmi\theta_3};q)Q_l(\cos\theta_2|q^{\Delta_w}\rme^{\pm\rmi\theta_4};q)}{(q,q^{2\Delta_w};q)_l}~,
    \end{aligned}
\end{equation}
where $Q_l(\cos\theta|X,Y;q)$ are the Al-Salam-Chihara polynomials \cite{al1976convolutions}, which are given by
\begin{equation}\label{eq:ASC pol}
    Q_l(\cos\theta|X,Y;q)=\frac{(XY;q)_l}{X^l}\sum_{n=0}^\infty\frac{(q^{-l},X \rme^{\pm\rmi\theta};q)_n}{(XY;q)_n}q^n~,
\end{equation}
and they satisfy $Q_{-1}=0$, $Q_{0}=1$ in the recurrence relation
\begin{equation}\label{eq:ASC recurrence relation}
\begin{aligned}
    2\cos\theta ~Q_n=Q_{n+1}+(X+Y)q^n Q_n+(1-XY q^{n-1})(1-q^n)Q_{n-1}~.
\end{aligned}
\end{equation}
It is known that the Fourier kernel of (\ref{eq:Fourier kernel Al Salam Chihara}) can be evaluated as \cite{askey1996general} (14.8):
\begin{equation}\label{eq:explicit kernel}
    \begin{aligned}
        &\sum_{l=0}^\infty q^{{\Delta_{\rm S}}l}\frac{Q_l(\cos\theta_1|q^{\Delta_w}\rme^{\pm\rmi\theta_3};q)Q_l(\cos\theta_2|q^{\Delta_w}\rme^{\pm\rmi\theta_4};q)}{(q,q^{2\Delta_w};q)_l}\\
    &=\frac{(q^{\Delta_{\rm S}}\rme^{-\rmi(\theta_3+\theta_4)},q^{{\Delta_{\rm S}}+\Delta_w}\rme^{\rmi(\theta_4\pm\theta_1)},q^{{\Delta_{\rm S}}+\Delta_w}\rme^{\rmi(\theta_3\pm\theta_2)};q)_\infty}{(q^{{\Delta_{\rm S}}+\Delta_w}\rme^{\rmi(\theta_3+\theta_4)},q^{\Delta_{\rm S}}\rme^{\rmi\qty(\pm\theta_1\pm\theta_2)};q)_\infty} \cdot\\
    &\quad \cdot{}_8W_7\qty(q^{{\Delta_{\rm S}}+2\Delta_w-1}\rme^{\rmi\qty(\theta_3+\theta_4)},q^{\Delta_{\rm S}}\rme^{\rmi\qty(\theta_3+\theta_4)},q^{\Delta_w}\rme^{\rmi\qty(\theta_1\pm\theta_3)},q\rme^{\rmi\qty(\theta_4\pm\theta_2)};q,q^{\Delta_{\rm S}}\rme^{-\rmi(\theta_3+\theta_4)})~,
    \end{aligned}
\end{equation}
where $_8W_7$ is a very-well-poised basic hypergeometric series
\begin{equation}\label{eq:very-well-poised basic hypergeometric series}
    {}_{r+1}W_r(a_1;a_4,\ldots,a_{r+1};q,z) =
\sum_{j=0}^\infty  \frac{1-a_1q^{2j}}{1-a_1}
\frac{(a_1,a_4,\ldots,a_{r+1};q)_jz^j }{(q,qa_1/a_4,\ldots,
qa_1/a_{r+1};q)_j}~.
\end{equation}
We can now combine 
(\ref{eq:set up GbetaLbetaR}, \ref{eq:Fourier kernel Al Salam Chihara},
\ref{eq:explicit kernel},
\ref{eq:very-well-poised basic hypergeometric series}) with the $_8W_7$ integral representation in (B.23) \cite{Berkooz:2018jqr} (also in \cite{gasper2011basic}).
The thermal two-point correlation function (\ref{eq:generating function 1 particle}) for the two-sided HH state (\ref{eq:HH state tL tR}) then takes the same form as (5.5) in  \cite{Berkooz:2018jqr} (i.e. a four-point function in $\mathcal{H}_{0}\otimes\mathcal{H}_{0}$ of the DSSYK model)\footnote{The connection between the two-point function in the one-particle chord space of the DSSYK model, and an four-point function in $\mathcal{H}^{L}_0\otimes\mathcal{H}_0^{R}$ has been noticed in \cite{Xu:2024gfm}; the point of this subsection is to derive (\ref{eq:Gdelta}) explicitly while allowing for two-sided finite temperature dependence.}
\begin{equation}\label{eq:Gdelta}
\begin{aligned}
{G}^{(\Delta_w)}_{\tilde{\Delta}}({\tau_L,\tau_R})=\frac{1}{Z_{\tilde{\Delta}}(\beta_L,\beta_R)}\int&\qty(\prod_{i=1}^4\rmd\mu(\theta_i)\rme^{-\tau_iE_i}) R_{\theta_2\theta_4}
^{(q)}\begin{pmatrix}
\theta_1&{\Delta_{\rm S}}\\
\theta_3&\Delta_w
\end{pmatrix} \cdot\\
& \cdot\gamma^{(q)}_{{\Delta_{\rm S}}}(\theta_1,\theta_2)\gamma^{(q)}_{{\Delta_{\rm S}}}(\theta_3,\theta_4)\gamma^{(q)}_{\Delta_w}(\theta_1,\theta_3)\gamma^{(q)}_{\Delta_w}(\theta_2,\theta_4)~,
\end{aligned}
\end{equation}
where we introduced a shorthand $E_i\equiv \frac{2J}{\lambda}\cos\theta_i$; and we also denoted 
\begin{equation}\label{eq:gammaq}
    \gamma^{(q)}_{\Delta}(\theta_L,\theta_R)\equiv\sqrt{\bra{\theta_L}q^{\Delta \hat{n}}\ket{\theta_R}}=\sqrt{\frac{(q^{2\Delta};q)_\infty}{(q^{\Delta}\rme^{\pm\rmi\theta_L\pm\rmi\theta_R};q)_\infty}}~.
\end{equation}
The R-matrix appearing in (\ref{eq:Gdelta}) is defined as \cite{Berkooz:2018jqr}
\begin{equation}\label{eq:Rmatrix}
\begin{split}
&R_{\theta_2\theta_4}
^{(q)}\begin{pmatrix}
\theta_1&\Delta_{\rm S}\\
\theta_2&\Delta_w
\end{pmatrix}= - \frac{1 }{(q;q)_{\infty } ^3 (1-q)^3} \frac{\Gamma _q(1+\Delta_{\rm S}-\Delta_w+\rmi y_1-\rmi y_4)\Gamma _q(\Delta_w-\Delta_{\rm S}+\rmi y_4-\rmi y_1)}{\Gamma (1+\Delta_{\rm S}-\Delta_w+\rmi y_1-\rmi y_4)\Gamma (\Delta_w-\Delta_{\rm S}+\rmi y_4-\rmi y_1)} \cdot \\
& \cdot \sqrt{\frac{\Gamma _q(\Delta_w-\rmi y_4\pm \rmi y_2) \Gamma _q(\Delta_w+\rmi y_1\pm \rmi y_3) \Gamma _q(\Delta_{\rm S}-\rmi y_1\pm \rmi y_2)\Gamma _q(\Delta_{\rm S}+\rmi y_4 \pm \rmi y_3)}{\Gamma _q(\Delta_w+\rmi y_4 \pm \rmi y_2) \Gamma _q(\Delta_w-\rmi y_1\pm \rmi y_3) \Gamma _q(\Delta_{\rm S}+\rmi y_1 \pm \rmi y_2) \Gamma _q(\Delta_{\rm S}-\rmi y_4\pm \rmi y_3)} } \cdot \\
& \cdot \int_{\mathcal{T}}  \frac{ds}{2\pi \rmi } q^{s-\Delta_{\rm S}+\rmi y_4-\rmi y_3} \frac{\Gamma (s+1-\Delta_{\rm S}+\rmi y_4-\rmi y_3)\Gamma (s+1-\Delta_w+\rmi y_1-\rmi y_3)}{\Gamma _q(s+1-\Delta_{\rm S}+\rmi y_4-\rmi y_3) \Gamma _q(s+1-\Delta_w+\rmi y_1-\rmi y_3)}  \cdot \\
&\quad \cdot \frac{\Gamma _q(s)\Gamma _q(s-2\rmi y_3)\Gamma _q(s+\rmi y_1+\rmi y_4-\rmi y_3 \pm \rmi y_2)\Gamma (\Delta_{\rm S}-s-\rmi y_4+\rmi y_3)\Gamma (\Delta_w-s+\rmi y_3-\rmi y_1)}{\Gamma _q(s+\Delta_w+\rmi y_1-\rmi y_3)\Gamma _q(s+\Delta_{\rm S}+\rmi y_4-\rmi y_3)}  ,
\end{split}
\end{equation}
where we denote $y_i=\qty(\pi-\theta_i)/\lambda$; the $\Gamma_q$ function is defined as
\begin{equation}
    \Gamma_q(x)\equiv(1-q)^{1-x}\frac{(q;q)_\infty}{(q^x;q)_\infty}~,
\end{equation}
and $\mathcal{T}$ is a contour in the complex plane passing through the imaginary axis of $s$, which can be deformed to avoid the poles of the different $\Gamma$ functions in the integrand (see Fig. \ref{fig:polesgamma}).

\subsection{Semiclassical evaluation for heavy or light operators}
So far, the expressions are valid for arbitrary $q\in[0,1)$, however, we are interested in the semiclassical regime $\lambda\rightarrow0$, and possibly in adding quantum corrections. {{Following \cite{Goel:2023svz}, we adopt an ansatz to evaluate the semiclassical correlation function (\ref{eq:generating function 1 particle}) when  $\Delta_w$ (appearing in $\gamma^{(q)}_{\Delta_w}(\theta_1,\theta_3)$, $\gamma^{(q)}_{\Delta_w}(\theta_2,\theta_4)$) is a fixed number as $\lambda\rightarrow0$}}: 
\begin{align}\label{eq:imp ansats}
    &\theta_1:=\theta_L+\lambda~\alpha_L~,\quad \theta_3:=\theta_L-\lambda~\alpha_L~,\\
    &\theta_2:=\theta_R+\lambda~\alpha_R~,\quad \theta_4:=\theta_R-\lambda~\alpha_R~,
\end{align}
{{where in this ansatz $\theta_{L/R}$, $\alpha_{L/R}$ are $\mathcal{O}(1)$ parameters as $\lambda\rightarrow0$.}} To leading order, the measure of integration in (\ref{eq:Gdelta}) can be then expressed
\begin{equation}
    \prod_{i=1}^4\rmd\mu(\theta_i)\eqlambda{16\pi^2}\rmd\theta_L\rmd\theta_R\rmd\alpha_L\rmd\alpha_R\rme^{2\qty(S(\theta_L)+S(\theta_R))}(4\sin\theta_L\sin\theta_R)^2~,
\end{equation}
where $S(\theta)=-\frac{2}{\lambda}\qty(\theta-\frac{\pi}{2})^2$, and we applied the relation (\ref{eq:approx (x;q)infty}).
Meanwhile, the Boltzmann weight in (\ref{eq:Gdelta}) becomes
\begin{equation}
    \rme^{-\sum_{i=1}^4\tau_iE(\theta_i)}=\rme^{-\beta_LE(\theta_L)-\beta_RE(\theta_R)+2\sum_{i=L,R}\alpha_i\sin\theta_i(2\tau_i-\beta_i)}~,
\end{equation}
where the relation between $\tau_{1\leq i\leq 4}$ with $t_{L/R}$ and $\beta_{L/R}$ is given in (\ref{eq:tau choices}).

We move on with the product of $\gamma^{(q)}_\Delta$ functions (\ref{eq:gammaq}):
\begin{equation}
    \begin{aligned}
        &\gamma^{(q)}_{{\Delta_{\rm S}}}(\theta_1,\theta_2)\gamma^{(q)}_{{\Delta_{\rm S}}}(\theta_3,\theta_4)\gamma^{(q)}_{\Delta_w}(\theta_1,\theta_3)\gamma^{(q)}_{\Delta_w}(\theta_2,\theta_4)\\
        &=\sqrt{\frac{(q^{2\Delta_{\rm S}};q)^2_\infty(q^{2\Delta_{w}};q)^2_\infty}{(q^{\Delta_{\rm S}\pm\rmi\alpha_L\pm\rmi\alpha_R}\rme^{\rmi\qty(\pm\theta_L\pm\theta_R)},q^{\Delta_w}\rme^{\pm2\rmi\theta_L},q^{\Delta_w}\rme^{\pm\rmi2\theta_R},q^{\Delta_w\pm2\rmi\alpha_L},q^{\Delta_w\pm2\rmi\alpha_R};q)_\infty}}\\
        &\eqlambda\mathcal{N}\frac{\sqrt{\Gamma(\Delta_w\pm2\rmi\alpha_L)\Gamma(\Delta_w\pm2\rmi\alpha_R)}}{\Gamma(2\Delta_w)}(q^{2\Delta_{\rm S}};q)_\infty\rme^{\frac{\pi^2}{6\lambda}-\frac{S(\theta_L)+S(\theta_R)}{2}}\cdot\\
        &\quad\cdot\rme^{\frac{1}{2\lambda}{\rm Li}_2\qty(q^{\Delta_{\rm S}}\rme^{\rmi(\pm\theta_L\pm\theta_R)})}\qty(4\sin(\theta_L)\sin(\theta_R))^{\Delta_w-1/2}\cdot\\
        &\quad\cdot\qty(\qty(1+q^{2\Delta_{\rm S}}-2\cos(\theta_L+\theta_R))\qty(1+q^{2\Delta_{\rm S}}-2\cos(\theta_L-\theta_R)))^{\Delta_{\rm S}-1/2}~,
    \end{aligned}
\end{equation}
where in the last equality we using the same notation as in (\ref{eq:partition one particle semiclassical}) for ${\rm Li_2}(\rme^{\pm x\pm y})$. We also applied (\ref{eq:approx (x;q)infty}), and the following expansions \cite{Blommaert:2024ymv,Xu:2024gfm};
\begin{equation}\label{eq:useful expansions}
\begin{aligned}
{\rm Li_2}(x(1-\lambda y))&={\rm Li_2}(x)+\lambda y\log(1-x)+\mathcal{O}(\lambda^2)~,\\
    \frac{(q^{2\Delta};q)_\infty}{(q^\Delta\rme^{\pm\rmi\theta};q)_\infty}&\eqlambda \mathcal{N}\frac{\Gamma(\Delta\pm\rmi\theta)}{\Gamma(2\Delta)}~.
\end{aligned}
\end{equation}
Note that $\mathcal{N}$ is a numerical factor (i.e. independent of $\Delta$ and $\theta$).

On the other hand, the R-matrix in (\ref{eq:generating function 1 particle}) with the Ansatz (\ref{eq:imp ansats}) at leading order becomes
\begin{equation}\label{eq:R matrix important}
\begin{aligned}
    &R_{\theta_2\theta_4}
^{(q)}\begin{pmatrix}
\theta_1&\Delta_{\rm S}\\
\theta_2&\Delta_w
\end{pmatrix}\eqlambda-\frac{1}{(q;q)_{\infty }^3 (1-q)^3}\prod_{i=L,R}\sqrt{\frac{\Gamma(\Delta_w-2\rmi\alpha_i)\Gamma_q(\Delta_w-\frac{2\rmi}{\lambda}\qty(\pi-\theta_i))}{\Gamma(\Delta_w+2\rmi\alpha_i)\Gamma_q(\Delta_w+\frac{2\rmi}{\lambda}\qty(\pi-\theta_i))}}\cdot\\
&~\cdot\sqrt{\frac{\Gamma_q(\Delta_{\rm S}+\frac{\rmi}{\lambda}(\theta_L-\theta_R)\pm\rmi(\alpha_L-\alpha_R))\Gamma_q(\Delta_{\rm S}\pm\frac{\rmi}{\lambda}(2\pi-\theta_L-\theta_R)+\rmi(\alpha_L+\alpha_R))}{\Gamma_q(\Delta_{\rm S}-\frac{\rmi}{\lambda}(\theta_L-\theta_R)\pm\rmi(\alpha_L-\alpha_R))\Gamma_q(\Delta_{\rm S}\pm\frac{\rmi}{\lambda}(2\pi-\theta_L-\theta_R)-\rmi(\alpha_L+\alpha_R))}}\cdot\\
&~\cdot\int_{\mathcal{T}}\rmd s~q^{s-\Delta_{\rm S}}\rme^{\rmi(\theta_L-\theta_R)}\frac{\Gamma(-s+\Delta_w+2\rmi\alpha_L)\Gamma_q\qty(-s+\Delta_{\rm S}+\rmi\qty(\alpha_L-\alpha_R)-\frac{\rmi}{\lambda}\qty(\theta_L-\theta_R))}{\Gamma(s+\Delta_w-2\rmi\alpha_L)\Gamma_q(s+\Delta_{\rm S}-\rmi(\alpha_L-\alpha_R)+\frac{\rmi}{\lambda}\qty(\theta_L-\theta_R))}\cdot\\
&~\cdot~~\Gamma_q(s)\Gamma_q\qty(s-\frac{2\rmi}{\lambda}(\pi-\theta_L)-2\rmi\alpha_L)\Gamma_q\qty(s+\frac{2\rmi}{\lambda}(\pi-\theta_R)-2\rmi\alpha_L)\Gamma(2\rmi(\alpha_L-\alpha_R))
~,
\end{aligned}
\end{equation}
where we have not yet taken the $\lambda\rightarrow0$ limit in the $\Gamma_q$ functions whose argument contains $\rmi\theta/\lambda$ terms. Before doing that, we should specify the scaling of $\Delta_{\rm S}$ as $\lambda\rightarrow0$. We distinguish two cases:
\begin{equation}\label{eq:heavy and light}
\begin{aligned}
    &\text{Heavy operators}~:\quad q^{\Delta_{\rm S}}=\rme^{-\lambda \Delta_{\rm S}}~\text{is fixed as}~\lambda\rightarrow0~;\\
   &\text{Light operators}~:\quad {\Delta_{\rm S}}~\text{is fixed as}~\lambda\rightarrow0~.
\end{aligned}
\end{equation}
There are useful relations that can help us to perform the evaluation of the $\Gamma_q(\rmi x/\lambda+y)$ terms in (\ref{eq:R matrix important}) as $\lambda\rightarrow0$. For instance, consider (A.5) in \cite{Mukhametzhanov:2023tcg}
\begin{equation}\label{eq:ratio Gammas q}
    \frac{\Gamma_q(\frac{\rmi x}{\lambda}+y_1)}{\Gamma_q(\frac{\rmi x}{\lambda}+y_2)}\eqlambda\qty(\frac{1-\rme^{-\rmi x}}{\lambda})^{y_1-y_2}~,
\end{equation}
with $x$, $y_{1,2}$ being fixed parameters as $\lambda\rightarrow0$. The necessary terms to simplify the ratio (\ref{eq:ratio Gammas q}) are not present in (\ref{eq:R matrix important}). One could multiply and divide by appropriate factors like $\Gamma_q(\pm\rmi\frac{\theta}{\lambda})$ to get a similar expression as (\ref{eq:ratio Gammas q}); however this produces additional terms that do not cancel out. Thus, we need an ansatz to be able to perform the full evaluation of the semiclassical R-matrix when $\hat{\mathcal{O}}_{\tilde{\Delta}}$ is a heavy composite operator.\footnote{{{In principle, one could proceed analogously to Sec. 5.3 in \cite{Goel:2023svz}.}}} Nevertheless, we have a prediction about the exact answer we should recover, i.e. (\ref{eq:generating function explicit DSSYK 1 particle}) (also included in App. \ref{app:results}). In the future, we would like to find the correct ansatz for $\theta_{L/R}$ that confirms our answer from classical phase space methods (See Sec. \ref{ssec:outlook}). We will now treat the light operator case in (\ref{eq:heavy and light}), where we indeed know what the correct ansatz is. We will show how to consistently recover the expected answer in the $\lambda\rightarrow0$ limit.

\subsection{Exact evaluation for light operators}
Once we let $\hat{\mathcal{O}}_{\tilde{\Delta}}$ be a single or composite light operator, {{we use a special case of the ansatz (\ref{eq:gammaq}) where we keep all the angles close to each other as $\lambda\rightarrow0$,}}
\begin{align}\label{eq:important ansats}
    &\theta_1=\theta+\lambda\qty(\alpha_L+\chi)~,\quad \theta_3=\theta+\lambda\qty(\alpha_L-\chi)~,\\
    &\theta_2=\theta+\lambda\qty(\alpha_R+\chi)~,\quad \theta_4=\theta+\lambda\qty(\alpha_R-\chi)~,
\end{align}
where $\theta,~\chi,~\alpha_{L/R}$ are parameters independent of $\lambda$. {{{In the following we verify that this ansatz reproduces the expected result for the two-sided two-point function with a light operator at leading order in the semiclassical limit.}}} We can proceed as we discussed in the more general case, including the Jacobian determinant, and using the approximations (\ref{eq:approx (x;q)infty}, \ref{eq:approx (x;q)infty}). We see that the term in the measure of integration together with the Boltzmann weight in parenthesis of (\ref{eq:Gdelta}) can be expressed as
\begin{equation}\label{eq:Jacobian determinant}
\begin{aligned}
    \prod_{i=1}^4\rmd\mu(\theta_i)\rme^{-\tau_iE_i}=&64\pi^2\lambda\rmd\theta\rmd\alpha_L\rmd\alpha_R\rmd\chi~(2\sin\theta)^4\cdot\\
    &\cdot\rme^{4 S(\theta)-\qty(\beta_L+\beta_R)E(\theta)+2J\sin\theta\qty(\chi(\beta_L-\beta_R)+\alpha_L\qty(2\tau_L-\beta_L)+\alpha_R\qty(2\tau_R-\beta_R))}~.
\end{aligned}
\end{equation}
The (arguably) next simplest term to evaluate is the product of $\gamma^{(q)}_{\Delta}(\theta_L,\theta_R)$ functions in (\ref{eq:Gdelta})
\begin{equation}\label{eq:prod gammas lambda small}
\begin{aligned}
    &\gamma^{(q)}_{{\Delta_{\rm S}}}(\theta_1,\theta_2)\gamma^{(q)}_{{\Delta_{\rm S}}}(\theta_3,\theta_4)\gamma^{(q)}_{\Delta_w}(\theta_1,\theta_3)\gamma^{(q)}_{\Delta_w}(\theta_2,\theta_4)\\
    &=\sqrt{\frac{(q^{2\Delta_{\rm S}};q)^2_\infty(q^{2\Delta_w};q)^2_\infty}{(q^{\Delta_{\rm S}\pm\rmi(\alpha_L+\alpha_R)}\rme^{\pm2\rmi\theta},q^{{\Delta_{\rm S}}\pm\rmi(\alpha_L-\alpha_R)\pm2\rmi\chi},q^{\Delta_w\pm2\rmi\chi}\rme^{\pm2\rmi\theta},q^{\Delta_w\pm2\rmi\alpha_R},q^{\Delta_w\pm2\rmi\alpha_L};q)_\infty}}\\
    &\eqlambda\mathcal{N}(\sin\theta)^{2(\Delta_{\rm S}+\Delta_w-1)}\rme^{-2S(\theta)}\tfrac{\sqrt{\Gamma(\Delta_{\rm S}\pm\rmi(\alpha_L-\alpha_R)\pm2\rmi\chi)\Gamma(\Delta_w\pm2\rmi\alpha_L)\Gamma(\Delta_w\pm2\rmi\alpha_R)}}{\Gamma(2\Delta_{\rm S})\Gamma(2\Delta_w)}~,
\end{aligned}
\end{equation}
where we used (\ref{eq:approx (x;q)infty}, \ref{eq:approx (x;q)infty}, \ref{eq:useful expansions}) in the last step.

We proceed with the evaluation of the R-matrix (\ref{eq:Rmatrix}) (which we simplified in  (\ref{eq:R matrix important})) using the ansatz (\ref{eq:important ansats}). In particular, the square root term is relatively straightforward; so we will first threat the integral on the complex plane,
\begin{equation}\label{eq:int prod Gamma functions}
\begin{aligned}
    &\int_{\mathcal{T}}\frac{\rmd s}{2\pi\rmi}q^{s-\Delta_{\rm S}-\rmi(\alpha_L-\alpha_R-2\chi)}\Gamma_q\qty(s\pm\frac{2\rmi}{\lambda}\qty(\pi-\theta)-2\rmi\qty(\alpha_L-\chi))\Gamma(s-2\rmi\qty(\alpha_L-\alpha_R))\cdot\\
    &\qquad\quad\cdot\Gamma(s)\frac{\Gamma(-s+\Delta_{\rm S}+\rmi\qty(\alpha_L-\alpha_R-2\chi))\Gamma(-s+\Delta_w+2\rmi\alpha_L))}{\Gamma(s+\Delta_w-2\rmi\alpha_L)\Gamma(s+\Delta_{\rm S}-\rmi\qty(\alpha_L-\alpha_R-2\chi))}~,
\end{aligned}
\end{equation}
where $\mathcal{T}$ is a contour in the imaginary axis of $s$. We need to identify the poles of the $\Gamma$ functions, which follow from the Weierstrass form
\begin{equation}\label{eq: W form}
    \Gamma(x)=\frac{\rme^{-\gamma_{EM}x}}{x}\prod_{n=1}^\infty\qty(\qty(1+\frac{x}{n})^{-1}\rme^{x/n})~,
\end{equation}
with $\gamma_{EM}$ being the Euler-Mascheroni constant; so that we only need to threat simple poles at $x=\mathbb{Z}_{\leq0}$. 

The product of $\Gamma$ functions in (\ref{eq:int prod Gamma functions}) has several poles, as illustrated in Fig. \ref{fig:polesgamma}. The dominant ones come from
\begin{equation}
    s=0~,\quad s=2\rmi(\alpha_L-\alpha_R)~.
\end{equation}
Meanwhile the poles at $s\rightarrow\mathbb{R}\pm\rmi\infty$ (as $\lambda\rightarrow0$), or with $n\neq0$ in (\ref{eq: W form}) are suppressed contributions since there are more $\Gamma$ terms in the numerator than in the denominator (as discussed in more detail in App. B of \cite{Mertens:2017mtv} for JT gravity).
\begin{figure}
\centering
\includegraphics[width=0.4\textwidth]{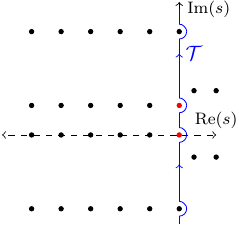}
\caption{Illustration of the poles (red and black dots) of the Gamma functions involved in the R-matrix integral (\ref{eq:int prod Gamma functions}) in the complex plane $s$, where the dominant poles at $s=0$ and $s=i(\alpha_R-\alpha_L)$ are showed in red. Figure based on \cite{Mertens:2017mtv}.}
\label{fig:polesgamma}
\end{figure}

Thus, we recover the total R-matrix in $\lambda\rightarrow0$ regime as
\begin{equation}\label{eq:integral Gammas Rmatrix}
\begin{aligned}
    &R_{\theta_2\theta_4}
^{(q)}\begin{pmatrix}
\theta_1&\Delta_{\rm S}\\
\theta_2&\Delta_w
\end{pmatrix}\eqlambda-\frac{\rme^{\frac{\pi^2}{2\lambda}}}{(2\pi\lambda)^{3/2}}\sqrt{\frac{\Gamma_q(\Delta_w\pm\frac{2\rmi}{\lambda}(\pi-\theta)-2\rmi\chi)}{\Gamma_q(\Delta_w\pm\frac{2\rmi}{\lambda}(\pi-\theta)+2\rmi\chi)}\prod_{i=L,R}\frac{\Gamma(\Delta_w-2\rmi\alpha_i)}{\Gamma(\Delta_w+2\rmi\alpha_i)}}\cdot\\
&\cdot
\sqrt{\frac{\Gamma(\Delta_{\rm S}\pm\rmi\qty(\alpha_L-\alpha_R+2\chi))\Gamma_q(\Delta_{\rm S}\pm\frac{2\rmi}{\lambda}(\pi-\theta)+\rmi(\alpha_L+\alpha_R))}{\Gamma(\Delta_{\rm S}\pm\rmi\qty(\alpha_L-\alpha_R-2\chi))\Gamma_q(\Delta_{\rm S}\pm\frac{2\rmi}{\lambda}(\pi-\theta)-\rmi(\alpha_L+\alpha_R))}}\biggl(\Gamma\qty(2\rmi(\alpha_R-\alpha_L))\cdot\\
&\cdot\frac{\Gamma_q\qty(\pm\frac{2\rmi}{\lambda}(\pi-\theta)-2\rmi(\alpha_L-\chi))\Gamma(\Delta_{\rm S}+\rmi\qty(\alpha_L-\alpha_R-2\chi))\Gamma(\Delta_w+2\rmi\alpha_L)}{\Gamma(\Delta_{\rm S}-\rmi\qty(\alpha_L-\alpha_R-2\chi))\Gamma(\Delta_w-2\rmi\alpha_L)}+\qty(L\leftrightarrow R)\biggr)~.
\end{aligned}
\end{equation}
Thus, combining our results thus far, the thermal correlation function can be expressed
\begin{equation}\label{eq:important intermediate integral}
    \begin{aligned}
        &\expval{q^{\Delta_w\hat{N}}}=-\mathcal{N}\int\rmd\theta\rmd\alpha_L\rmd\alpha_R\rmd\chi(2\sin\theta)^{2(\Delta_{\rm S}+\Delta_w+1)}\rme^{2S(\theta)-(\beta_L+\beta_R)E(\theta)+2J\chi\sin\theta\qty(\beta_L-\beta_R)}\cdot\\
    &\cdot\rme^{2J\sin\theta\sum_{i=L,R}\alpha_i(2\tau_i-\beta_i)}\qty(\prod_{i=L,R}\Gamma(\Delta_w-2\rmi\alpha_i))\sqrt{\frac{\Gamma(\Delta_{\rm S}\pm\rmi\qty(\alpha_L-\alpha_R)\pm2\rmi\chi)}{\Gamma(2\Delta_w)^2\Gamma(2\Delta_{\rm S})^2}}\cdot\\
    &\cdot\sqrt{\frac{\Gamma_q(\Delta_w\pm\frac{2\rmi(\pi-\theta)}{\lambda}-2\rmi\chi)\Gamma(\Delta_{\rm S}\pm\rmi(\alpha_L-\alpha_R)+2\rmi\chi)\Gamma(\Delta_{\rm S}\pm\frac{2\rmi}{\lambda}(\pi-\theta)+\rmi(\alpha_L+\alpha_R))}{\Gamma_q(\Delta_w\pm\frac{2\rmi(\pi-\theta)}{\lambda}+2\rmi\chi)\Gamma(\Delta_{\rm S}\pm\rmi(\alpha_L-\alpha_R)-2\rmi\chi)\Gamma(\Delta_{\rm S}\pm\frac{2\rmi}{\lambda}(\pi-\theta)-\rmi(\alpha_L+\alpha_R))}}\cdot\\
&\cdot\biggl(\frac{\Gamma\qty(2\rmi(\alpha_R-\alpha_L))\Gamma_q\qty(\pm\frac{2\rmi(\pi-\theta)}{\lambda}-2\rmi(\alpha_L-\chi))\Gamma(\Delta_{\rm S}+\rmi\qty(\alpha_L-\alpha_R-2\chi))\Gamma(\Delta_w+2\rmi\alpha_L)}{\Gamma(\Delta_{\rm S}-\rmi\qty(\alpha_L-\alpha_R-2\chi))\Gamma(\Delta_w-2\rmi\alpha_L)}\\
&\qquad+\qty(L\leftrightarrow R)\biggr)~.
    \end{aligned}
\end{equation}
Since $\Gamma(2\rmi(\alpha_L-\alpha_R))$ has a simple pole when $\alpha_L=\alpha_R$ (besides the $n\neq0$ poles in (\ref{eq: W form}) which are subleading contributions in (\ref{eq:important intermediate integral})), we can perform either of the $\alpha_L$ or $\alpha_R$ integrals along its real line from the residue theorem, {{e.g.
\begin{equation}\label{eq:residue new}
\begin{aligned}
    \int_{-\infty}^\infty\rmd\alpha_{R}~\Gamma(2\rmi(\alpha_R-\alpha_L))=&2\pi\rmi\lim_{\alpha_R\rightarrow\alpha_L}\qty[(\alpha_R-\alpha_L)\Gamma(2\rmi(\alpha_R-\alpha_L))]=\pi~,
\end{aligned}
\end{equation}}}
and we relabel the remaining variable $\alpha_{R/L}\rightarrow\alpha$. Thus, (\ref{eq:important intermediate integral}) then transforms into
\begin{equation}\label{eq:important intermediate integral2}
    \begin{aligned}
        \expval{q^{\Delta_w\hat{N}}}=&\mathcal{N}\int\rmd\theta\rmd\alpha\rmd\chi(2\sin\theta)^{2(\Delta_{\rm S}+\Delta_w+1)}\cdot\\
    &\cdot\rme^{2S(\theta)-(\beta_L+\beta_R)E(\theta)+2J\chi\sin\theta\qty(\beta_L-\beta_R)+2J\sin\theta\alpha(2(\tau_L+\tau_R)-\beta_L-\beta_R)}\cdot\\
    &\cdot\frac{\Gamma(\Delta_w\pm2\rmi\alpha)\Gamma(\Delta_{\rm S}\pm2\rmi\chi)}{\Gamma(2\Delta_w)\Gamma(2\Delta_{\rm S})}\Gamma_q\qty(\pm\frac{2\rmi(\pi-\theta)}{\lambda}-2\rmi(\alpha-\chi))\cdot\\
    &\cdot\sqrt{\frac{\Gamma_q(\Delta_w\pm\frac{2\rmi(\pi-\theta)}{\lambda}-2\rmi\chi)\Gamma_q(\Delta_{\rm S}\pm\frac{2\rmi(\pi-\theta)}{\lambda}+2\rmi\alpha)}{\Gamma_q(\Delta_w\pm\frac{2\rmi(\pi-\theta)}{\lambda}+2\rmi\chi)\Gamma_q(\Delta_{\rm S}\pm\frac{2\rmi(\pi-\theta)}{\lambda}-2\rmi\alpha)}}~.
    \end{aligned}
\end{equation}
{{where we have absorbed the residue (\ref{eq:residue new}) in the overall normalization $\mathcal{N}$.}} We can significantly simplify the remaining terms using the leading order relation (\ref{eq:ratio Gammas q}),
\begin{equation}
    \begin{aligned}
        &\frac{\Gamma_q(\Delta_w\pm\frac{2\rmi(\pi-\theta)}{\lambda}-2\rmi\chi)\Gamma_q(\Delta_{\rm S}\pm\frac{2\rmi}{\lambda}(\pi-\theta)+2\rmi\alpha)}{\Gamma_q(\Delta_w\pm\frac{2\rmi}{\lambda}(\pi-\theta)+2\rmi\chi)\Gamma_q(\Delta_{\rm S}\pm\frac{2\rmi}{\lambda}(\pi-\theta)-2\rmi\alpha)}\eqlambda\qty(\frac{2\sin\theta}{\lambda})^{4\rmi(\gamma-\alpha)}~,\\
        &\Gamma_q\qty(\pm\frac{2\rmi(\pi-\theta)}{\lambda}-2\rmi(\alpha-\chi))=\qty(\frac{2\sin\theta}{\lambda})^{4\rmi(\alpha-\gamma)}{\Gamma_q\qty(\pm\frac{2\rmi(\pi-\theta)}{\lambda})}~,
    \end{aligned}
\end{equation}
where we multiplied and divided by $\Gamma_q\qty(\pm\frac{2\rmi(\pi-\theta)}{\lambda})$. Using the definition of the $\Gamma_q$ function in (\ref{eq: Gamma q}), we can express
\begin{equation}
   \Gamma_q\qty(\pm\frac{2\rmi(\pi-\theta)}{\lambda})=\frac{(q;q)_\infty^3(1-q)^2}{(q,\rme^{\pm2\rmi(\pi-\theta)};q)_\infty}~.\label{eq: Gamma q}
\end{equation}
Now, we can recognize the term in the denominator as the measure of integration in the energy basis (\ref{eq:norm theta}) (up to a factor $2\pi$). From (\ref{eq:approx (x;q)infty}) the semiclassical limit of (\ref{eq: Gamma q}) reduces to 
\begin{equation}
   \Gamma_q\qty(\pm\frac{2\rmi(\pi-\theta)}{\lambda})\eqlambda\frac{\pi\lambda\rme^{-\frac{\pi^2}{2\lambda}-S(\theta)}}{\sin\theta}~.
\end{equation}
Thus, combining all terms in (\ref{eq:important intermediate integral2}), we get
\begin{equation}
\begin{aligned}\label{eq:almost evaluated correlator}
    \expval{q^{\Delta_w\hat{N}}}=\mathcal{N}&\int\rmd\theta\rmd\alpha\rmd\chi(2\sin\theta)^{2(\Delta_{\rm S}+\Delta_w)+1}\frac{\Gamma(\Delta_w\pm2\rmi\alpha)\Gamma(\Delta_{\rm S}\pm2\rmi\chi)}{\Gamma(2\Delta_w)\Gamma(2\Delta_{\rm S})}\cdot\\
    &\cdot\rme^{S(\theta)-(\beta_L+\beta_R)E(\theta)+2J\chi\sin\theta\qty(\beta_L-\beta_R)+2J\sin\theta\alpha(2(\tau_L+\tau_R)-\beta_L-\beta_R)}~.
\end{aligned}
\end{equation}
We will perform the $\theta$ integral using saddle points methods, so that
\begin{equation}
\begin{aligned}\label{eq:almost done correlator2}
    &G_{\tilde{\Delta}}^{(\Delta_w)}(\tau_L,\tau_R)=\frac{\expval{q^{\Delta_w\hat{N}}}}{Z_{\tilde{\Delta}}(\beta_L,\beta_R)}=\frac{\expval{q^{\Delta_w\hat{N}}}}{\frac{\mathcal{N}}{4\pi}\int_0^\pi\rmd\theta(\sin\theta)^{2\Delta_S+1}\rme^{S(\theta)-(\beta_L+\beta_R)E(\theta)}}\\
    &=(2\sin\theta)^{2\Delta_w}2^{2\Delta_{\rm S}}4\pi\int{\rmd\alpha}\frac{\Gamma(\Delta_w\pm{2\rmi\alpha})}{\Gamma(2\Delta_w)}\rme^{4\rmi J\sin\theta\alpha(t_L+t_R)}\int\rmd\chi\frac{\Gamma(\Delta_{\rm S}\pm{2\rmi\chi})}{\Gamma(2\Delta_{\rm S})}~,
\end{aligned}
\end{equation}
where in the third equality we combined (\ref{eq:Partition function light fields final}) with (\ref{eq:almost evaluated correlator}). In the fourth equality we carried out an analytic continuation $\tau_{L/R}\rightarrow\rmi t_{L/R}+\beta_{L/R}/2$, and we noticed that the saddle point evaluation fixes $\beta_L=\beta_R=\beta(\theta)/2$, where $J\beta(\theta)=\frac{2\theta-\pi}{\sin\theta}$ (\ref{eq:entropy temp}). We know how to perform the remaining integrals, namely \cite{Mertens:2017mtv}
\begin{equation}
    \int_{-\infty}^\infty\rmd u\frac{\rme^{\tau u}\Gamma(\Delta\pm\rmi\frac{u}{2\pi})}{\Gamma(2\Delta)}=(2\cos(\pi\tau))^{-2\Delta}~,
\end{equation}
{which follows from (6.6) of \cite{Lam:2018pvp}.}

The final result from (\ref{eq:almost done correlator2}) after taking the $\lambda\rightarrow0$ limit is then given by
\begin{equation}\label{eq:result chord amplitude}
    G^{(\Delta_w)}_{\tilde{\Delta}}(\tau_L,\tau_R)\eqlambda\qty(\frac{\sin\theta}{\cosh\qty(J\sin\theta(t_L+t_R))})^{2\Delta_w}~.
\end{equation}
The explicit correlation function agrees with the classical phase space answer derived in (\ref{eq:generating function explicit DSSYK 1 particle}) in the expected regime of validity of the ansatz (\ref{eq:important ansats}) (for technical convenience), namely at leading order in the $\lambda\rightarrow0$ limit and with $\Delta_{\rm S}\sim\mathcal{O}(1)$ as $\lambda\rightarrow0$. This means that $\hat{\mathcal{O}}_{\tilde{\Delta}}$ is treated as a single or composite light operator. {In the future, it is important to develop techniques to recover the full result in the phase space formalism (\ref{eq:womrhole semiclassical answer many}) from the chord-diagram two-point correlation function, when $\hat{\mathcal{O}}_{\Delta_1},\dots\hat{\mathcal{O}}_{\Delta_m}$ is a heavy single or composite operator.} Nevertheless, the result (\ref{eq:result chord amplitude}) provides a non-trivial consistency check of (\ref{eq:womrhole semiclassical answer many}).

\section{Smooth wormholes}\label{app:conversion}
In this section, we search for the conditions for the two-sided HH state (\ref{eq:HH state tL tR}) or more general states, constructed from the double-scaled operators acting on $\ket{0}$, to have a smooth geometric description.

Two-sided two-point correlation function have been argued to be a useful measure of spacetime connectedness \cite{Berkooz:2022fso}. In terms of the one-particle chord space, we can recover these types of correlation functions, for instance from \cite{Xu:2024gfm},
\begin{equation}\label{eq:2point correlation function m particles}
    G^{(\Delta_w)}_{\tilde{\Delta}}(\tau_L,\tau_R)\equiv\frac{\bra{\Psi_{\tilde{\Delta}}(\tau_L,\tau_R)}q^{\Delta_w \hat{N}}\ket{\Psi_{\tilde{\Delta}}(\tau_L,\tau_R)}}{Z_{\tilde{\Delta}}(\beta_L,\beta_R)}~,
\end{equation}
where $\hat{N}$ is the total chord number in the $\mathcal{H}_1$ irrep (i.e. when $\hat{\mathcal{O}}_{\tilde{\Delta}}$ is a single or composite operator).

\paragraph{Conversion property}Consider two decoupled boundary theories that might be entangled between each other. It was argued in \cite{Berkooz:2022fso} that in general, one should study whether a simple operator can be automatically transported between each of the dual boundary theories. Intuitively, this indicates that simple quantum correlations can be used to describe sooth semiclassical geometries. Before deriving the criterion in the DSSYK case, we briefly explain the general definition of \emph{conversion property} in \cite{Berkooz:2022fso}.

To be more specific, consider a two-sided bulk system dual to a quantum system described by a density matrix made of operators in the left/right boundary theories,
\begin{equation}\label{eq:B density matrix}
    \hat{\rho}=\sum_{ab}C_{ab}\hat{\mathcal{O}}^a_L\hat{\mathcal{O}}^b_R~.
\end{equation}
Here $\hat{\mathcal{O}}^{L/R}_a$ are endomorphism acting on $\mathcal{H}_{L}^\dagger\otimes \mathcal{H}_{R}$ respectively.\footnote{These are abstract Hilbert spaces in this discussion do not need to be related to the chord Hilbert space.} The authors in \cite{Berkooz:2022fso} noted that an operator from one of the sides can be converted to an equivalent operator on the opposite side. This means that (\ref{eq:B density matrix}) serves as a map of the type $\hat{\rho}$: $\hat{\mathcal{O}}^{R}\rightarrow \hat{\mathcal{O}}^{L}$.\footnote{It would be interesting to generalize this notion for discrete holography, e.g.\cite{Basteiro:2024crz}.} The conversion is most optimal (meaning that any simple operator can be transported from one side to the other one) when $C_{ab}=\delta_{ab}$, since
\begin{equation}\label{eq:conversion property}
    \hat{\mathcal{O}}^{L}=\sum_a\expval{\hat{\mathcal{O}}^{R}\hat{\mathcal{O}}^{R}_a}_R\hat{\mathcal{O}}^{L}_a~,
\end{equation}
\paragraph{Double-scaled case}We would like an equivalent criterion, specifically for the DSSYK model that can serve as a guiding principle for generating smooth wormhole geometries directly from the auxiliary chord space with particles $\mathcal{H}_m$. First notice that the operator basis of the two-sided HH state (\ref{eq:HH state tL tR}) is recovered from the double-scaled operators without the tracial state $\ket{0}$ as
\begin{equation}
    \hat{\rho}_{HH}=\rme^{-\tau_R^*\hH_L}\hat{\mathcal{O}}^{L}_{\Delta}\rme^{-\tau^*_L\hH_L}\rme^{-\tau_R\hH_R}\hat{\mathcal{O}}^{R}_{\Delta}\rme^{-\tau_L\hH_R}~.
\end{equation}
We can then define the expectation value acting only on one of the sides ($L/R$) by taking traces on the respective subspace, i.e.
\begin{subequations}
    \begin{align}\label{eq: new conversion}
        \expval{\hat{\mathcal{O}}_{\Delta}^{L}}_L&\equiv\Tr_L(\hat{\mathcal{O}}_{\Delta}^{L}~\hat{\rho}_{HH})\\
        &=\bra{0}\otimes\mathbb{1}\rme^{-\tau_L^*\hH_R}\hat{\mathcal{O}}_{\Delta}^{R}\rme^{-\tau_R^*\hH_R}\hat{\mathcal{O}}_{\Delta}^{L}\rme^{-\tau_L\hH_L}\hat{\mathcal{O}}_{\Delta}^{L}\rme^{-\tau_R\hH_L}\ket{0}\otimes\mathbb{1}\\
        &=\rme^{-\tau_L^*\hH_R}\hat{\mathcal{O}}_{\Delta}^{R}\rme^{-\tau_L^*\hH_R}\bra{0}\hat{\mathcal{O}}_{\Delta}^{L}\rme^{-\tau_L}\hat{\mathcal{O}}_{\Delta}^{L}\rme^{-\tau_R\hH_L}\ket{0},\label{eq: resulting property}
    \end{align}
\end{subequations}
Then, (\ref{eq: resulting property}) has the same structure as the conversion property for abstract operators in (\ref{eq:conversion property}) acting on a doubled Hilbert space without matter $\mathcal{H}_0^L\otimes\mathcal{H}_0^R$ through the isometric linear map $\hF$ (\ref{eq:isometric map}). This means that the two-sided HH PETS (\ref{eq:HH state tL tR}) may have a dual smooth wormhole interpretation (consistent with our findings in Sec. \ref{sec:bulk}). This schematically illustrated in Fig. \ref{fig:conversion}.
\begin{figure}
    \centering
    \subfloat[]{\includegraphics[height=0.33\textwidth]{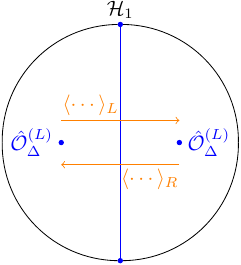}}\hfill\subfloat[]{\includegraphics[height=0.33\textwidth]{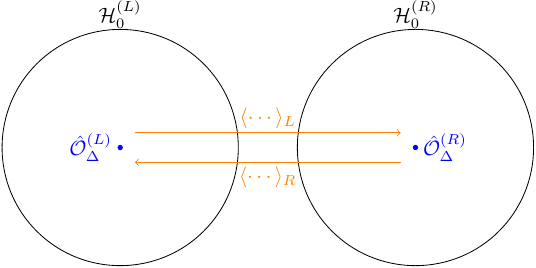}}
    \caption{Conversion property (\ref{eq: new conversion}) for operators $\hat{\mathcal{O}}^{L/R}_\Delta$in the double-scaled algebra acting on (a) $\mathcal{H}_1$, or (b) $\mathcal{H}_0^{L}\otimes\mathcal{H}_0^{R}$, which are related thought the isometric map $\hat{\mathcal{F}}$ (\ref{eq:isometric map}).}
    \label{fig:conversion}
\end{figure}

\section{Properties of the OTOC}\label{app:properties OTOC}
In this appendix we discuss different properties of the semiclassical OTOC (\ref{eq:twosided twopoint OTOC}) that we derived in Sec. \ref{ssec:twopoints} in relation with related literature.

\paragraph{A bound on OTOC}
It is interesting to compare our explicit results with bounds on the growth of OTOCs in the literature. In particular, \cite{Milekhin:2024vbb} found that, considering fermionic systems at infinite temperature, the following OTOCs can be bounded in terms of two-point functions at infinite temperature
\begin{align}\label{eq:MilenkinXu}
    \abs{\expval{\hat{\mathcal{O}}_\Delta(t)\hat{\mathcal{O}}_\Delta(0)\hat{\mathcal{O}}_\Delta(t)\hat{\mathcal{O}}_\Delta(0)}}\geq 2{G}^{(\Delta)}(t)^2-1~,\\
    \abs{\expval{\hat{\mathcal{O}}_\Delta(t)\hat{\mathcal{O}}_\Delta(0)\hat{\mathcal{O}}_\Delta(t)\hat{\mathcal{O}}_\Delta(0)}}\geq G^{(\Delta)}(t)^2~.\label{eq:MilenkinXu_other}
\end{align}
where $G^{(\Delta)}(t)$ is a two-point correlation function. The first and second inequalities above were denoted as weak and strong bounds on OTOC growth by \cite{Milekhin:2024vbb}.

One can explicitly check the strong version of the bound by inserting both the crossed four-point function (\ref{eq:generating function explicit DSSYK 1 particle}) at leading order in the $\lambda\rightarrow0$ limit, and the thermal two-point function $G^{(\Delta)}(t)$ (\ref{eq:correlator DSSYK}) in $\mathcal{H}_0$ into (\ref{eq:MilenkinXu_other}):
\begin{equation}
    \qty(\frac{1}{\cosh^2(Jt)-q^{\Delta_{\rm S}}\sinh^2(Jt)})^{2\Delta_w}\geq\qty(\frac{1}{\cosh^2(Jt)})^{2\Delta_w}~.
\end{equation}
The bound above is always explicitly satisfied since $q^{\Delta_{\rm S}}>0$. Similarly for (\ref{eq:MilenkinXu_other}),
\begin{equation}\label{eq:explicit weaker bound}
    \qty(\frac{1}{\cosh^2(Jt)-q^{\Delta_{\rm S}}\sinh^2(Jt)})^{2\Delta_w}\geq2\qty(\frac{1}{\cosh^2(Jt)})^{2\Delta_w}-1~.,\end{equation}
since both sides in (\ref{eq:explicit weaker bound}) are monotonically decreasing, and the bound is saturated at $t=0$.\footnote{\label{fnt:MX bound}One could check if the bound is no longer satisfied at late times once we add quantum corrections to (\ref{eq:generating function explicit DSSYK 1 particle}).} Thus, we see that the bound (\ref{eq:MilenkinXu_other}) is explicitly obeyed at all times in this system. Meanwhile, \cite{Milekhin:2024vbb} found that (\ref{eq:MilenkinXu_other}) holds at early times in generic fermionic systems (based on numerical evidence).

\paragraph{(Sub-)maximal chaos}On the other hand, we are interested in finite temperature effects in the evolution of the OTOC (\ref{eq:generating function explicit DSSYK 1 particle}). Considering the same time regimes as (\ref{eq:early KOC}, \ref{eq:late time COp}) for the OTOC (\ref{eq:generating function explicit DSSYK 1 particle}), we recover
\begin{equation}\label{eq:classical phase space OTOC}
\begin{aligned}
    &\eval{G_{\tilde{\Delta}}^{(\Delta_w)}(-\tau,\tau)}_{\tau=\frac{\beta(\theta)}{2}+\rmi t}\\
    &=\wick[offset=1.2em,sep=0.4em]{\langle\rme^{\frac{\beta(\theta)}{2}\hH_L}\hat{\mathcal{O}}_{\Delta_{\rm S}}(t_L)\rme^{-\frac{\beta(\theta)}{2}\hH_L}\c{\hat{\mathcal{O}}_{\Delta_w}}(0)\rme^{-\frac{\beta(\theta)}{2}\hH_L}\hat{\mathcal{O}}_{\Delta_{\rm S}}(t_L)\rme^{\frac{\beta(\theta)}{2}\hH_L}\c{\hat{\mathcal{O}}_{\Delta_w}}(0)\rangle}\\
    =&\qty(A(\theta,-q^\Delta)+B(\theta,-q^\Delta)\cosh(2J\sin\theta t))^{-2\Delta_w}\\
\simeq&\begin{cases}
    (A(\theta,-q^\Delta)+B(\theta,-q^\Delta))^{-2\Delta_w}\qty(1-\frac{\Delta_wB(\theta,-q^\Delta)}{A(\theta,-q^\Delta)+B(\theta,-q^\Delta)}\qty(\frac{2\pi}{{\beta}_{\rm fake}}~t)^2)~,&t\ll\tfrac{1}{J\sin\theta}~,\\
    A(\theta,-q^\Delta)^{-2\Delta_w}\qty(1-\frac{\Delta_wB(\theta,-q^\Delta)}{A(\theta,-q^\Delta)}\exp(\frac{2\pi}{{\beta}_{\rm fake}}t))~,&\tfrac{1}{J\sin\theta}\ll t\ll t_{\rm sc}~,\\
    \qty(\frac{1}{2}B(\theta,-q^\Delta))^{-2\Delta_w}\exp(-\frac{4\pi\Delta_wt}{{\beta}_{\rm fake}})~,&t\gg t_{\rm sc}~,
\end{cases}
\end{aligned}
\end{equation}
where $\beta_{\rm fake}=\frac{\pi}{J\sin\theta}$, and $A(\theta,q^\Delta)$, $B(\theta,q^{\Delta})$ and $t_{\rm sc}$ are in (\ref{eq:A theta}), (\ref{eq:B theta}) and (\ref{eq:scrambling time}) respectively. Let us compare with the chaos bound conjecture in \cite{Maldacena:2015waa}, which states that the maximal Lyapunov exponent of an OTOC in generic quantum systems is given by the factor $\frac{\beta}{2\pi}$.\footnote{\label{fnt:TTbar}It has been suggested that the chaos bound \cite{Maldacena:2015waa} could be violated in Krylov operator complexity by T$\overline{\text{T}}$ deformed conformal field theory (CFT) \cite{Chattopadhyay:2024pdj}. One might test if the bound on OTOC growth (\ref{eq:classical phase space OTOC}) might also be modified once we include the one-dimensional version of T$\overline{\text{T}}$ deformation in the DSSYK model \cite{Aguilar-Gutierrez:2024oea}.} In contrast, (\ref{eq:classical phase space OTOC}) is sub-maximally chaotic in the OTOC sense \cite{Maldacena:2015waa} when we compare the exponent at $t\in[(J\sin\theta)^{-1},~t_{\rm sc}]$ with respect to the physical temperature (\ref{eq:entropy temp}) $\beta(\theta)=(\frac{2\theta}{\pi}-1)\beta_{\rm fake}$. Similar observations for two-point functions in $\mathcal{H}_0$ appeared in \cite{Blommaert:2024ymv}.

\paragraph{Hyperfast growth}
We remark that  (\ref{eq:scrambling time}) displays a hyperfast growth, in the sense of \cite{Anegawa:2024yia} (and it also agrees with their estimate). Namely, the scrambling time of the DSSYK model with matter chords is independent of the size of the system, which would correspond to the total number of fermions in the original DSSYK Hamiltonian (\ref{eq:DDSSYK Hamiltonian}) (which is implicit in $q$). This can be contrasted with the fast-scrambling conjecture in \cite{Sekino:2008he}. It was argued that the fastest scrambling time for \emph{information recovery} scales is of the order $\log(S)$, where $S$ is the entropy of the system. One might argue that the definitions of scrambling time we are comparing are different and thus it is not surprising they can scale differently. However, as we have shown, the scrambling time in Krylov operator complexity (\ref{eq:Krylov complexity Operator many}) is determined by an OTOC in (\ref{eq:rule 1particle}).\footnote{See \cite{Milekhin:2023bjv} for hyperfast scrambling in the Brownian DSSYK.} This implies that the scrambling time in Krylov complexity, in this particular system, coincides with the scrambling time of an OTOC (defined as the time for the exponential growth to become small). Thus, the DSSYK model with matter behaves as a hyperfast scrambler of information in a closely related notion to the one proposed in \cite{Sekino:2008he}.

\section{Krylov basis and Lanczos coefficients in the one-particle chord space}\label{app:details Lanczos Krylov}
In this appendix, we fill in some of the technical details in the Lanczos algorithm (\ref{eq:Lanczos algorithm eta}). We find the Krylov basis and the Lanczos coefficients in the semiclassical limit, taking similar steps as \cite{Ambrosini:2024sre}.

\paragraph{Lanczos coefficients and the Krylov basis}
We begin proposing an Ansatz for the Krylov basis. Based on the work by \cite{Ambrosini:2024sre}, we search for a decomposition of the Krylov basis as $\ket{K_n^{(\eta)}}=\ket{k_n^{(\eta)}}+$additional state, where $\ket{k_n^{(\eta)}}$ appears in (\ref{eq:decomposition Krylov}), and it indeed solves the algorithm (\ref{eq:Lanczos eta}) exactly up to $n=3$. The additional state is determined by tridiagonalizing the Liouvillian $\mathcal{L}_\eta\propto(\hat{a}_R+\eta\hat{a}_L+\hat{a}^\dagger_R+\eta\hat{a}^\dagger_L)$ (\ref{eq:two-sided new}). For this reason, we evaluate
\begin{equation}
    \begin{aligned}
    \mathcal{L}_\eta\ket{k_n^{(\eta)}}=&\frac{J~c^{(\eta)}_n}{\sqrt{\lambda}}(\hat{a}_R+\eta\hat{a}_L+\hat{a}^\dagger_R+\eta\hat{a}^\dagger_L)\sum_{k=0}^n\eta^k\begin{pmatrix}
        n\\
        k
    \end{pmatrix}\ket{\Delta_{\rm S};k,n-k}~,
    \end{aligned}
\end{equation}
where
\begin{align}
    &(\hat{a}^\dagger_R+\eta\hat{a}^\dagger_L)\sum_{k=0}^n\eta^k\begin{pmatrix}
        n\\
        k
    \end{pmatrix}\ket{\Delta_{\rm S};k,n-k}=\sum_{k=0}^{n+1}\eta^k\begin{pmatrix}
        n+1\\
        k
    \end{pmatrix}\ket{\Delta_{\rm S};k,n+1-k}~,\\
    &(\hat{a}_R+\eta\hat{a}_L)\sum_{k=0}^n\eta^k\begin{pmatrix}
        n\\
        k
    \end{pmatrix}\ket{\Delta_{\rm S};k,n-k}=\sum_{k=0}^{n-1}Z^{(\eta)}_k(n)\eta^k\begin{pmatrix}
        n-1\\
        k
    \end{pmatrix}\ket{\Delta_{\rm S};k,n-1-k}\\
    &\text{with}\quad Z^{(\eta)}_k(n)=\frac{n[n-k]_q}{n-k}(1+\eta q^{\Delta_{\rm S}+k})+\frac{n[k+1]_q}{k+1}(1+\eta q^{\Delta_{\rm S}+n-1-k})~.\nonumber
\end{align}
In the first equality, we used $\begin{pmatrix}
        n\\
        k
    \end{pmatrix}+\begin{pmatrix}
        n\\
        k-1
    \end{pmatrix}=\begin{pmatrix}
        n+1\\
        k
    \end{pmatrix}$, and in the second one (\ref{eq:aLdagger,aL}, \ref{eq:aRdagger,aR}):
    \begin{equation}
        \hat{a}_R+\eta\hat{a}_L=\hat{\alpha}_L[n_L]_q(1+\eta q^{\Delta_{\rm S}+n_R})+\hat{\alpha}_R[n_R]_q(1+\eta q^{\Delta_{\rm S}+n_L})~.
    \end{equation}
    Collecting the relations, we can then write:
\begin{equation}
    \begin{aligned}\label{eq:almost tridiag}
        \mathcal{L}_\eta\ket{k_n^{(\eta)}}=&b_{n+1}\ket{k_{n+1}^{(\eta)}}+b_n\ket{k_{n-1}^{(\eta)}}\\
        &+\sum_{k=0}^{n-1}\qty(Z^{(\eta)}_k(n)-Z^{(\eta)}_{n/2}(n))\begin{pmatrix}
        n-1\\
        k
    \end{pmatrix}\eta^k\ket{\Delta_{\rm S};k,n-1-k}~,
    \end{aligned}
\end{equation}
where $\ket{k_{n-1}^{(\eta)}}=c_n^{(\eta)}\sum_{k=0}^n\eta^k\begin{pmatrix}
        n\\
        k
    \end{pmatrix}\ket{\Delta_{\rm S};k,n-k}$, and
\begin{equation}\label{eq:results Lanczos eta}
    b^{(\eta)}_n\eqlambda J\sqrt{Z^{(\eta)}_{n/2}(n)}~,\quad c_n^{(\eta)}={J^n}\prod_{i=1}^n(b_n^{(\eta)})^{-1}~.
\end{equation}
In the semiclassical limit (i.e. $\lambda n$ fixed as $\lambda\rightarrow0$ \cite{Lin:2022rbf,Lin:2023trc,Rabinovici:2023yex}), the binomial factor $\begin{pmatrix}
        n\\
        k
    \end{pmatrix}$ localizes at $k=n/2$ when $\eta=\pm1$; while for $\eta\in\mathbb{R}$ $k=\eta n/(1+\eta)$ \cite{Ambrosini:2024sre}. This means that the last term in (\ref{eq:almost tridiag}) is negligible compared to $\qty{\ket{k_n^{(\eta)}}}$ in the semiclassical limit when $\eta=\pm1$. For more general $\eta$, this term does not cancel out, and thus we would not solve the two-sided Lanczos algorithm. In principle, there might be a different ansatz instead of (\ref{eq:decomposition Krylov}) that one should use to work on the more general case.

Thus, from (\ref{eq:results Lanczos eta}) we find the Lanczos coefficients in the algorithm (\ref{eq:Lanczos eta})
\begin{equation}\label{eq:Lanczos many}
    a^{(\eta)}_n=0~,\quad b^{(\eta)}_n\eqlambda\frac{2J}{\sqrt{\lambda(1-q)}}\sqrt{(1-q^{n/2})(1+\eta q^{n/2+\Delta_{\rm S}})}~,
\end{equation}
and the normalization constants of $\ket{k_n^{(\eta)}}$,
\begin{equation}\label{eq:normalization Krylov}
\begin{aligned}
    \lim_{\lambda\rightarrow0}c^{(\eta)}_n=\lim_{\lambda\rightarrow0}\sqrt{\frac{\lambda^n(1-q)^n(1+\eta)}{2^n(q^{1/2};q^{1/2})_n(-\eta;q^{1/2})_{n+1}}}=\sqrt{\frac{\lambda^n}{n!(1+\eta)^n}}~,\quad \text{for }\eta\neq-1~.
\end{aligned}
\end{equation}
The case $\eta=-1$ is studied in \cite{Aguilar-Gutierrez:2025mxf}.

\paragraph{Generalized wavefunctions}
We would like to find explicit expressions for the wavefunctions of the form \footnote{This is equivalent to (\ref{eq:decomposition Krylov}) in the $\lambda\rightarrow0$ limit.}
\begin{equation}
    {\Psi}^{(\eta)}_n(\tau)=\bra{k^{(\eta)}_n}\rme^{-\mathcal{L}_\eta}\ket{\Psi_{\tilde{\Delta}}(\tau_L=\eta\tau,\tau_R=\tau)}~,
\end{equation}
which are the coefficients of the total chord number basis $\qty{\ket{k_{n}^{(\eta)}}}$ (\ref{eq:normal ordered total chord number}). In the $\lambda\rightarrow0$ limit, the Krylov space of the two-sided HH state (\ref{eq:HH state tL tR}) becomes:
\begin{align}
    \Psi_n^{(\eta)}(\tau)=&c_n^{(\eta)}\bra{\Psi_{\tilde{\Delta}}(\eta\tau,\tau)}\sum_k\eta^k\begin{pmatrix}
        n\\
        k
    \end{pmatrix}\ket{\Delta_{\rm S};k,n-k}\\
    =&c_n^{(\eta)}\sum_k\eta^k\begin{pmatrix}
        n\\
        k
    \end{pmatrix}\bra{\Delta_{\rm S};0,0}\rme^{-\tau(\hH_R+\eta\hH_L)}\ket{\Delta_{\rm S};k,n-k}\\
    =&c_n^{(\eta)}\int\rmd\mu(\theta_L)\rmd\mu(\theta_R)\frac{(q^{\Delta_{\rm S}};q)_\infty}{(q^{\Delta_{\rm S}}\rme^{ \pm\rmi\theta_L \pm\rmi\theta_R};q)_\infty}\rme^{-\tau(\eta E(\theta_L)+ E(\theta_R))}\cdot\label{eq:Explicit amplitude}\\
    &\qquad\cdot\sum_k\eta^k\begin{pmatrix}
        n\\
        k
    \end{pmatrix}\frac{H_k(\cos\theta_L|q)H_{n-k}(\cos\theta_R|q)}{\sqrt{(q;q)_k(q;q)_{n-k}}}~,\nonumber
\end{align}
where $c_n^{(\eta)}$ is a normalization constant (\ref{eq:normal ordered total chord number}), and we used (\ref{eq: proj E0 theta}) to express
\begin{equation}
(\theta_L,\theta_R|k,n-k)=\frac{H_k(\cos\theta_L|q)H_{n-k}(\cos\theta_R|q)}{\sqrt{(q;q)_k(q;q)_{n-k}}}~.    
\end{equation}
As a future direction, this form of the correlation function could be evaluated numerically in the $\lambda\rightarrow0$ limit to test our results from classical phase space methods (\ref{eq:womrhole semiclassical answer many}). Namely, to determine Krylov complexity for $\eta=\pm1$, as
\begin{equation}
\mathcal{C}^{(\eta)}(t)=\frac{\sum_nn\abs{\Psi_n^{(\eta)}(\frac{\beta(\theta)}{2})+\rmi t)}^2}{\sum_n\abs{\Psi_n^{\eta}(\frac{\beta(\theta)}{2})}^2}~.
\end{equation}

\section{Non-composite many particle chord states}\label{app:more general states}
In this appendix we construct PETS (\ref{eq:PETS general precursor}) for non-composite many particle chord operators acting on the tracial state $\ket{0}$, states on $\mathcal{H}_{m\in\mathbb{Z}_{\geq0}}$. This allows to generalize our discussion on composite operators acting on the tracial state to evaluate dynamical observables in thermal ensembles, such as (\ref{eq:HH state tL tR}). We are particularly interested in states of the form
\begin{equation}\label{eq:general PETS}
    \rme^{-\tau_0\hH_L}\hat{\mathcal{O}}_{\Delta_1}\rme^{-\tau_1\hH_L}\hat{\mathcal{O}}_{\Delta_2}\cdots \rme^{-\tau_{m-1}\hH_L}\hat{\mathcal{O}}_{\Delta_m}\rme^{-\tau_{m}\hH_L}\ket{0}~,
\end{equation}
to derive an explicit decomposition in the $\mathcal{H}_{0}$ and $\mathcal{H}_{1}$ irreps. 

We start from $\hat{\mathcal{O}}_{\Delta}=b^\dagger_\Delta+b_\Delta$ with $b^\dagger_\Delta$, $b_\Delta$ in (\ref{eq:property operator}, \ref{eq:property operator2}). The simplest non-trivial example of (\ref{eq:general PETS}) is $m=1$, where
\begin{equation}
    \ket{\Psi_{\Delta_1}(\tau_0,\tau_1)}=\rme^{-\tau_0\hH_L}\hat{\mathcal{O}}^{L}_{\Delta_1}\rme^{-\tau_1\hH_L}\ket{0}=\sum_{n_0,n_1}\phi_{n_0}(\tau_0)\phi_{n_1}(\tau_1)\ket{\Delta;n_0,n_1}~,
\end{equation}
where $\phi_n(\tau)=\bra{n}\rme^{-\tau\hH}\ket{0}$, and in the last step we used the isometric map $\hat{\mathcal{F}}$ (\ref{eq:isometric map}) (similar to (5.6) in \cite{Xu:2024gfm}). 

Moving on to $m=2$ case in (\ref{eq:general PETS}),
\begin{equation}
    \ket{\Psi_{\Delta_1,\Delta_2}(\tau_0,\tau_1,\tau_2)}=\rme^{-\tau_0\hH_L}\hat{\mathcal{O}}_{\Delta}\rme^{-\tau_1\hH_L}\hat{\mathcal{O}}_ {\Delta}\rme^{-\tau_2\hH_L}\ket{0}~.
\end{equation}
We need to distinguish between two cases. For $\Delta_1=\Delta_2=\Delta$:
    \begin{equation}\label{eq:Hm to H1}
\begin{aligned}
    &\ket{\Psi_{\Delta,\Delta}(\tau_0,\tau_1,\tau_2)}\\
    &=\sum_{n_1,n_2}\phi_{n_1}(\tau_1)\phi_{n_2}(\tau_2)\qty(q^{n_1}\rme^{-\tau_0\hH}\ket{n_1+n_2}+\sum_{n_0}\phi_{n_0}(\tau_0)\ket{\Delta^{(2)};n_0,n_1,n_2})~,\\
    &=\sum_{n_0,n_1,n_2=0}^\infty\biggl(q^{n_1}\phi^{(n_1+n_2)}_{n_0}(\tau_0)\ket{n_0}\\
    &\qquad+\phi_{n_1}(\tau_1)\phi_{n_2}(\tau_2)\sum_{l+m_L+m_R=n_1}\chi_{l, m_L, m_R}\ket{2\Delta +l;n_0+m_L,n_2+m_R}\biggr)~,
\end{aligned}
\end{equation}
where $\phi^{(n_1+n_2)}_{n_0}(\tau_0)\equiv\bra{n_0}\rme^{-\tau_0\hH}\ket{n_1+n_2}$, $\Delta^{(2)}=\qty{\Delta,\Delta}$, and we have applied (\ref{eq:LinStanfordIdentity}) in the last line. Meanwhile for $\Delta_1\neq\Delta_2$:
    \begin{equation}\label{eq:Hm to H1_neq}
\begin{aligned}
    \ket{\Psi_{\Delta_1,\Delta_2}(\tau_0,\tau_1,\tau_2)}=&\sum_{n_0,n_1,n_2}\phi_{n_0}(\tau_0)\phi_{n_1}(\tau_1)\phi_{n_2}(\tau_2)\ket{\Delta_1,\Delta_2;n_0,n_1,n_2}\\
    =\sum_{n_0,n_1,n_2=0}^\infty\phi_{n_0}(\tau_0)&\phi_{n_1}(\tau_1)\phi_{n_2}(\tau_2)\cdot\\
    \cdot\sum_{l+m_L+ m_R=n_1}&\chi_{l, m_L, m_R}\ket{\Delta_1+\Delta_2+l;m_L+n_0, m_R+n_2}~.
\end{aligned}
\end{equation}
Now that we have recovered an explicit form of the non-composite two particle PETS. One may continue recursively to express the most general double-scaled PETS (\ref{eq:PETS general precursor}) into the zero and one-particle irreps. The price to pay is the addition of the Clebsch–Gordan coefficients. A key advantage of our procedure is that the integration measure in energy basis is known for the zero and one-particle irreps. One can therefore use (\ref{eq:Hm to H1} or \ref{eq:Hm to H1_neq}) to explicitly evaluate partition functions and two (or higher)-point correlation functions (analogous to App. \ref{app:partition1p} and \ref{app:twopoint oneparticle} respectively), which we leave for future directions.

\bibliographystyle{JHEP}
\bibliography{references.bib}
\end{document}